\documentclass{article}

\usepackage{arxiv}

\usepackage[utf8]{inputenc} 
\usepackage[T1]{fontenc}    
\usepackage{hyperref}       
\usepackage{url}            
\usepackage{booktabs}       
\usepackage{amsfonts}       
\usepackage{nicefrac}       
\usepackage{microtype}      
\usepackage{lipsum}

\usepackage{amsmath,graphicx}
\usepackage{booktabs}
\usepackage{multirow}
\usepackage{color, soul}
\usepackage{url}
\usepackage{mathrsfs} 
\usepackage{amsmath}

\title{Generative-based Airway and Vessel Morphology Quantification on Chest CT Images}

\author{
	Pietro Nardelli \\
	Applied Chest Imaging Laboratory\\
	Brigham and Women's Hospital\\
	Harvard Medical School\\
	Boston, MA, 02115 USA \\
	\texttt{pnardelli@bwh.harvard.edu} \\
	\And
	James C. Ross \\
	Applied Chest Imaging Laboratory\\
	Brigham and Women's Hospital\\
	Harvard Medical School\\
	Boston, MA, 02115 USA \\
	\texttt{jross@bwh.harvard.edu} \\
	\And
	Ra\'{u}l San Jos\'{e} Est\'{e}par \\
	Applied Chest Imaging Laboratory\\
	Brigham and Women's Hospital\\
	Harvard Medical School\\
	Boston, MA, 02115 USA \\
	\texttt{rsanjose@bwh.harvard.edu} \\
}

\begin{document}
	\maketitle
	
	\begin{abstract}
		Accurately and precisely characterizing the morphology of small pulmonary structures from Computed Tomography (CT) images, such as airways and vessels, is becoming of great importance for diagnosis of pulmonary diseases. The smaller conducting airways are the major site of increased airflow resistance in chronic obstructive pulmonary disease (COPD), while accurately sizing vessels can help identify arterial and venous changes in lung regions that may determine future disorders. However, traditional methods are often limited due to image resolution and artifacts.
		
		We propose a Convolutional Neural Regressor (CNR) that provides cross-sectional measurement of airway lumen, airway wall thickness, and vessel radius. CNR is trained with data created by a generative model of synthetic structures which is used in combination with Simulated and Unsupervised Generative Adversarial Network (SimGAN) to create simulated and refined airways and vessels with known ground-truth.
		
		For validation, we first use synthetically generated airways and vessels produced by the proposed generative model to compute the relative error and directly evaluate the accuracy of CNR in comparison with traditional methods. Then, in-vivo validation is performed by analyzing the association between the percentage of the predicted forced expiratory volume in one second (FEV1\%) and the value of the Pi10 parameter, two well-known measures of lung function and airway disease, for airways. For vessels, we assess the correlation between our estimate of the small-vessel blood volume and the lungs' diffusing capacity for carbon monoxide (DLCO). 
		
		The results demonstrate that Convolutional Neural Networks (CNNs) provide a promising direction for accurately measuring vessels and airways on chest CT images with physiological correlates.
	\end{abstract}
	

	\section{Introduction}	
	
	In the last decades, changes in peripheral lung airways and vessels have been shown to be key elements of the pathophysiological development of lung diseases associated to chronic tobacco exposure, among others \cite{niimi2000airway,santos2003enhanced,barr2007impaired,hogg2013small}. Therefore, the ability to detect small changes in the morphology of those structures is crucial to the early diagnosis of the disease and the development of new therapies that can reserve those early changes. The development of those therapies is hampered by the lack of accurate biomarkers related to small airways and vessels morphological changes that can assess and monitor the therapeutical response \cite{barr2007impaired,hogg2013small}. For example, chronic obstructive pulmonary disease (COPD) is mostly affecting peripheral airways \cite{hogg2013small} and airway wall thickness predicts airflow obstruction and physical impairment \cite{charbonnier:2019fw}. In asthma, several studies have shown that CT-measured airway wall thickness predicts the severity and duration of attacks \cite{awadh1998airway,niimi2000airway}.
	
	Additionally, several studies have demonstrated that small pulmonary arteries become smaller and shrink at subsegmental levels in patients with COPD \cite{jacobson1967vascular,cordasco1968newer}, and endothelial dysfunction may be caused by both pulmonary and extra-pulmonary vascular alterations in COPD \cite{santos2003enhanced,barr2007impaired}. Therefore, having an automated method for airway and vessel morphology assessment will help  precise measurements of the geometrical properties of bronchial and venous trees which, in turn, may lead to improved diagnosis and open the door to new studies on lung disorders.
	
	While in the past several methods have been proposed with the aim of helping physicians accurately locate small pulmonary airways and veins on chest CT images \cite{rudyanto2014comparing,bian2018small}, up to date not much work has been proposed for sub-voxel morphology assessment. Traditional approaches for airway wall thickness detection are based on edge-detection methods, that, although limited by the Nyquist theorem, use the reconstructed CT signal to analyze properties of the structure directly. Among them, the full width at half max (FWHM) \cite{schwab1993dynamic}, for which the true edge of an ideal step function undergoing low-pass filtering is located at the FWHM location, is one of the most typical algorithms. In \cite{hackx2015chronic}, a modified FWHM, as proposed in \cite{nakano2000,nakano2002development} is used as available in the Virtual Bronchoscopy program (Siemens Medical Solution), to determine the effect of bronchodilation on airway metrics derived from airway wall thickness reflecting airway disease in patients with COPD.

	The main drawback of the FWHM method is that if the airway wall is considered as a laminated structure whose scale is close to the scanner scale resolution, the edge principle the FWHM is based on would be violated. Therefore, as reported in \cite{reinhardt1997accurate}, the measurements provided by FWHM are biased towards under- or over-estimation of the inner or outer boundary, respectively. For this reason, in \cite{reinhardt1997accurate} a model-based fitting method, assuming a Gaussian point spread function (PSF) and an idealized step-like model for the airway, was proposed. However, this method is computationally very expensive and does not take into account deviations from the proposed model, such as areas where the airway wall is in contact with a vessel.
	
	Another popular approach to measure airway walls involves the use of the zero crossings of the second order derivative (ZCSD) and the phase congruency of the local phase \cite{estepar2006accurate,conradi2010measuring,lutey2013accurate}, which characterizes lumen-to-wall and wall-to-parenchyma transitions. While shown to provide better localization of the airway wall than FWHM and to be less sensitive to different reconstruction kernels and radiation doses, this technique is still computationally expensive and, as admitted by the authors, still sensitive to overestimation.
	
	A method based on 2D dynamic programming approaches was also proposed to detect both the inner and outer boundaries in cross-sectional images, utilizing cost functions combining the first and second derivatives \cite{tschirren2005intrathoracic}. However, validation was perfomed only on phantom CT images and in-vivo results were not provided.
	
	More recently, a new algorithm for airway wall segmentation that combines coarse airway segmentation and optimal graph construction was proposed \cite{petersen2011optimal}. Comparison to manual annotation and correlation of Inner Volume (IV) and Wall Volume Percentage (WV\%) with predicted forced expiratory volume in one second (FEV1\%) were used to validate the method. However, the proposed technique requires a complex and time consuming parameter tuning and it still computationally expensive. Moreover, no comparison with FWHM and ZCSD is provided. A similar approach was also proposed in \cite{li2005optimal,liu2012optimal}, where a more accurate validation was presented. Nevertheless, as claimed by the authors, the suggested technique may not always be achievable as it requires the pre-segmentation step to define complete airway trees, which is not always feasible.
	
	While different methods have been proposed in the literature for measuring airway wall thickness and lumen, just a few studies have tried to accurately measure the vessels radius with high precision, making the proposed technique very relevant in the field. Some approaches are mainly based on edge detectors and the relation between scale and physical radius based on an idealized Gaussian model for vessels \cite{estepar2012computational}. In a recent work, a technique for vascular morphology quantification based on FWHM was also proposed \cite{zhai2019automatic}. However, the method uses the centerline and its distance to the vessel border (previously segmented) to compute the vessel radius. While acceptable for big vessels, this method is prone to high errors for peripheral vessels. Also, no validation of the vessel morphology quantification is provided.
	
	The main drawback of all traditional methods is that they suffer from over- and under-estimation errors, especially for small stuctures at the scanning resolution \cite{reinhardt1997accurate}. 
	
	To overcome these issues when measuring small airways and vessels, we propose the development of a convolutional neural regressor (CNR) \cite{lecun2015deep}, that accurately learns the main characteristics of the structures and automatically regresses the airway wall thickness, airway lumen, and vessel radius regradless of the parameters used for CT acquisition. This is accomplished by using small 2D patches extracted on the orthogonal plane along the main axis of the structure of interest.

   \section{Paper Contributions}
   
	Since manually measuring airways and vessels on clinical CT images is a complicated process and traditional methods are not reliable in providing highly accurate measures of small structures, in this paper we propose to use a generative approach based on synthetic models for airways and vessels that aim at mimicking their main distinguishing features with known physical dimensions for training the network. A Simulated and Unsupervised Generative Adversarial Network (SimGAN) \cite{shrivastava2017learning} is then used to refine the generative model. We use the data created by means of the proposed generator and refined with SimGAN to train the CNR that, through a specifically designed loss function aimed at jointly optimizing both accuracy and precision, regresses the size of the structures of interest.
	
	For validation, we first demonstrate on synthetic generated data that the same technique can be successfully applied both for airway and for vessel assessment with very accurate results, extending our workshop paper \cite{nardelli2018accurate}, in which our CNR and model generator were introduced, but testing was limited to show only preliminary outcomes.
	
	Then, three experiments are performed on clinical cases that show very encouraging results and demonstrate that the technique is very accurate and precise even when the initial conditions, scanner brand, and scanner protocols are modified.
	
	Finally, we compute an indirect in-vivo validation via a correlative analysis with physiological factors for both airways and vessels. For the bronchial tree, we used 3,038 subjects to assess the association between Pi10 and FEV1\% and a comparison to the 3D airway measurement software package Pulmonary Workstation (VIDA Diagnostics, Inc., Iowa City, IA, USA) \cite{tschirren2005intrathoracic} is provided. Also, linear models were created to look at the association between our measurements and functional small airway disease (fSAD) using the parametric response mapping (PRM) method, a non-invasive imaging biomarker that identifies small airway loss, narrowing, and obstruction \cite{galban2012computed}. Then, to carry out a physiological evaluation of our ability to accurately measure vessel lumen radius, we also analyze how three metrics of blood volume distribution (total blood volume, TBV, blood volume of vessels of less than 5 $mm^2$, BV5, and blood volume of vessels of less than 10 $mm^2$, BV10) correlate to lungs' diffusing capacity for carbon monoxide (DLCO) adjusted by site altitude. 
	
	While demonstrating that the proposed technique outperforms state-of-the-art methods for bronchial morphology assessment, in this paper we also propose an accurate approach to vessel sizing for which up to date no relevant methods have been presented in the literature.
		
	\section{Material and Methods}
		
\begin{figure*}[t!]
		\centering
		\includegraphics[width=1.\textwidth]{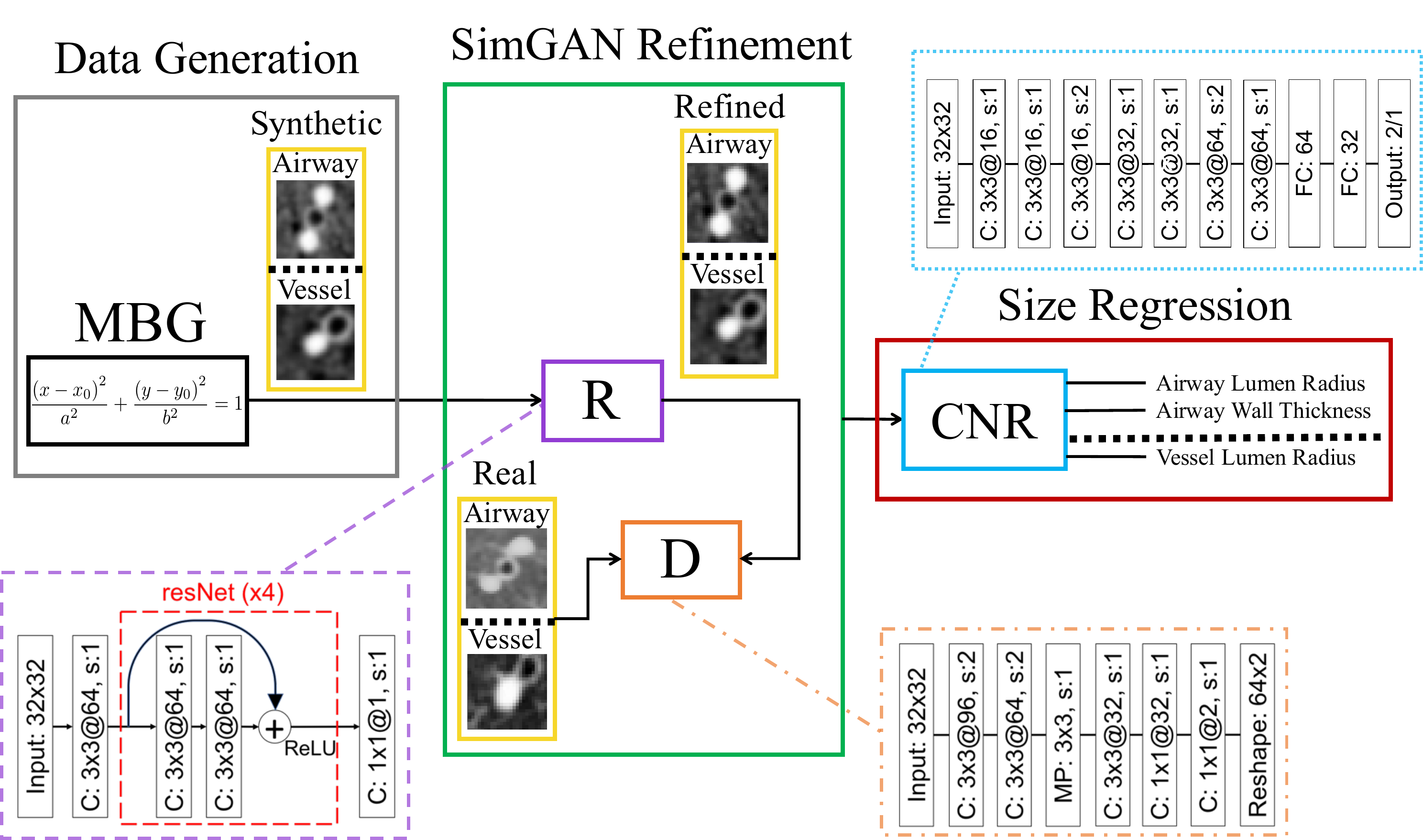} 
		\caption{Color-coded scheme of the proposed method for airway and vessel assessment with neural networks. The model-based generator (MGB, grey square) create the synthetic patches, that are then passed to the SimGAN refinement process (green). In this step, images are refined by a pre-trained refiner (R, purple) and used as input of a discriminator (D, orange) along with real patches, that is trained to distinguish the two images. A minimax game between R and D is used to "fool" D and make the generated images indistinguishable from real ones. As a last step, the refined images are used to train a convolutional neural regressor (CNR) for airway or vessel measurement. All networks involved in the workflow are identical for the analysis of both structure. In this figure, a black dash line is used to separate the airways and vessel.}
		\label{cnn_scheme} 
	\end{figure*}

	A color-coded scheme of the proposed method for both airway and vessel assessment is presented in Fig. \ref{cnn_scheme}, whereas the different parts of the workflow are detailed in the following sections.
	
	\subsection{Airway and Vessel Model-based Generator}
	
	A model-based generator (MBG, black box in Fig. \ref{cnn_scheme}) was developed to synthesize patches that simulate the main characteristics of the structure of interest as well as the CT scanner attributes, including resolution, PSF, imposed noise and blurring.
	
\begin{figure}[t!]
		\centering
		\begin{tabular}{cc}
			\centering
			\includegraphics[width=0.4\columnwidth]{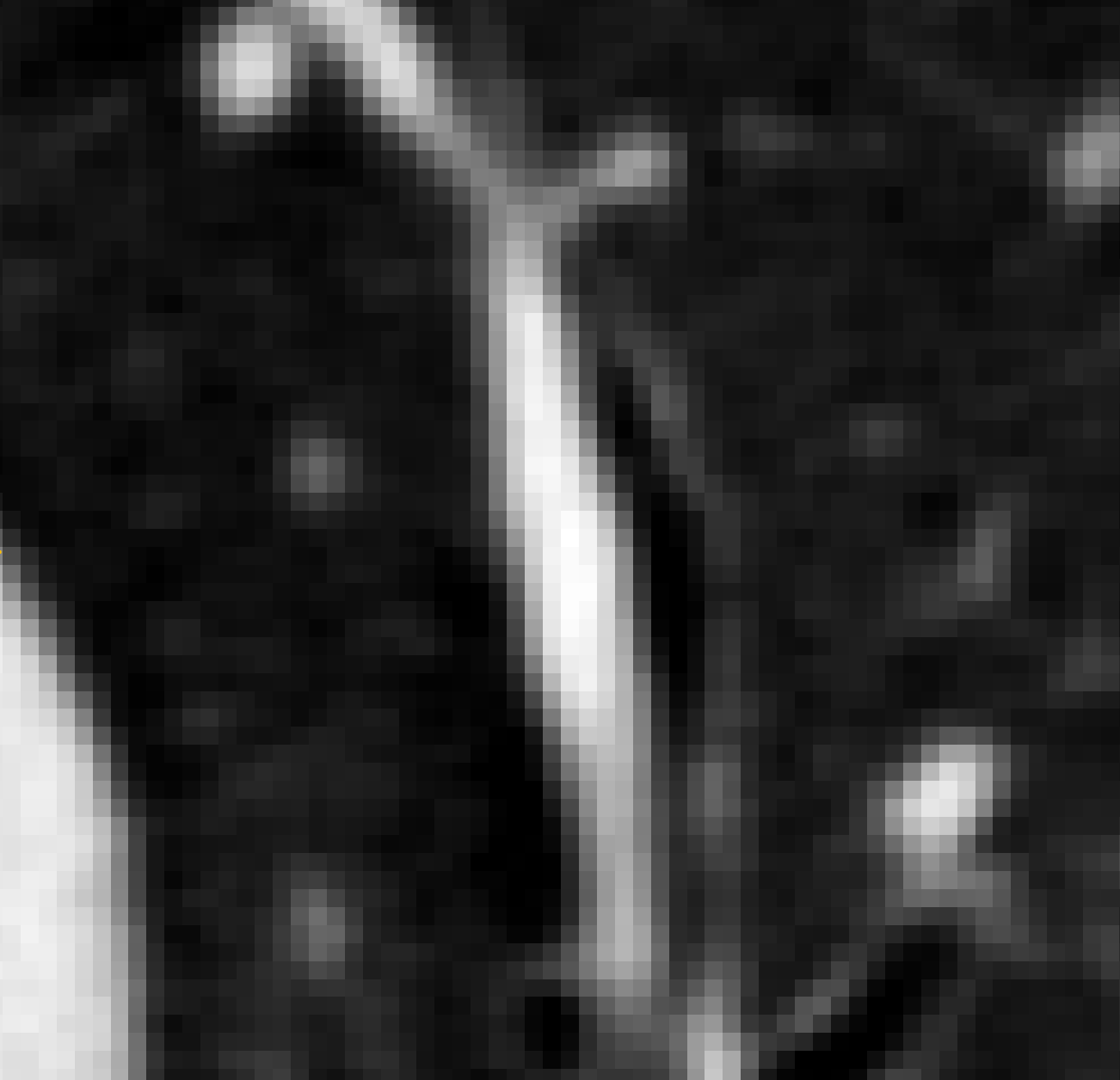}
			&
			\includegraphics[width=0.4\columnwidth]{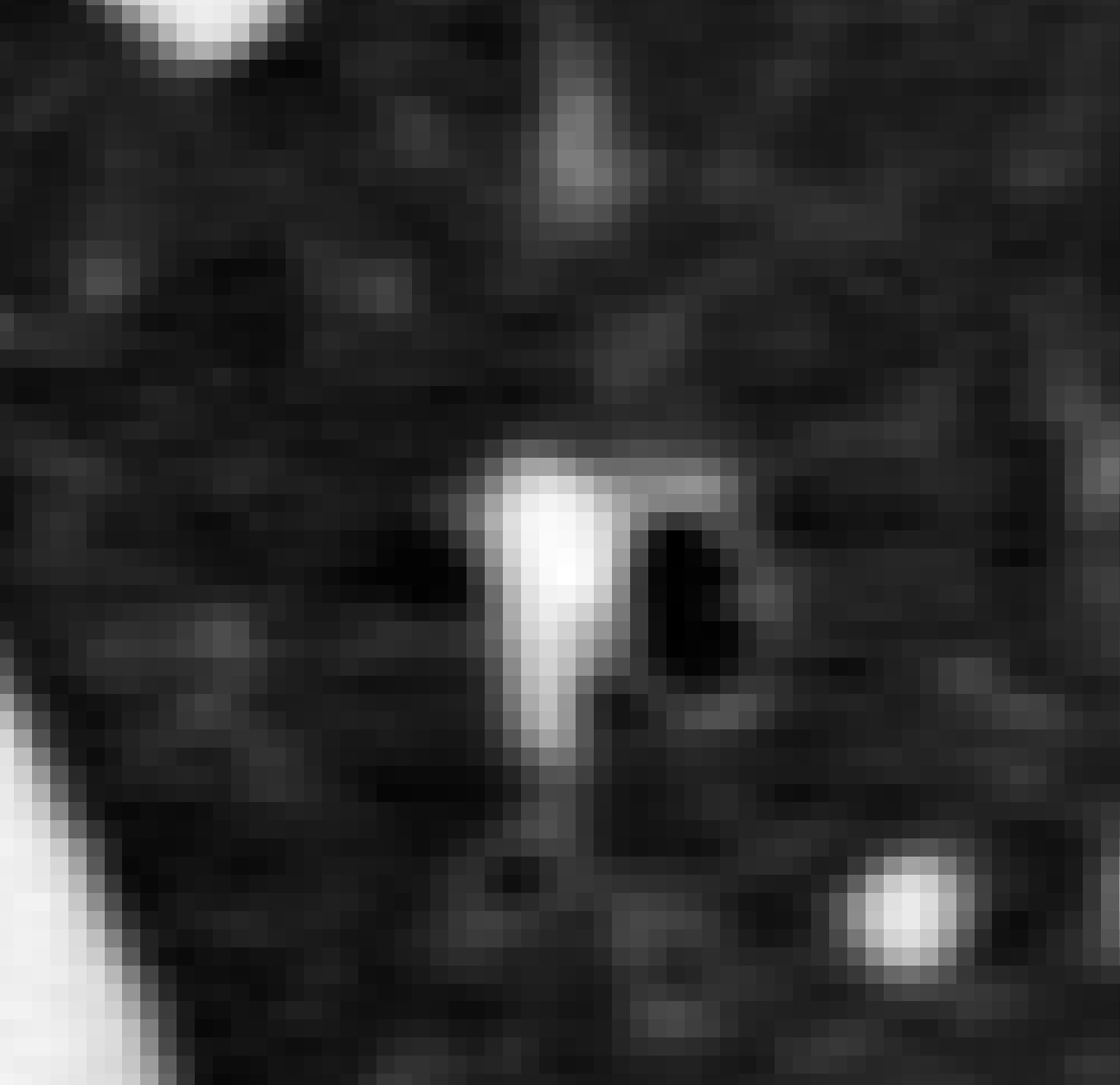} \\
			(a) & (b) \\
		\end{tabular}
		\caption{Example of (a) vessel patch on a CT axial slice and (b) its reformatted version along its main axis, defined by the first eigenvector of the Hessian matrix. }
		\label{ReformattedPatch} 
	\end{figure}

	To avoid possible issues in measuring the structures due to their orientation, we simulated airways and vessels on a reformatted axial plane. This is a reasonable simulation, as when considering in-vivo CT images the first eigenvector of the Hessian matrix (or the structure's centerline) can be used to extract the patches along the airway/vessel's main axis. An example of a vessel reformatted on its main axis can be seen in Fig. \ref{ReformattedPatch}.
	
	Based on the structures anatomy and their surrounding morphology \cite{weibel1965morphometry}, the airway geometrical model consisted of two bright ellipses (inner and outer walls) with a dark central zone (airway lumen) and zero, one or two tangent vessels, represented by bright ellipses rotated around the airway. Conversely, to simulate vessels, a bright central ellipse was generated, with zero, one or two simulated airways in the surrounding. Since only arteries are tangent to bronchi, the simulated airways might also be separated from the vessel. 
	
	Since the structures are simulated as ellipses, for the measurement of airway and vessel lumen we consider the nominal radius of the shape, given by:
	\begin{equation}
	R_{n}= \sqrt{\frac{D_{max}}{2} * \frac{D_{min}}{2}}
	\end{equation}
	where $D_{max}$ and $D_{min}$ represent the maximum and minimum diameters of the ellipse.
	
	In Tab. \ref{paramTab}, all parameter values (randomly chosen based on the physiological values proposed in \cite{weibel1965morphometry}) used to create the structures are reported. To mimic the parenchyma and its texture of multiple structures, we used a Gaussian distributed noise to which a Gaussian smoothing (standard deviation = 5) was applied. The correlated noise was then altered to have a mean intensity of -900 HU and a standard deviation of 150. All values were chosen from nominal values. 
	
	\begin{table*}[t!]
		\caption{Value ranges of the parameters for airway and vessel patch generation. All values are uniformly distributed within the ranges, chosen in accordance to the physiological values proposed in \cite{weibel1965morphometry}.}
		\centering
		\resizebox{1.\textwidth}{!}{%
			\begin{tabular}{ccc}
				\hline
				\textbf{Parameter} & \textbf{Airway} & \textbf{Vessel}\\
				\hline
				Lumen radius (LR) & [0.5, 6.0] mm & [VR, VR + 0.8] mm\\
				Airway wall thickness & [0.1*LR + 0.2, 0.3*LR + 1.5] mm & [0.1*LR + 0.2, 0.3*LR + 1.5] mm\\
				Number of vessels & [0, 2] & [1, 3]\\
				Number of airways & 1 & [0, 2]\\
				Vessel radius (VR) & [LR, LR + 0.8] mm & [0.5, 4.5] mm\\
				Skewness of reconstruction & [-25, 25] degrees &  [-25, 25] degrees\\
				Airway Lumen Intensity & [-1150,-1050] HU & [-1150, -1050] HU \\
				Airway Wall Intensity & [-500, 50] HU & [-500, 50] HU\\
				Vessel Intensity & [-50, 50] HU & [-50, 50] HU \\
				Noise Level & 25 HU & 25 HU \\
				Smoothness Level & [0.5, 0.875] mm & [0.5, 0.875] mm \\
				\hline
			\end{tabular}
		}
		\label{paramTab}
	\end{table*}
	
	From an accurate analysis of CT images, it is also possible to notice that some peripheral vessels may be located close to the chest wall, the presence of which may affect the measure of the CNR. To deal with this potential issue, we randomly added one or two curved regions at the two opposite corners or borders of the patch of some vessels with radius lower than 1.5 mm. For this added region, we used uniformly distributed values in a range $\sigma_n \in [-100, 200]$ HU. 
	
	For training the CNR, we used patches of 32$\times$32 pixels, as enough neighborhood information can be included for big structures without losing specificity for thin and small characteristics. A resolution of 0.05 mm/pixel in a sampling grid of 640$\times$640 pixels is initially used to generate the images. Then, a down-sampling to 0.5 mm/pixel is applied and a simulated PSF is applied to mimic the CT blurring caused by an image reconstruction process. 
	
	Estimating a PSF can be challenging due to its complexity and variance across manufactures \cite{san2008three}. However, experimental measurements of the PSF of a CT scanner demonstrated that it can be approximated by a 3D Gaussian function with the assumption that it is locally space invariant and separable \cite{schwarzband2005point}. Due to the small size of the generated patch, this assumption is valid and we approximate the PSF by a spatially invariant Gaussian function with standard deviation randomly chosen in a range $\sigma_{s} \in [0.4, 0.9] mm$ that simulates the PSF variation across different CT scanners and manufacturers.
	
	Then, we apply a Gaussian random noise smoothed with a Gaussian filter (standard deviation = 2), with zero mean and a standard deviation $\sigma_{n} = 25$, values chosen based on nominal noise characteristics in high dose CT scans \cite{VegasSanchezFerrero:2017vt}. 
	
\begin{figure*}[t!]
		\centering
		\begin{tabular}{cccccc}
			\centering
			\includegraphics[width=0.14\textwidth]{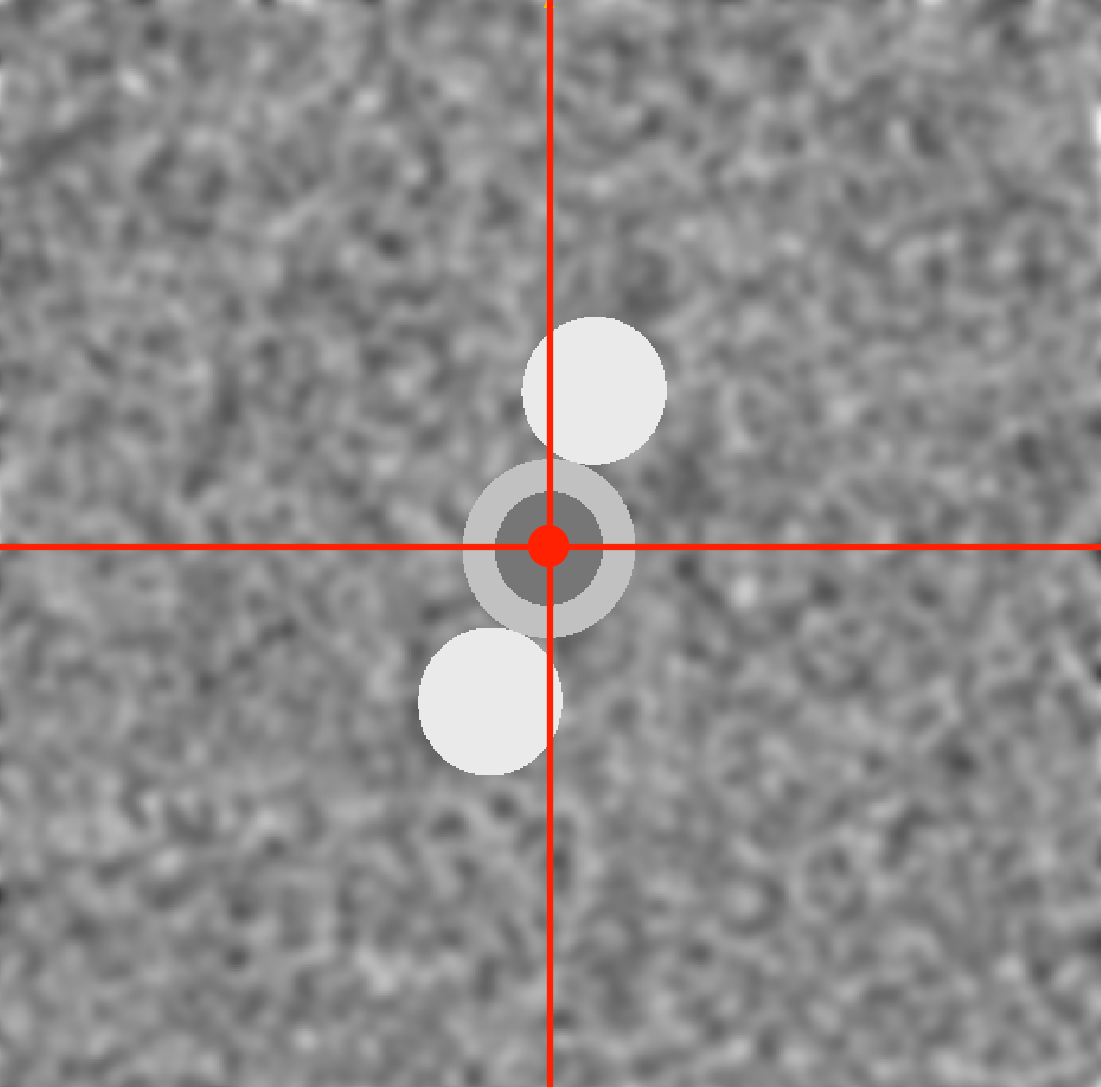}
			&
			\includegraphics[width=0.14\textwidth]{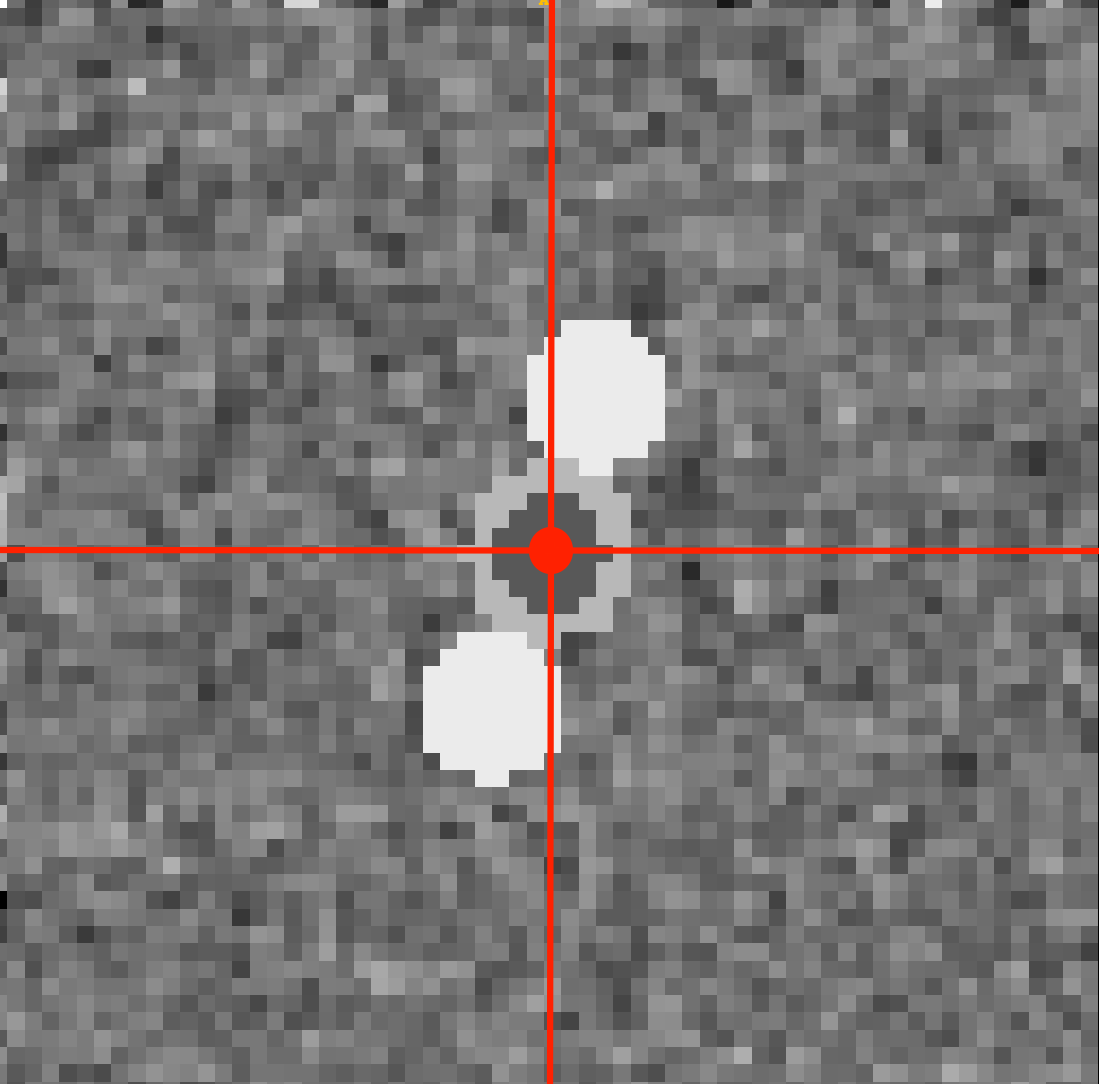}
			&
			\includegraphics[width=0.14\textwidth]{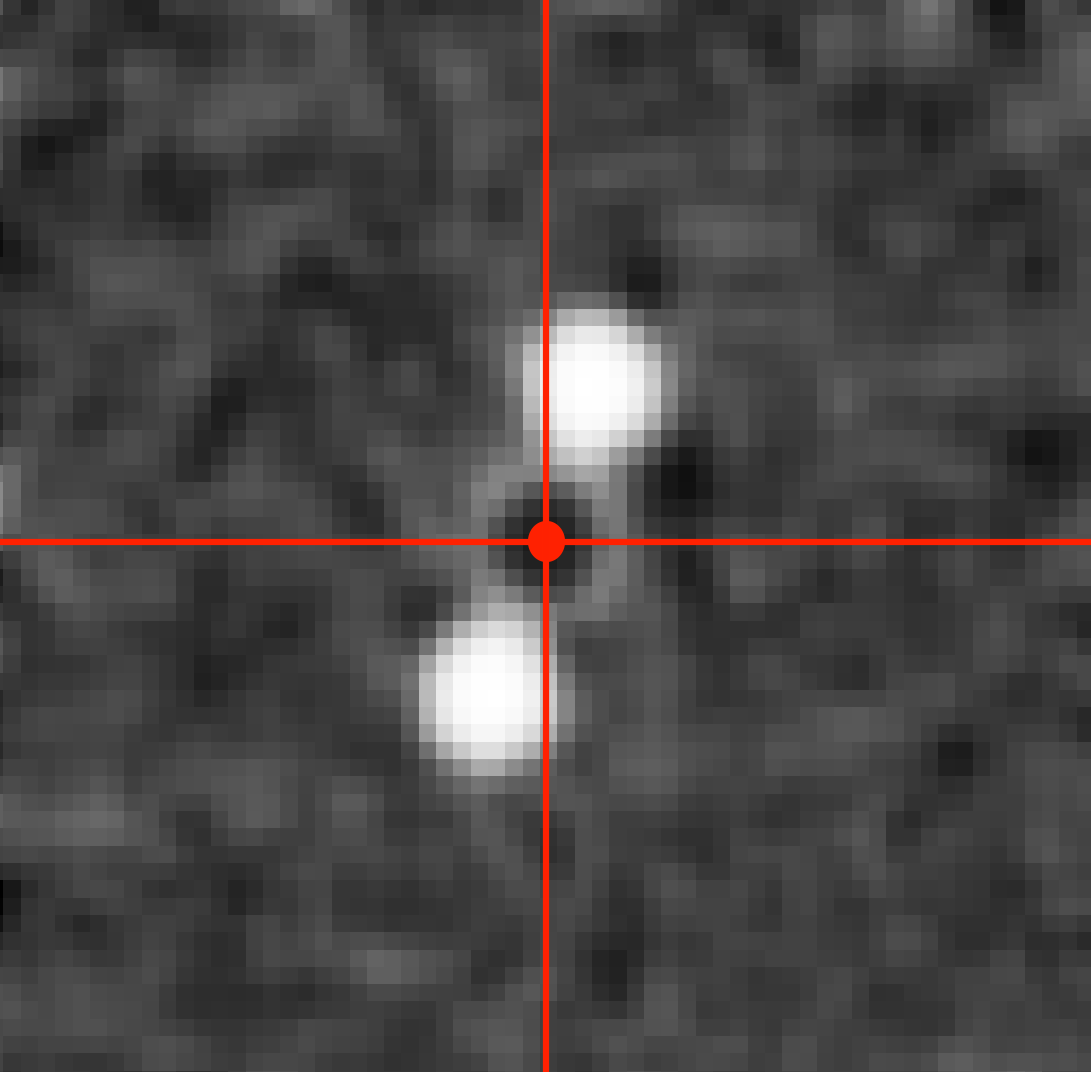}
			&
			\includegraphics[width=0.14\textwidth]{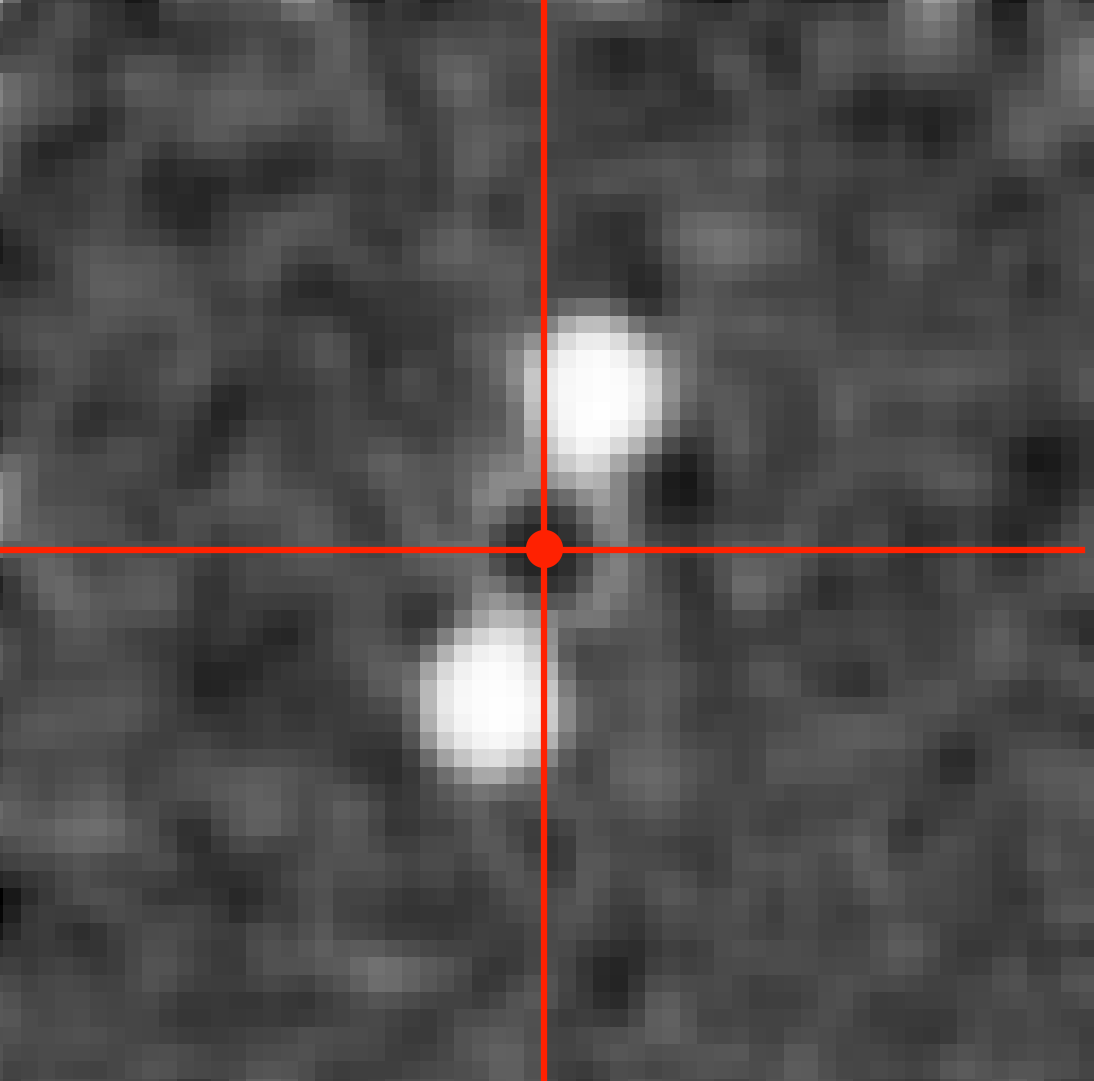}
			& 
			\includegraphics[width=0.14\textwidth]{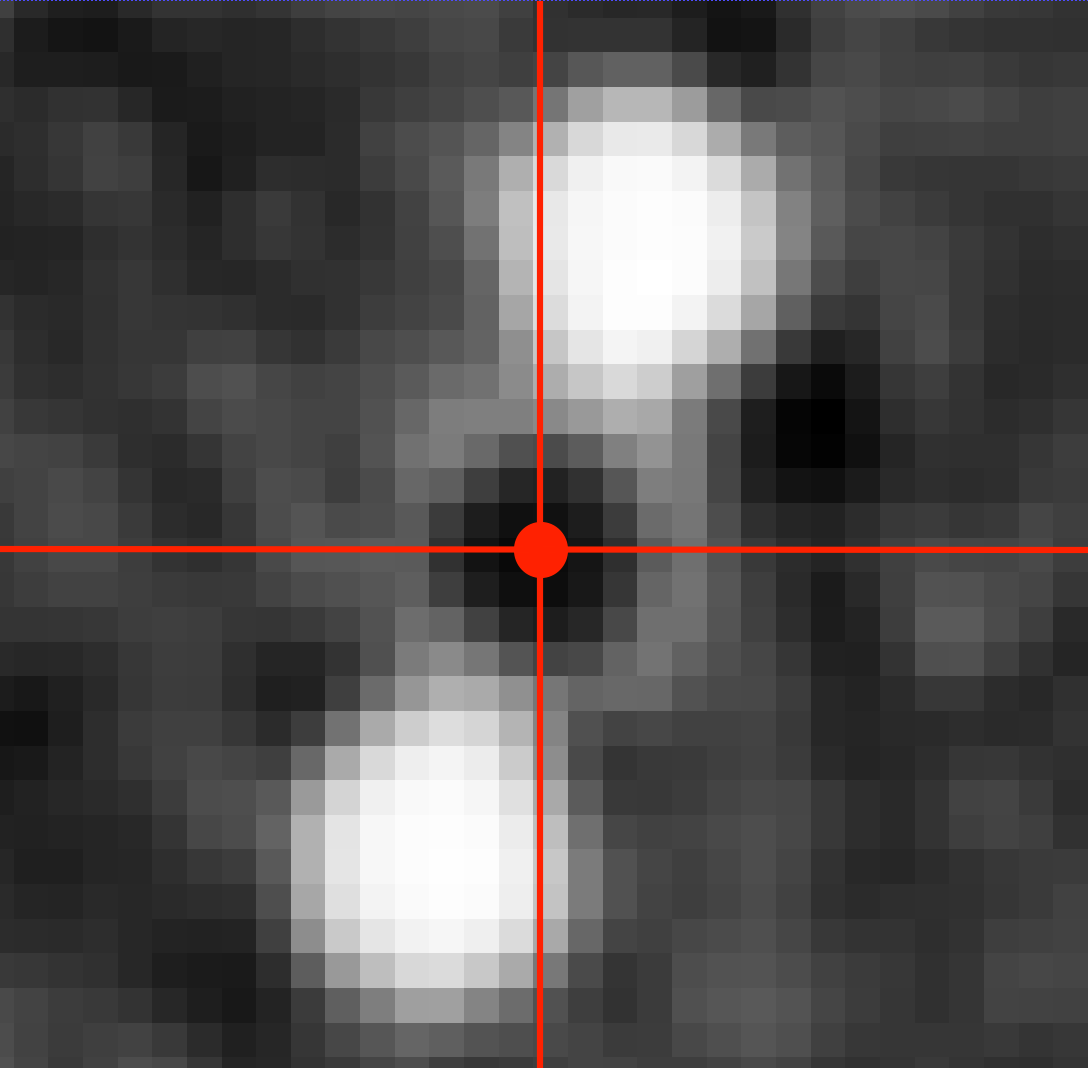}
			&
			\includegraphics[width=0.14\textwidth]{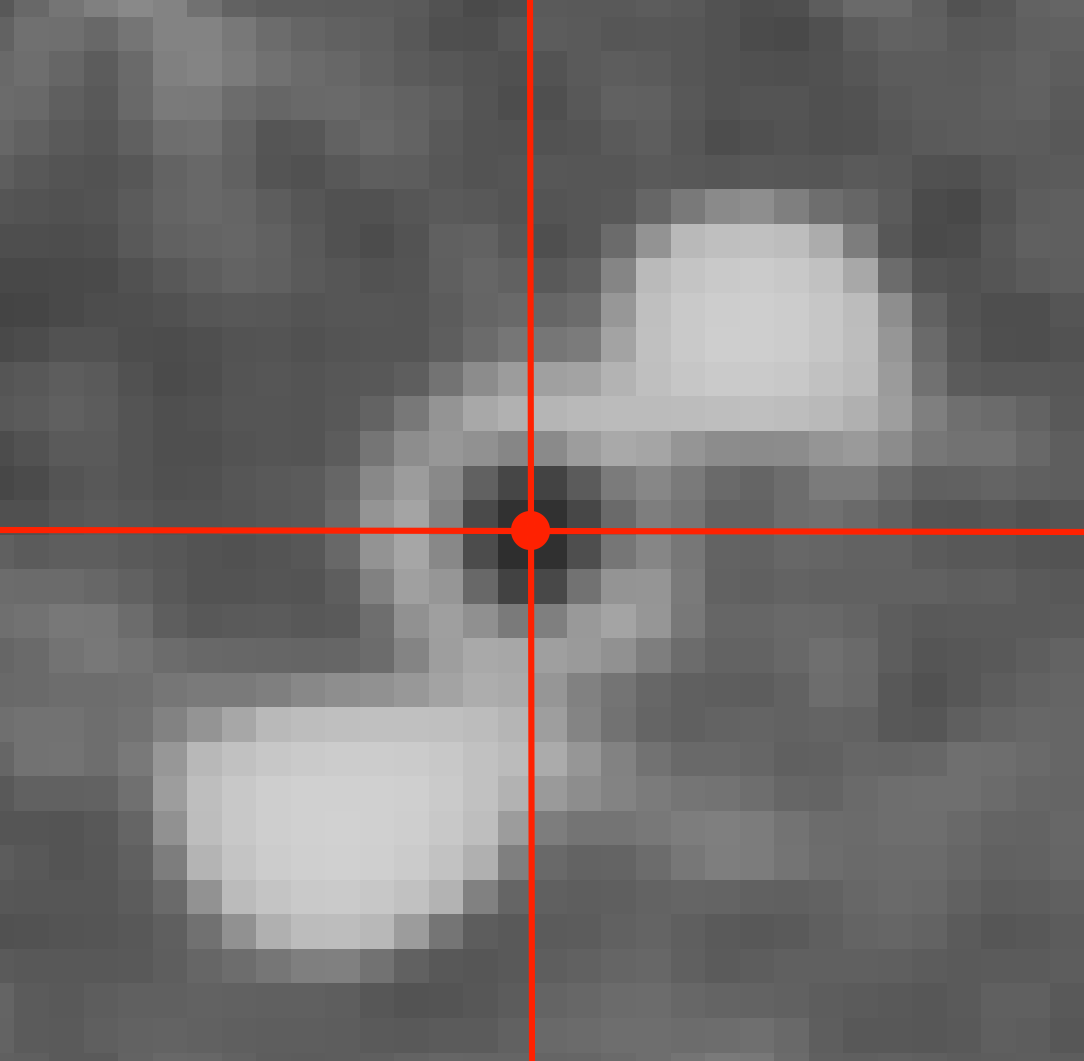}\\
			& & & & & \\
			\includegraphics[width=0.14\textwidth]{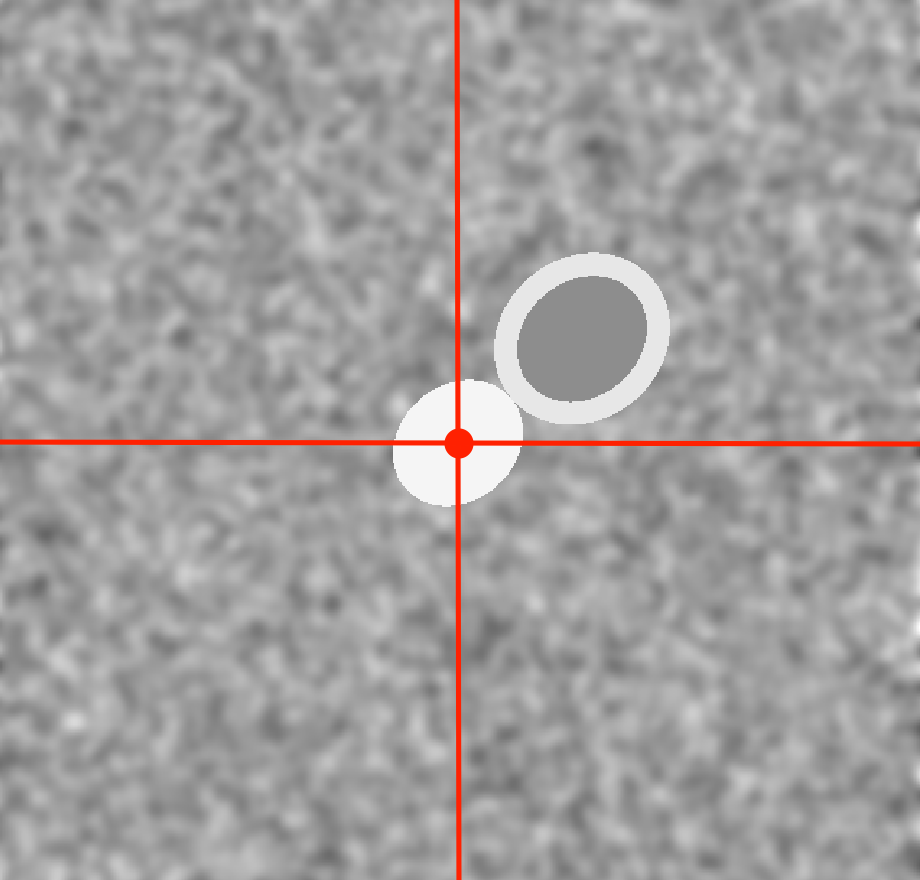}
			&
			\includegraphics[width=0.14\textwidth]{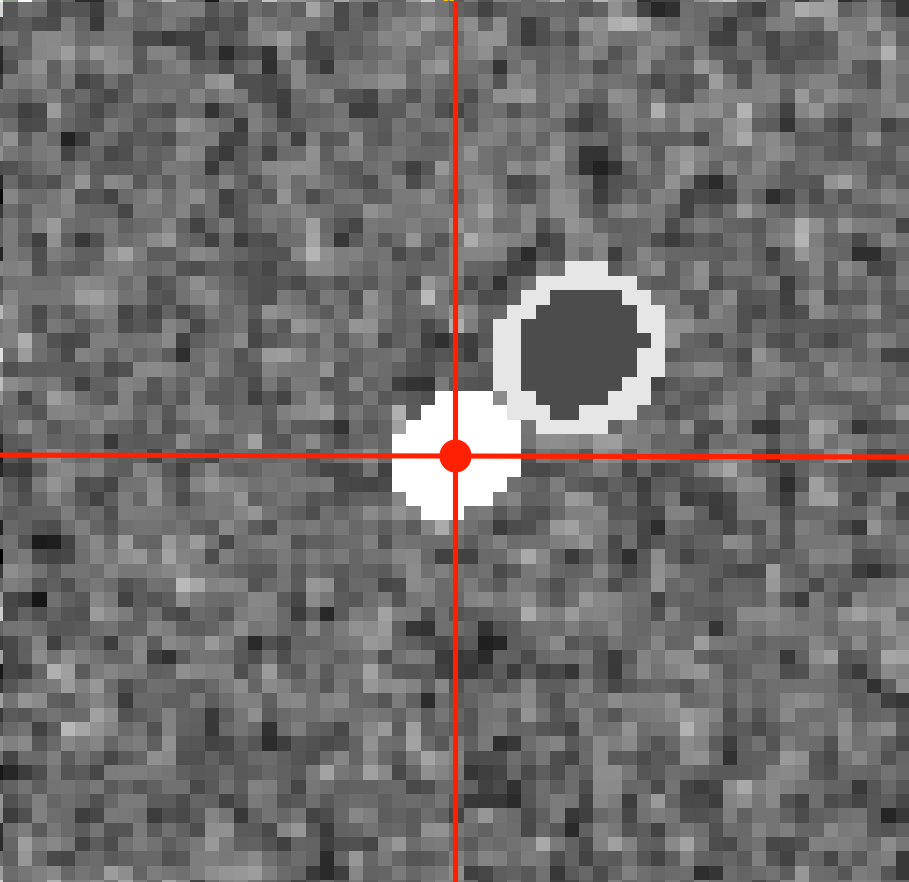}
			&
			\includegraphics[width=0.14\textwidth]{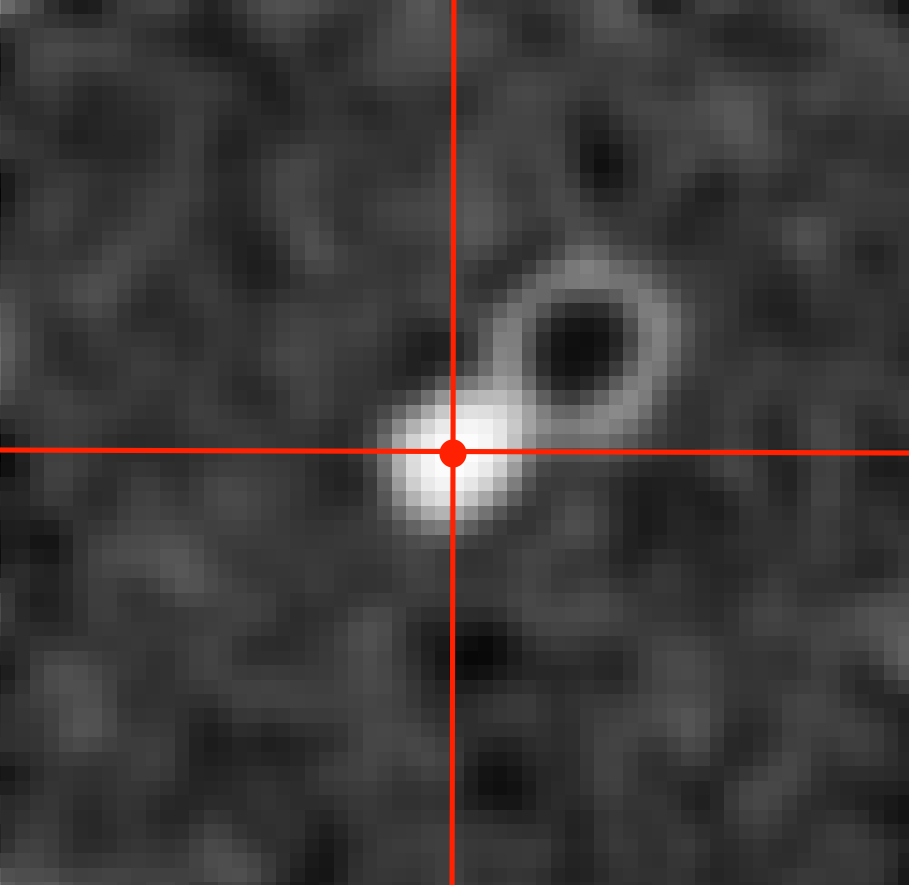}
			&
			\includegraphics[width=0.14\textwidth]{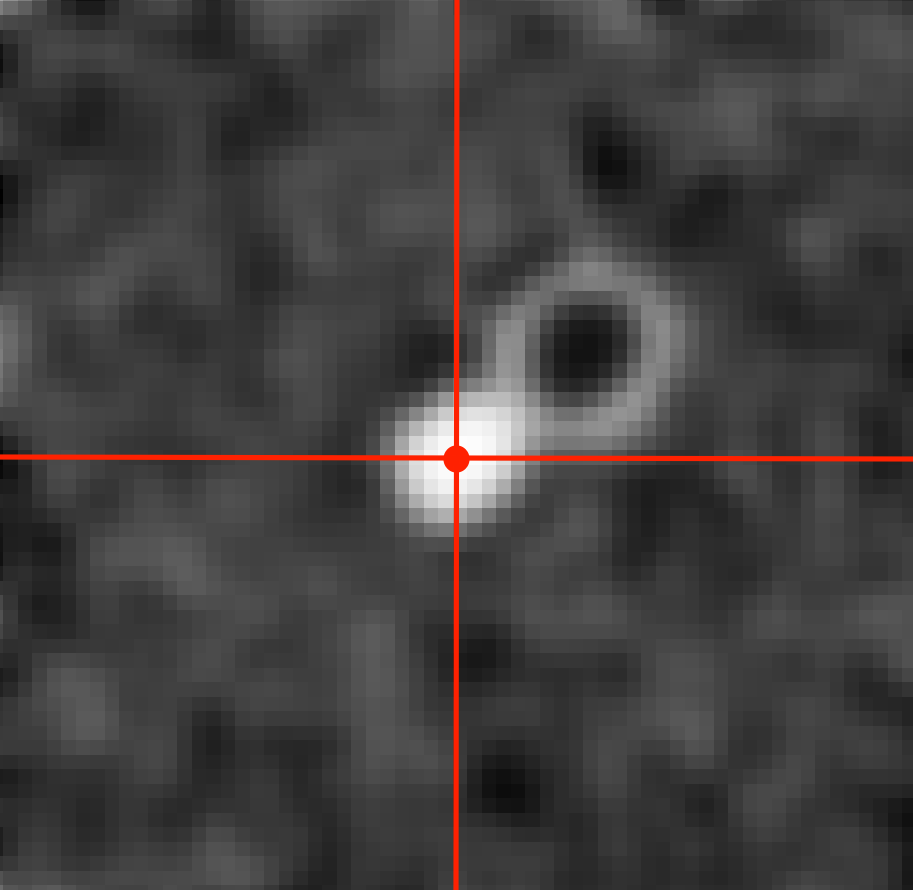}
			& 
			\includegraphics[width=0.14\textwidth]{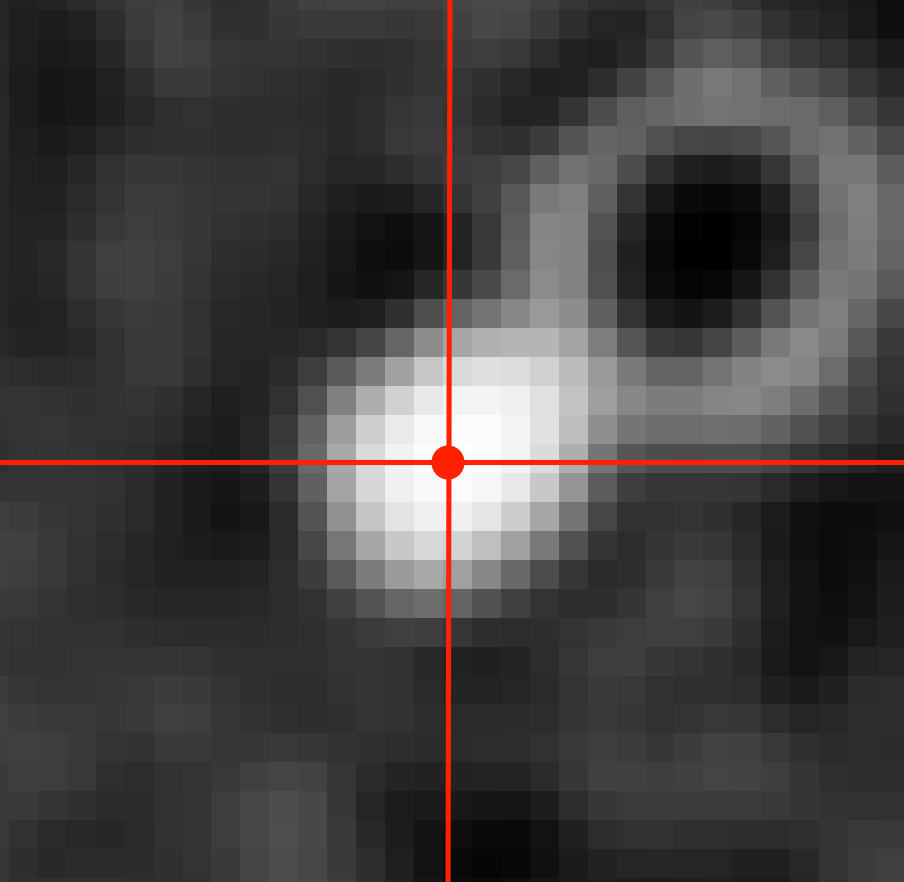}
			&
			\includegraphics[width=0.145\textwidth]{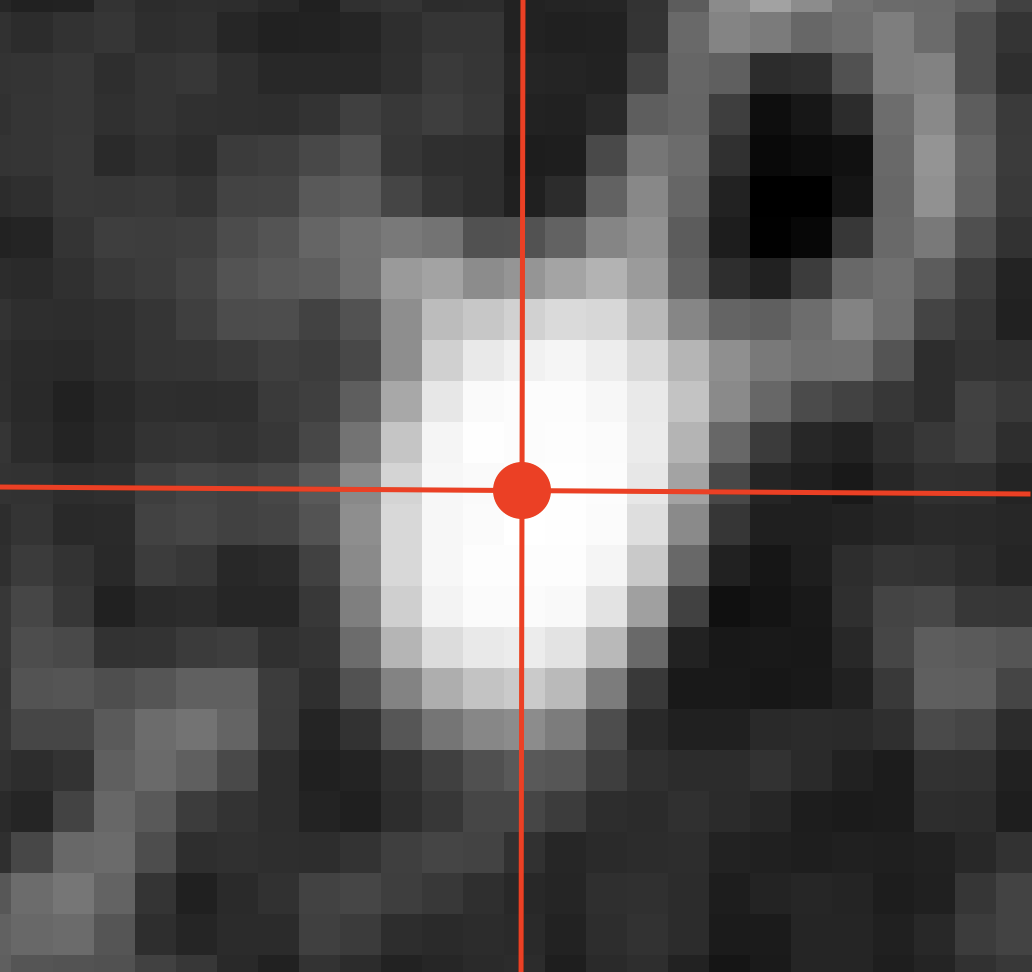}\\
			(a) & (b) & (c) & (d) & (e) & (f) \\
		\end{tabular}
		\caption{Examples of generation of a synthetic airway (top) and a synthetic vessel (bottom). (a) The initial geometric model; (b) down-sampling of the model; (c) blurring of the patch; (d) noise addition; (e) final cropped and refined structure; (f) a similar patch extracted from the reformatted axial plane of in-vivo CT images. The red dot and lines show the patch center.}
		\label{PatchGeneration} 
	\end{figure*}

	As a last step, an image cropping is implemented to obtain the final 32$\times$32 pixels patch. In Fig \ref{PatchGeneration}a-d, an example of airway (top) and vessel (bottom) patch as created by the MBG at each step is shown.
	
	\subsection{SimGAN Refinement}
	\label{SimGANrefinement}
	
	The proposed MBG simulates reasonably well the geometrical aspects of the structure of interest. However, although patch values were chosen to be as accurate as possible, differences from real anatomy may still appear. We thus implemented a SimGAN refinement, similar to \cite{shrivastava2017learning}, that makes use of simulated and unsupervised learning through a generative adversarial network (GAN) and was originally implemented to specifically avoid the introduction of artifacts and preserve annotation information on the refined images, which for synthetic structures is automatically available by definition (in the presented paper, lumen and wall thickness size).
	
	The SimGAN refinement is utilized to adjust CT intensity values and improve texture and affinity of the patches to in-vivo structures. To the best of our knowledge this is the first time a SimGAN approach is applied to a medical image problem.
	
	\subsubsection{In-vivo Airway/Vessel Patches Extraction for SimGAN Training}
	\label{realpatches}
	To generate the dataset of real 2D patches for SimGAN training, 32$\times$32 images were extracted around airway and vessel candidates from in-vivo CT images. To this end, the lung region was first segmented using the method in \cite{ross2009lung} and a probability map of the different structures of interest was extracted using the algorithm proposed in \cite{nardelli2017ct}. A thresholding operation (threshold = 0.7) and a binary skeletonization were then used to define the initial candidate locations.
	
	Since the in-vivo patches have to be extracted with the same axial-oriented appearance as in the simulated images, the main axis of the structure of interest is considered. To this end, the skeletonized mask is used as input to the scale-space particles sampling method presented in \cite{kindlmann2009sampling}, that starts from an airway/vessel mask to identify an "oriented" bronchial/vascular tree centerline by means of the second-order local information of the image (Hessian matrix). The centerline is stored in the form of a collection of points (called particles) that contain information about scale, orientation (Hessian eigenvectors and eigenvalues), and central pixel intensity. This approach capitalizes on the multi-scale self-similarity of the considered tree, making it more robust to noise in the smaller structures than standard methods \cite{estepar2012computational}.  
	
	\subsubsection{SimGAN Training Implementation}
	
	To implement a SimGAN training, a refiner, R (purple box in Fig. \ref{cnn_scheme}), is trained to create a refined patch that includes the main aspects of real structures, while at the same time a discriminator, D (orange box in Fig. \ref{cnn_scheme}), is used discriminate between real and synthetic patches. This starts a minimax game between R and D (with the weights of two networks updated alternately) that continues until D is not any more capable to distinguish between the two different domains.
	
	The network R consists of a pixel-to-pixel fully convolutional neural network with ResNet blocks, while 4 convolutional layers separated by max pooling and a last layer that outputs the probability of the patch of being a refined image are used to build D.
	
	For D, a cross-entropy loss for a two class classification problem is used, while, as in \cite{shrivastava2017learning}, the cost function of R is given by:
	\begin{equation}
	\label{simGAN_label}
	\mathcal{L}_{R}(\theta) = \sum_{i}\bigg(\mathcal{L}_{real}(\theta; x_{i}, Y) + \alpha\cdot\mathcal{L}_{reg}(\theta; x_{i})\bigg)
	\end{equation}
	where $x_{i}$ is the i-th synthetic training patch, Y is the set of real patches,and $\theta$ are the function parameters. In this work, $\alpha$ has been empirically set to 0.01. 
	
	In function \ref{simGAN_label}, $\mathcal{L}_{real}$ adds realism to the synthetic images forcing D to fail classifying the refined images as synthetic, whereas $\mathcal{L}_{reg}$ is a self-regularization loss that is used to minimize per-pixel difference between a feature transform of the synthetic and refined images and thus preserve the annotation information of the MBG and avoid that, for example, the structure change shape or orientation. More details about the cost functions can be found in \cite{shrivastava2017learning}.
	
	As in \cite{shrivastava2017learning}, the receptive field of D is limited to local regions, so that multiple local adversarial losses per patch are considered. Moreover, to improve the stability of the network, a mini-batch of refined images (randomly selected from a buffer of refined images generated on previous iterations) are included into the training batch.
	
	For the implementation of this network, R is first pre-trained on synthetic images for 1,000 steps. Then, D is also pre-trained for 200 steps, as suggested in \cite{shrivastava2017learning}, using real patches (extracted as described in section \ref{realpatches}) and refined ones, obtained from the pre-trained R. As a final step, an adversarial training of the SimGAN is executed for 10,000 steps. All networks are trained with a constant 0.001 learning rate and 256 batch size.
	
	For the training of the airway/vessel SimGANs, a total of 165,640 in-vivo airway patches and 2,320,000 in-vivo vessel patches were extracted from 30 clinical cases of the COPDGene study (phase 1) \cite{regan2011genetic}. In order to get balanced datasets, we generated 200,000 and 2,500,000 airway and vessel synthetic patches, respectively. The refined output from the SimGANs is depicted in Fig. \ref{PatchGeneration}e. It is worth noting the resemblance between a simulated airway/vessel (Fig. \ref{PatchGeneration}e) and a real airway/vessel extracted from a CT scan corresponding to the same nominal structure size (Fig. \ref{PatchGeneration}f). 
	
	\subsection{Airway/Vessel Measurement Regression}
	The airway/vessel measurements were computed by means of two separate CNRs  having the same basic design (light blue box in Fig. \ref{cnn_scheme}). A 9-layer 2D network, consisting of 7 convolutional layers, 5 with stride one and 2 with stride two, and 2 fully-connected layers was implemented. While the network for airway measurement has two outputs (lumen radius and wall thickness), for vessels a single output (lumen radius) is used. 
	
	The main goal of the network is to regress the size of the structure centered in the 32$\times$32 pixels patches. An Adam update ($\beta_1$=0.9, $\beta_2$=0.999, $\epsilon$=$1e^{-08}$) with a loss function defined by a combination of absolute RE and precision of the measure, as described in section \ref{lossFunction}, was used to train both networks. To improve invariance of the network, we also applied data augmentation that adds random noise (range $\sigma_n \in [1,20]$ HU), randomly inverts intensity values inside the patches, and slightly shifts and flips the images.
	
	Training was performed on a NVIDIA Titan X GPU machine, using the Keras framework \cite{chollet2015keras} on top of TensorFlow \cite{abadi2016tensorflow}, for 300 epochs at a learning rate of 0.001 (batch size of 1000).  
	
	\subsubsection{Loss Function for Airway and Vessel Measurement}
	\label{lossFunction}
	One of the main issues of typical approaches for airway and vessel morphology assessment, is that they tend to under- or over-estimate the measurement of small airways and vessels (especially with size at the image resolution level). Therefore, in this paper we use a new loss function, similar to \cite{nardelli2018accurate}, that combines the loss of the RE over all images, $\mathcal{L}_{\mu}$, and the precision of the measure over a set of replicas of the exactly same structure, $\mathcal{L}_{\sigma}$:
	\begin{equation}
		\label{lgeneralLoss}
		\mathcal{L}(\boldsymbol{y, \widehat{y}}) = \mathcal{L}_{\mu}(\boldsymbol{y, \widehat{y}})  + \lambda \cdot \mathcal{L}_{\sigma}(\boldsymbol{y, \widehat{y}})
	\end{equation}
	where $\boldsymbol{y}$ is the true measure, $\boldsymbol{\widehat{y}}$ the predicted size, and $\lambda$ determines the weight of $\mathcal{L}_{\sigma}$ with respect to $\mathcal{L}_{\mu}$ (in this work, $\lambda=2$ has been empirically chosen). $\mathcal{L}_{\mu}$ is given by: 
	\begin{equation}
		\mathcal{L}_{\mu}(\boldsymbol{y, \widehat{y}}) = \frac{1}{N \times M}\sum_{i=0, j=0}^{N, M} \frac{|y_{i, j} - \widehat{y}_{i, j}|}{y_{i, j}}
	\end{equation}
	where N is the total number of original patches, and M is the number of replicas used (here, we use M=25). 
	
	Conversely, the precision loss term, $\mathcal{L}_{\sigma}$, is calculated over the M replicas of the same geometric model (with fixed physical dimensions) to which varying PSFs as well as a different number of airways and vessels with various locations and rotations are added. This way, the network learns to precisely measure the structure of interest regardless of possible confounding factors. The definition of $\mathcal{L}_{\sigma}$ is given by:
	\begin{equation}
		\mathcal{L}_{\sigma}(\boldsymbol{y, \widehat{y}}) = \frac{1}{N} \sum_{i=0}^{N} \Bigg(\frac{1}{M}\sum_{j=0}^{M} \Big(y_{i,j} - \widehat{y}_{i,j}\Big)^2 - \Big(\frac{1}{M} \sum_{j=0}^{M} (y_{i,j} - \widehat{y}_{i,j}) \Big)^2\Bigg)
	\end{equation}
	
	Since airway lumen and wall thickness are measured simultaneously, $\mathcal{L}_{\mu}$ and $\mathcal{L}_{\sigma}$ for airway assessment are given by the sum of the corresponding loss computed independently for the two measures.
	
	Since we noticed that the standard deviation was higher for the assessment of small structures (radius $\leq$ 1.0 mm), we also assigned a higher weight to $\mathcal{L}_{\sigma}$ of these structures to give them more importance during training. Therefore, equation \ref{lgeneralLoss} for vessels becomes:
	\begin{equation}
		\mathcal{L}_{v}(\boldsymbol{y, \widehat{y}}) = \mathcal{L}_{\mu}(\boldsymbol{y, \widehat{y}})  + \lambda \cdot \omega_{v} \cdot \mathcal{L}_{\sigma}(\boldsymbol{y, \widehat{y}})
	\end{equation}
	where 
	\begin{equation}
		\omega_{v} = 
		\begin{cases}
			3.0, & \text{if vessel radius} < \text{1.0 mm} \\
			1.0, & \text{otherwise} \\
		\end{cases}
	\end{equation}
	while for airways, it becomes:
	\begin{equation}
		\mathcal{L}_{a}(\boldsymbol{y, \widehat{y}}) = \mathcal{L}_{\mu}(\boldsymbol{y, \widehat{y}})  + \lambda \cdot \Big( \omega_{\text{l}} \cdot \mathcal{L}_{\sigma, \text{l}}(\boldsymbol{y, \widehat{y}}) + \omega_{\text{wt}} \cdot \mathcal{L}_{\sigma, \text{wt}}(\boldsymbol{y, \widehat{y}})   \Big) 
	\end{equation}
	where l indicates the airway lumen, wt stands for wall thickness, and
	\begin{equation}
		\omega_{\text{l}} = 
		\begin{cases}
			1.5, & \text{if airway lumen} < \text{1.0 mm} \\
			1.0, & \text{otherwise} \\
		\end{cases}
	\end{equation}
	\begin{equation}
		\omega_{\text{wt}} = 
		\begin{cases}
			3.0, & \text{if wall thickness} < \text{1.0 mm} \\
			1.0, & \text{otherwise} \\
		\end{cases}
	\end{equation}
	
	\subsubsection{Training Set Definition}
	Training data for both structures consisted of 100,000 geometric models, each replicated 25 times by varying PSFs as well as adding a different number of airways and vessels at various locations and rotations. On the other hand, the validation set was generated with 1,000,000 patches (40,000 geometric models, each replicated 25 times) for both structures.	
	
	\subsection{Experimental Analysis}
	Both synthetic and in-vivo experiments were performed to evaluate the two algorithms for airway and vessel assessment. 
	
	For synthetic validation, we first generated a dataset of 200,000 patches with random values that we used to compute the absolute RE across all cases, compare results to state-of-the-art methods - considering airways (vessels) with wall thickness (vessel radius) of 1.0 mm and 0.5 mm (image resolution) -, and compare wall thickness measurements to FWHM and ZCSD. As a metric, the mean RE was computed for three separate groups, generated based on the known wall thickness value (WT $\leq$ 0.7 mm, 0.7 mm $<$ WT $\leq$ 1.5 mm, WT $>$ 1.5 mm).
	
	As a second synthetic validation, we computed the accuracy of the algorithm by calculating the RE obtained comparing the CNR measurements and the geometrical model ground truth when varying lumen (wall thickness = 1.2 mm) and wall thickness (lumen size = 2.0 mm) for airways, and the radius for vessels. To this end, we computed the mean RE across 100 patches generated for each size value.
	
	In an attempt to demonstrate the reliability of the method regardless of the presence of noise and smoothness, as a further experiment we generated a dataset by first varying the level of noise ($\sigma_n \in [0,40]$ HU, $\sigma_s=1.3$ mm) and generating 100 synthetic patches for each noise value. Then, we created a second dataset by fixing $\sigma_n=25$ HU and changing the applied smoothness ($\sigma_s \in [0.4,0.9]$ mm) to generate 100 synthetic images per smoothing level. We finally computed the mean RE across the 100 patches for each level of noise and smoothness. 
	
	For airways, we first fixed the wall thickness at 1.5 mm and used three different airway lumens (small: 0.5 mm; medium: 2.5 mm; large: 4.5 mm), and then we fixed the lumen at 1.5 mm and used three wall thickness values (small: 0.5 mm; medium: 1.2 mm; large: 2.0 mm). For vessels, we fixed three different radius values (small: 0.5 mm, medium: 2.0 mm, and large: 3.5 mm).
	
	In order to compare the proposed algorithm to state-of-the-art methods, we also validated the performance of the algorithm a CT airway phantom consisting of 8 tubes with different wall thickness and lumen diameter. The tubes were constructed using Nylon66 and were inserted into polystyrene, in an attempt to simulate the lung parenchyma. 
	
\begin{figure}[t!]
		\centering
		\includegraphics[width=0.8\textwidth]{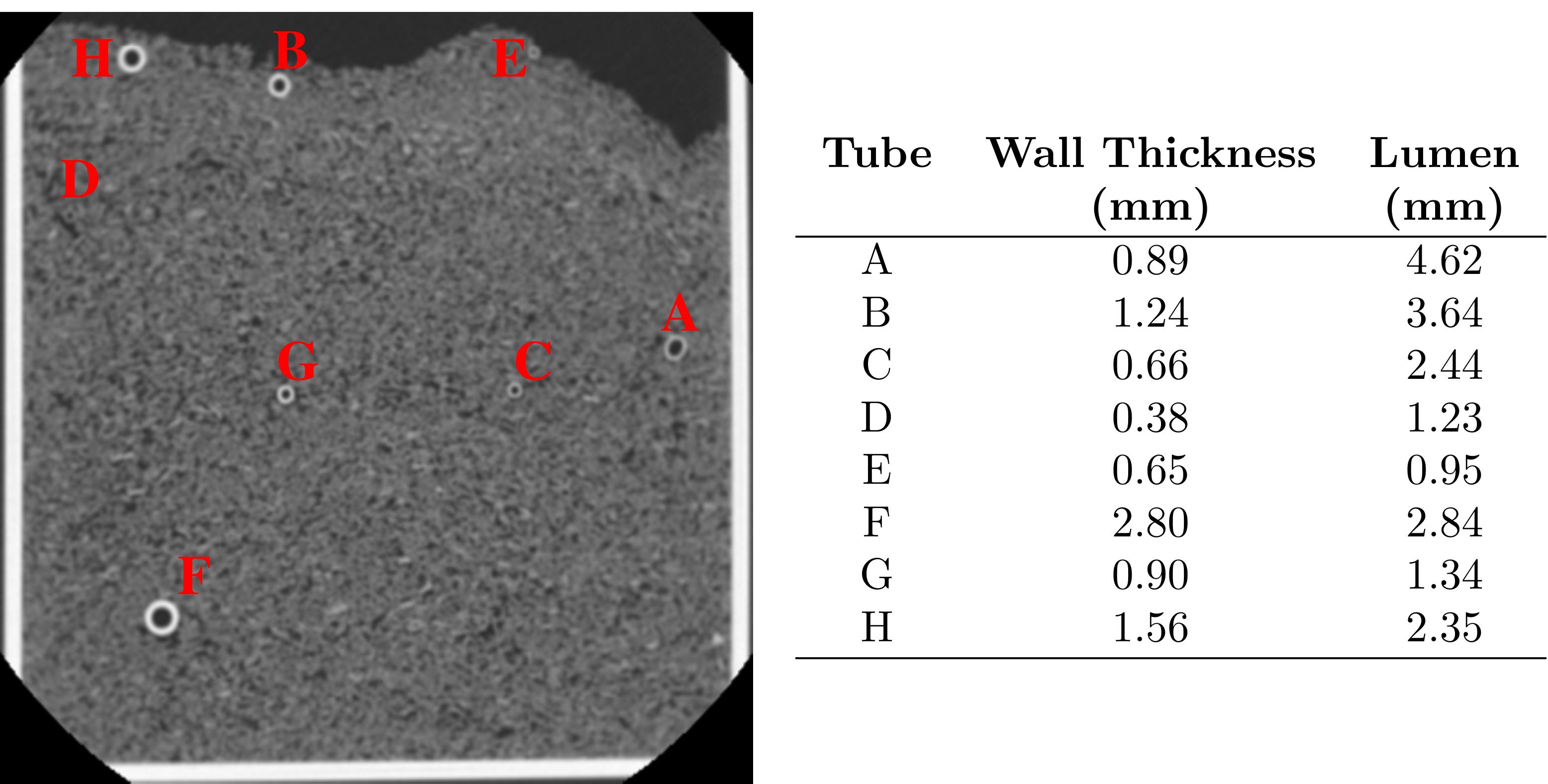} 
		\caption{A sample slice of the phantom CT scan (left) with wall thickness and lumen diameter values (in mm) of the eight synthetic tubes (right).}
		\label{phantom_image} 
	\end{figure}

	The CT image of the phantom was taken using a GE Siemens Sensation 64 CT scanner, with a field of view (FOV) of 40 cm and reconstructed with a standard reconstruction kernel to obtain non-overlapping, 0.6 collimation images. An example of a CT slice of the phantom with wall thickness and lumen values of the eight tubes (as measured by a caliper) are presented in Fig. \ref{phantom_image}.

	As additional experiments, since an accurate and reliable in-vivo ground-truth is very complicated to obtain, we performed two indirect validations on in-vivo clinical cases. 
	
    First, we evaluated the reliability of the proposed method using 50 inspiratory thoracic cases from the COPDGene Phase 2 study cohort \cite{regan2011genetic}, for which 6 scans per subject acquired with a different dose and reconstructed with varying parameters were available. All scans for a subject were acquired using the same scanner, just varying the dose or using a different reconstruction approach. Of the 50 cases, 25 were taken using a GE scanner and 25 using a Siemens scanner. Each subject was scanned with two high dose reconstructions (standard, vs. sharp kernel), two field of views (bigger FOV vs. smaller FOV, both in standard high dose), and low dose with two reconstructions (standard and iterative). To make sure that patches extracted from one subject are the same for all scans, we computed the scale-space particles from the high dose image reconstructed with standard kernel and big FOV (STD), and we then registered the created particles to the other scans, using the advanced normalization tools (ANTs) technique described in \cite{avants2009advanced}. 
    
    For the statistical analysis, we compared the CNR measurements of each scan to those obtained for the reference image (STD) by means of intra-class correlation coefficient (ICC) and concordance correlation coefficient (CCC) as well as analysis of the distribution of the difference with Bland-Altman (BA) analysis, box-plots, and violin-plots.
    
    Finally, as a second in-vivo validation, we computed a physiological evaluation of the measurement. For the bronchial tree, we compared the correlation between Pi10 and the FEV1\% predicted using our approach and the 3D airway measurement software package Pulmonary Workstation (VIDA Diagnostics, Inc., Iowa City, IA, USA) on 3,038 clinical cases from the COPDGene Phase 1 study \cite{regan2011genetic}. Pi10 is a metric of airway thickness that is computed measuring the square root of the wall area across the whole airway tree and regressing the value at a hypothetical airway with an internal perimeter of 10 mm. The wall area is found by subtracting the area of the lumen from the airway area, while Pi is computed from the lumen radius. 
    
    Linear models were created to look at the association between our measurements and a metric of  functional small airway disease (fSAD) using the PRM method, a non-invasive technique that measures lung density during inhalation and exhalation, processes the resulting images, and classifies each point in three-dimensional space as normal lung parenchyma, functional small airway disease, or emphysema \cite{galban2012computed}. 
    
    For vessel measurement, we analyzed how total blood volume (TBV), blood volume of vessels of less than $5 mm^2$ (BV5), and blood volume of vessels of less than $10 mm^2$ (BV10) correlate to DLCO adjusted by geographical altitude, based on the study in \cite{estepar2013computed}. For this experiment, we compared the correlation obtained using the radius measure provided by our technique and the scale computed by the particle method \cite{kindlmann2009sampling} on 1,958 clinical cases from the COPDGene Phase 2 study \cite{regan2011genetic}. 
	
	\section{Results}
	
	\subsection{Synthetic Validation}
	\begin{table*}[t!]
		\caption{Mean relative error (RE) (in \%) obtained when measuring the wall thickness (WT) on 200,000 synthetically generated patches by using the proposed method (CNR) in comparison to Full Width Half Max (FWHM), and Zero Crossing Standard Deviation (ZCSD)). Three groups based on the known wall thickness value were considered.}
		\centering
		\resizebox{1.\columnwidth}{!}{%
			\begin{tabular}{c ccc ccc ccc}
				\hline
				& \multicolumn{3}{c}{\textbf{WT $\leq$ 0.7 mm}} & \multicolumn{3}{c}{\textbf{0.7 mm $<$ WT $\leq$ 1.5 mm}} & \multicolumn{3}{c}{\textbf{WT $>$ 1.5 mm}} \\
				& CNR & FWHM & ZCSD & CNR & FWHM & ZCSD & CNR & FWHM & ZCSD\\
				\hline
				RE (\%) & -7.6$\pm$16.7 & -1153.9$\pm$61185.4 & -1034.2$\pm$18297.8 & 1.04$\pm$7.5 & -582.9$\pm$17409.4 & -895.6$\pm$64770.8 & 0.28$\pm$2.8 & -450.8$\pm$50765.9 & -200.3$\pm$1274.3 \\
				\hline
			\end{tabular}
		}
		\label{REComparison}
	\end{table*}
	
	\begin{figure*}[t!]
		\centering
		\begin{tabular}{ccc}
			\centering
			\includegraphics[width=0.32\textwidth]{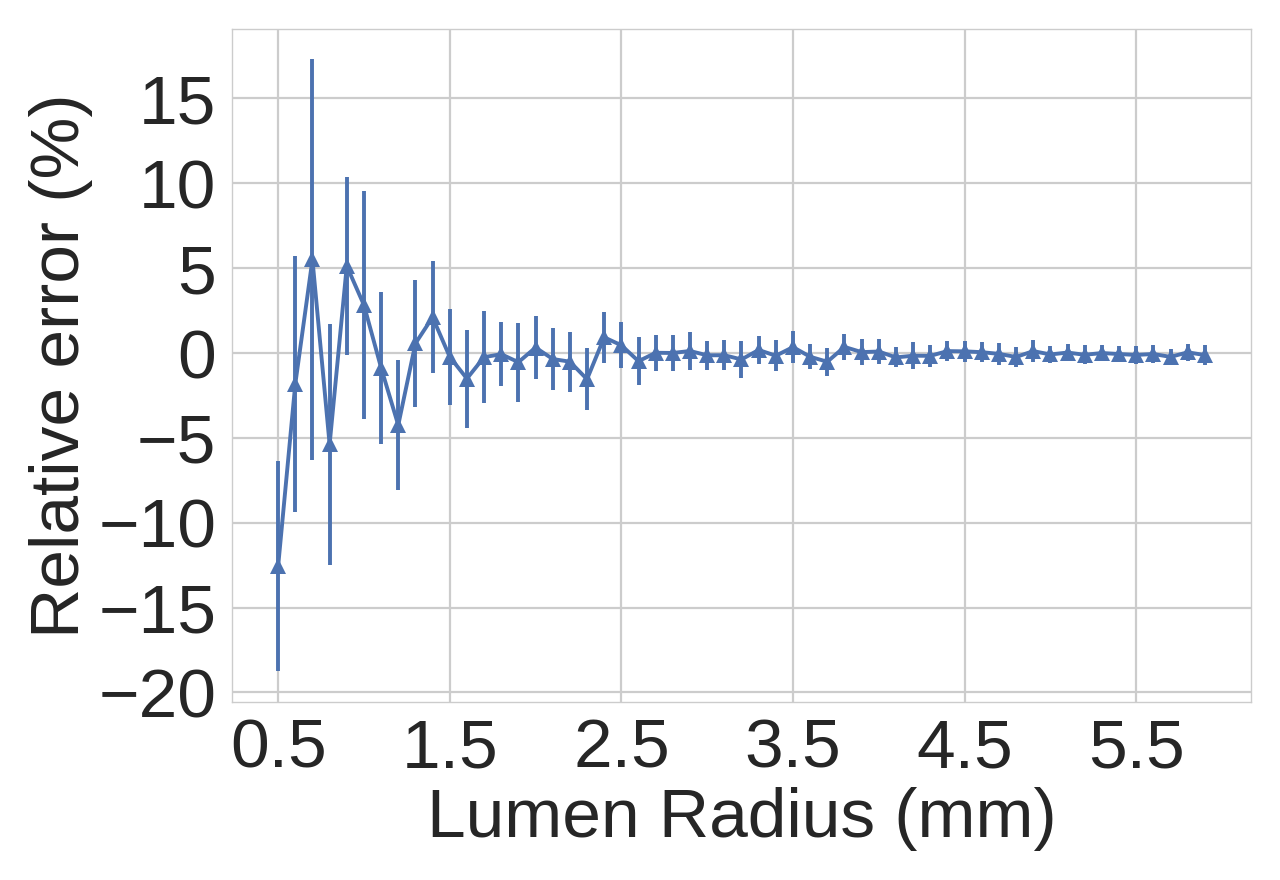}
			&
			\includegraphics[width=0.32\textwidth]{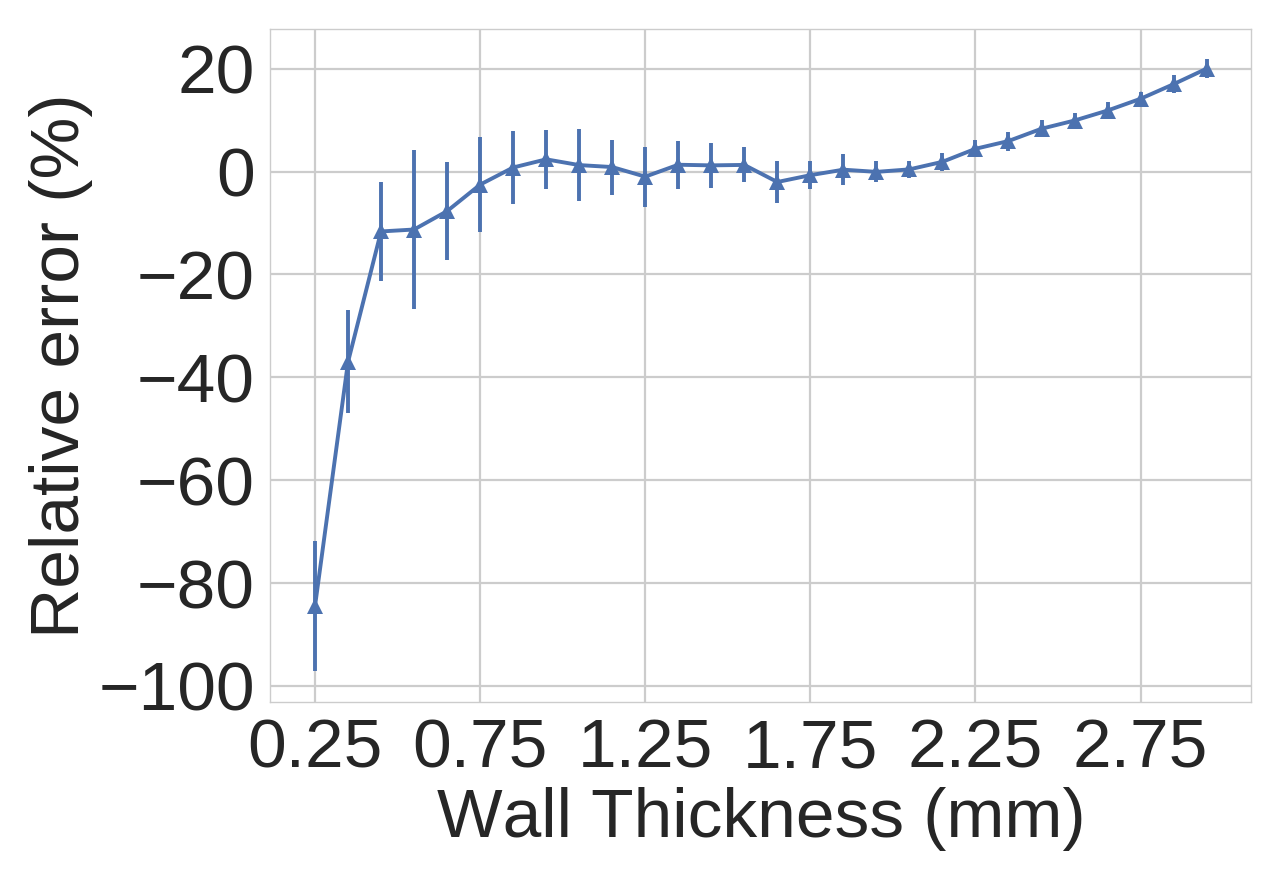} 
			&
			\includegraphics[width=0.32\textwidth]{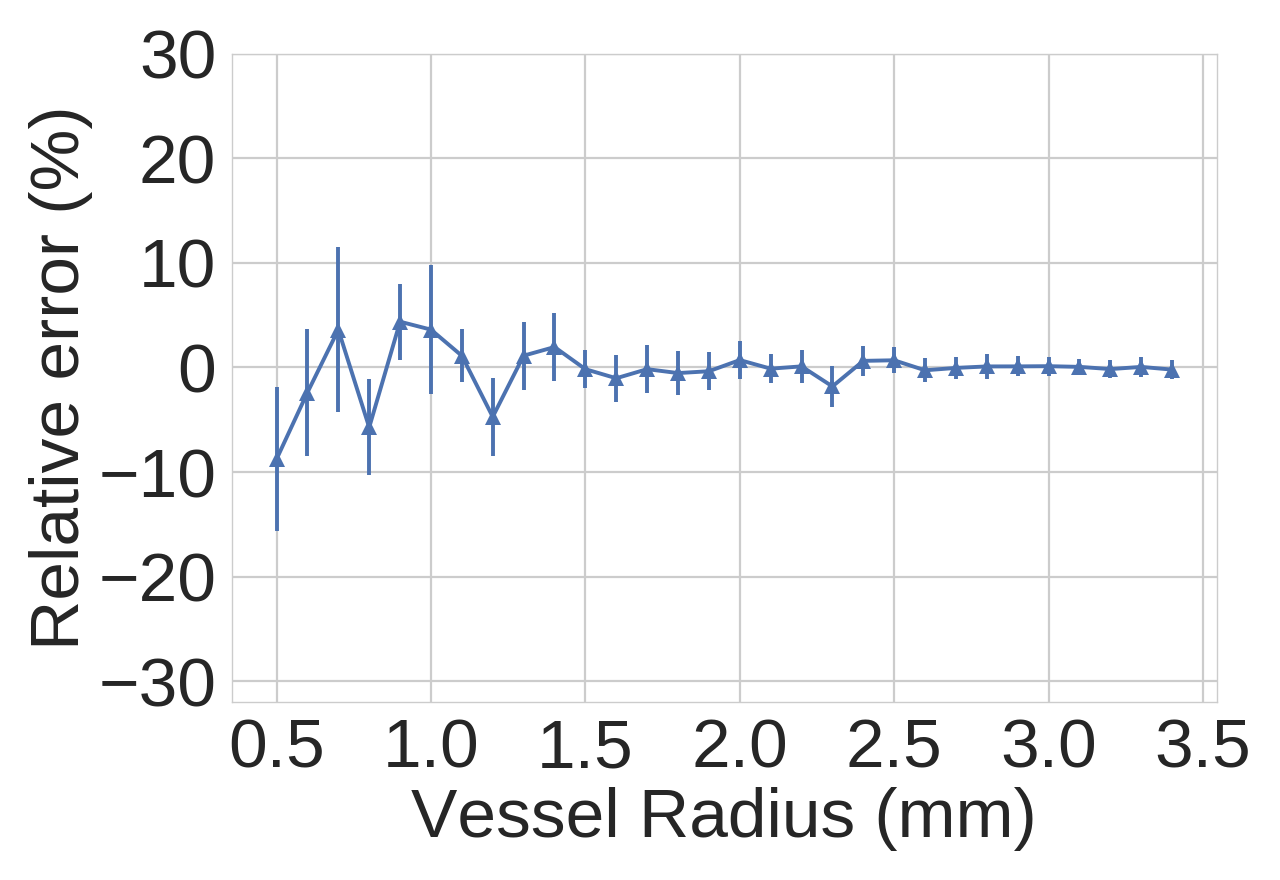} 
			\\
			(a) & (b) & (c) \\
		\end{tabular}
		\caption{Tendency of relative error (RE) obtained with the proposed CNR when varying (a) airway lumen, (b) wall thickness, and (c) vessel radius. For each size value, the mean RE is computed across 100 synethic patches generated with the proposed generator and refined with SimGAN.}
		\label{RelativeErrorSize} 
	\end{figure*}
	
	\begin{figure*}[th!]
		\centering
		\begin{tabular}{ccc}
			\centering
			\includegraphics[width=0.32\textwidth]{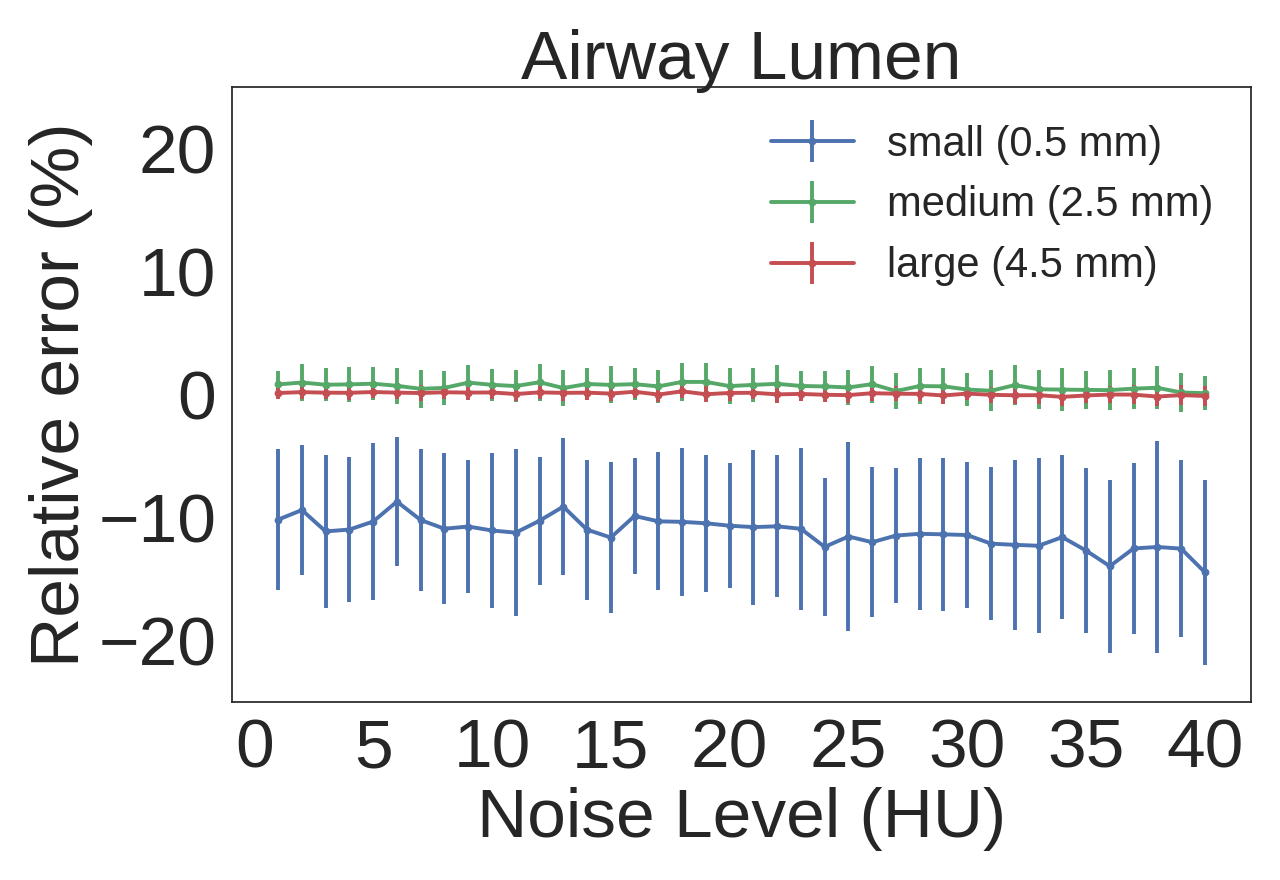}
			&
			\includegraphics[width=0.32\textwidth]{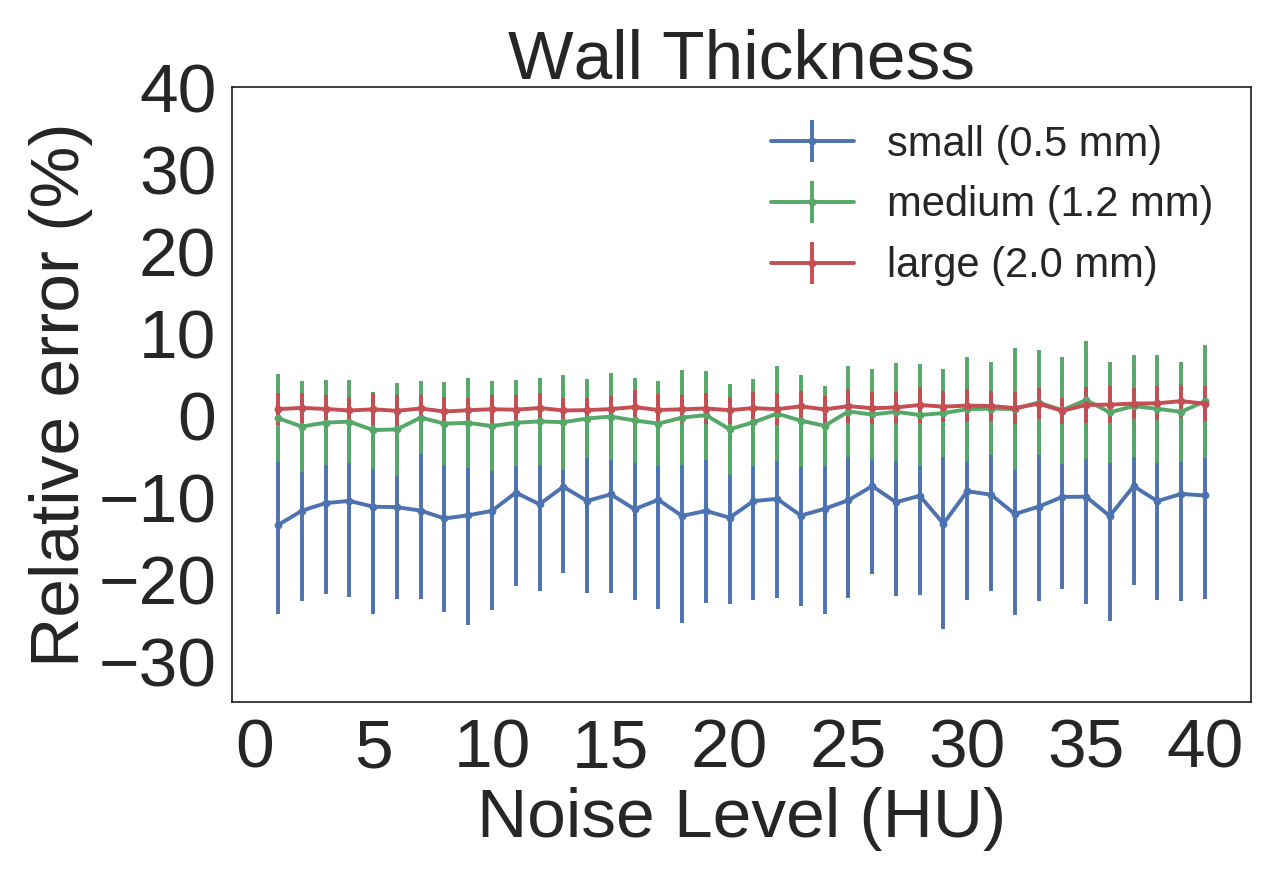} 
			&
			\includegraphics[width=0.32\textwidth]{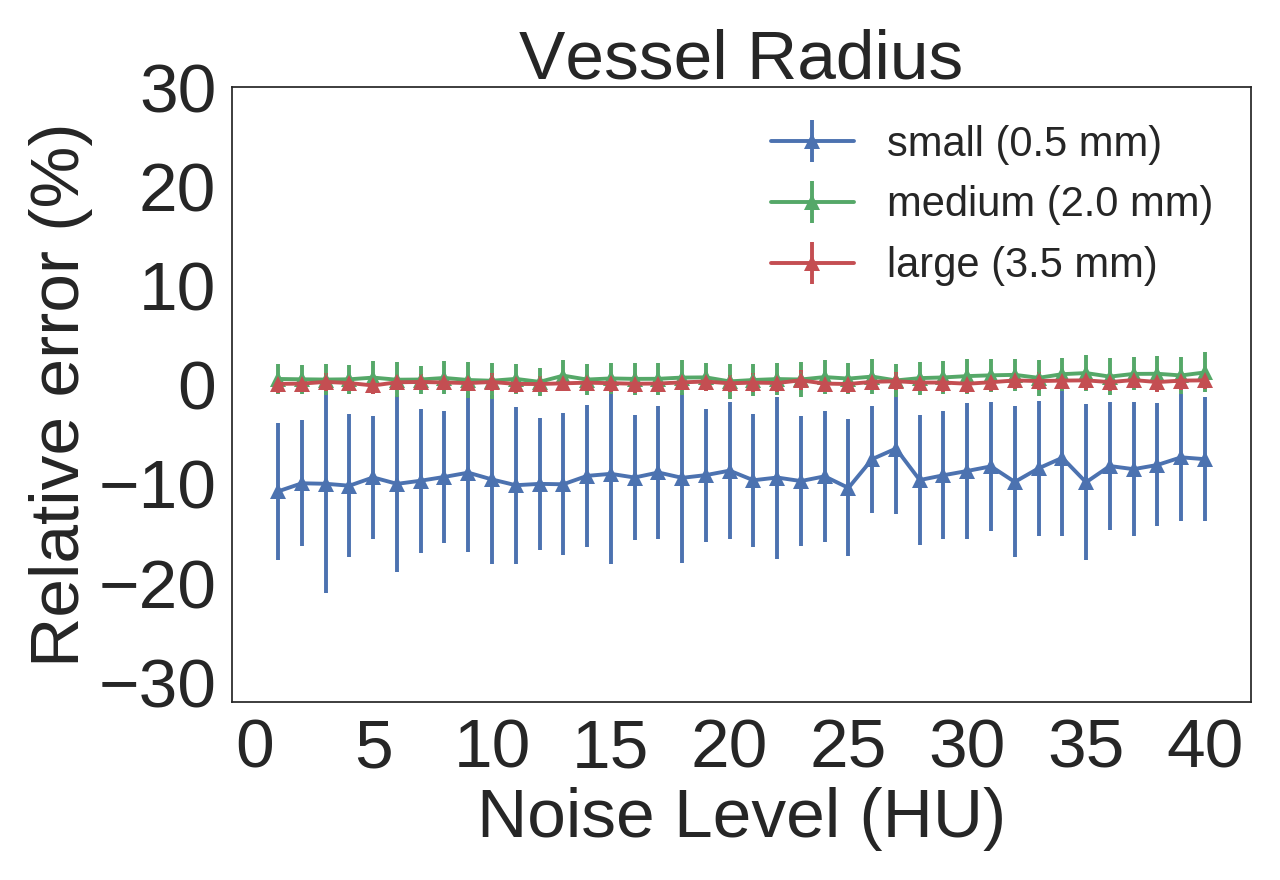} \\
			\includegraphics[width=0.32\textwidth]{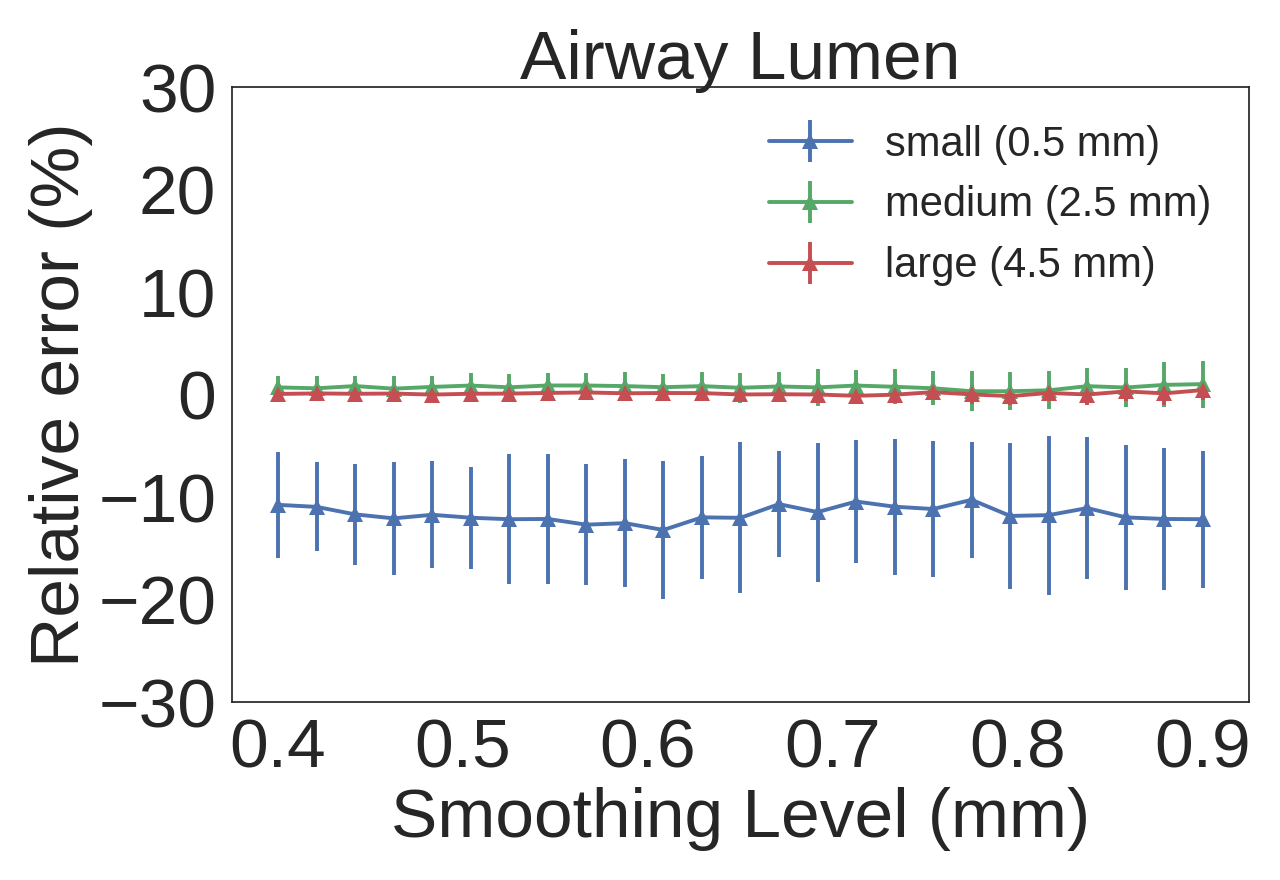}
			&
			\includegraphics[width=0.32\textwidth]{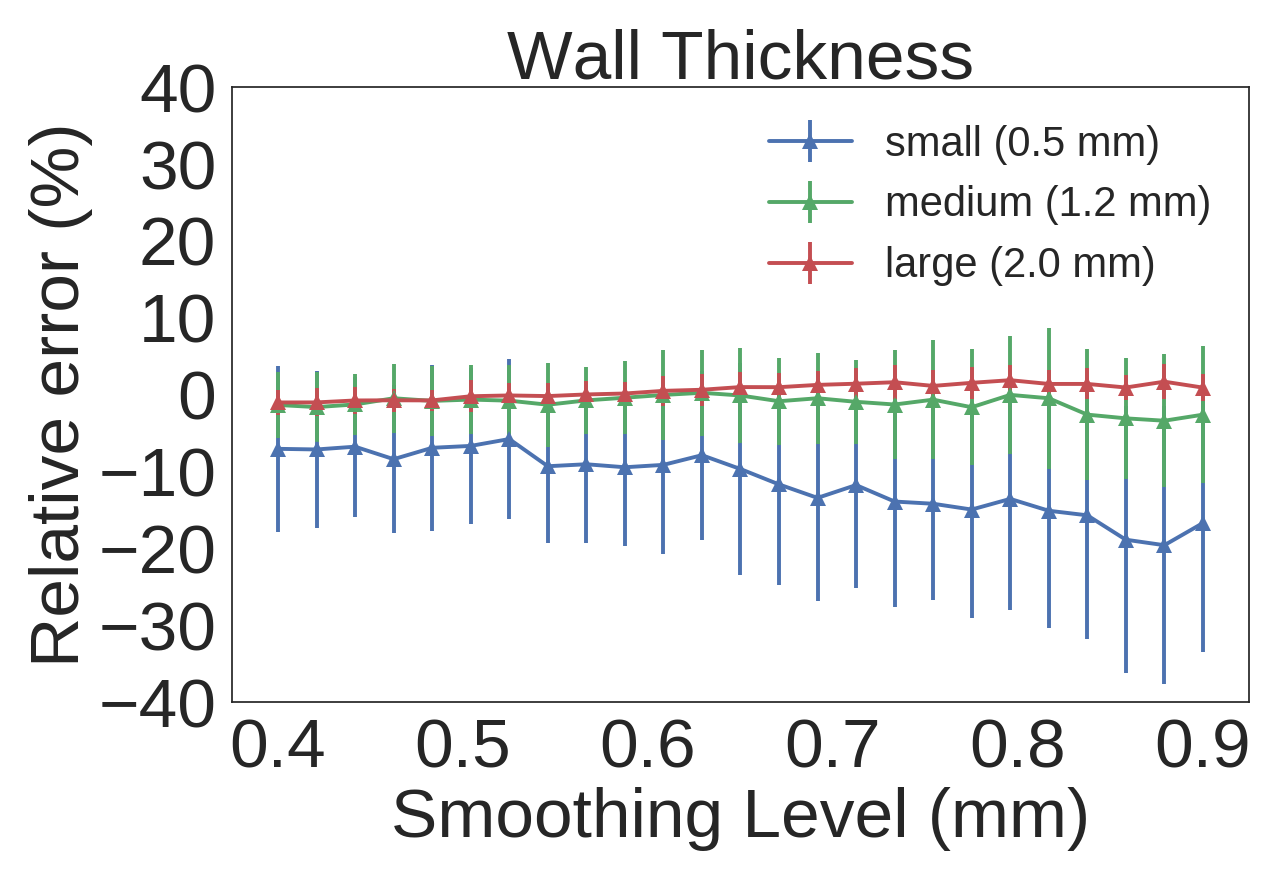} 
			& 
			\includegraphics[width=0.32\textwidth]{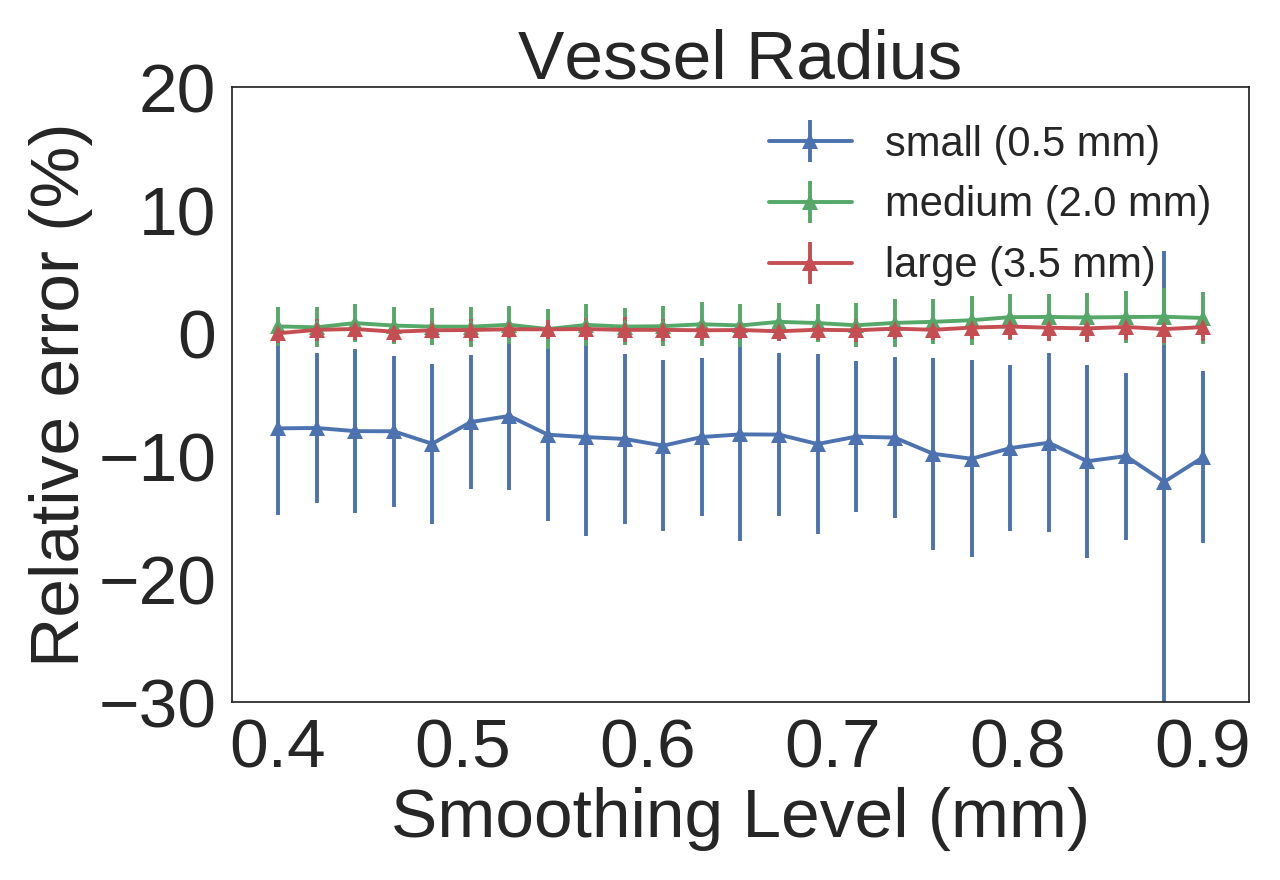} \\
			(a) & (b) & (c) \\
		\end{tabular}
		\caption{Effect of varying noise (top) and smoothing level (bottom) on (a) lumen (wall thickness=1.2 mm), (b) wall thickness (lumen=2.0 mm), and (c) vessel radius predictions. All results are reported in \%. For each noise level, smoothing was fixed at 1.3 mm and 100 syntethic images generated. Conversely, for each smoothing level the noise was fixed at 25 HU to generate 100 synthetic patches and compute the relative error.}
		\label{RENoiseSmooth} 
	\end{figure*}
	
	The mean absolute RE obtained for wall thickness, airway lumen, and vessel radius across the 200,000 generated patches was 4.9\%, 2.02\%, and 2.25\%, respectively. When considering airways with a wall thickness of 1.0 mm, a mean absolute RE of 6.3\% is obtained, while for wall thickness the image resolution level (0.5 mm) the mean absolute RE is 13.09\%. 
	
	These REs are significantly lower than those previously reported in the literature for structures of similar sizes \cite{reinhardt1997accurate,estepar2006accurate}. Regarding vessels, a mean absolute RE of 6.09\% is obtained for structures of 1.0 mm, while considering vessels with a radius of 0.5 mm the mean absolute RE was 11.3\%.
	
	In Tab. \ref{REComparison} the mean RE grouped by wall thickness size ($\leq$0.7 mm, $\in(0.7, 1.5]$ mm, and $>$1.5 mm) using CNR in comparison to ZCSD and FWHM on 200,000 testing patches is presented. As shown, while traditional methods yield a very high RE, especially for small airways, a small RE and a very high accuracy is obtained with the proposed method.
	
	In Fig. \ref{RelativeErrorSize}(a-c) the tendency of the RE obtained on synthetic patches when varying airway lumen (Fig. \ref{RelativeErrorSize}a), airway wall thickness (Fig. \ref{RelativeErrorSize}b), and the vessel lumen (Fig. \ref{RelativeErrorSize}c) is presented. As expected, a small RE is obtained for big airways, while a tendency to under-estimate the measure (although always below a 10\%) appears for small airways (lumen $<1.0$ mm). 
	
	For wall thickness results (Fig. \ref{RelativeErrorSize}b), at sub-pixels levels (wall thickness $<0.5$ mm) a significant under-estimation error is obtained, while for thicker walls (wall thickness $>2.0$ mm) the network tends to over-estimate the measurement. Conversely, for vessels (Fig. \ref{RelativeErrorSize}c), the RE obtained is close to zero, with just a small tendency to under-estimation for vessels of 0.5 mm.
	
	Fig. \ref{RENoiseSmooth} presents the results obtained when varying the noise and smoothing level applied to the generated patch and using three fixed values of airway lumen size (0.5, 2.5, and 4.5 mm), wall thickness (0.5, 1.2, and 2.5 mm), and vessel lumen radius (0.5, 2.0, and 3.5 mm). As shown, for all structures a stable RE across the varying levels of noise and smoothness is obtained. The CNR yields a very high accuracy (RE $~0\%$) for medium and large structures, while a slightly higher RE with a bigger standard deviation is obtained when airway lumen, airway wall thickness, and vessel lumen size are at the image resolution of 0.5 mm. 
	
	For the three structures, CNR yields a stable RE when varying noise and smoothing level with only a small bias introduced determined by a little under-estimation of the small structures, as expected. For airways with a small wall thickness of 0.5 mm, if the level of smoothing is below 0.6 mm a very small RE is obtained, while if higher levels of smoothing are introduced to the patches the RE slightly increases. 
	
	\subsubsection{SimGAN Validation}
	One of the key aspects of the proposed algorithm is the usage of an adversarial mechanism that allows a refinement of synthetic patches into more realistic images. 
	
	We validate the output of the SimGAN by testing the performance of the trained discriminator to distinguish between real and synthetic patches. To this end, we used the MBG to generate 300,000 new synthetic airway patches and vessel patches with varying parameters and we extracted 31,316 real airway images and 155,000 real vessel patches (randomly selected) from 10 subjects of the COPDGene Phase 2 study that were not used for training the discriminator. 
	
	When passed to the discriminator, 98.772\% of synthetic airway patches and 98.983\% synthetic vessel patches were classified as coming from the real domain, indicating the fidelity of the SimGAN results in terms of discrimination. As for real patches, 99.901\% airways and 99.985\% vessels were correctly classified by the discriminator.

	\subsection{Phantom Evaluation}
	In Tab. \ref{phantom_results} the RE obtained when measuring the wall thickness of the eight tubes of the phantom using both the proposed method (CNR) and traditional techniques is presented. Even though the lumen size is not provided by traditional methods, for completness the RE obtained when measuring the airway lumen with our CNR is also reported in Tab. \ref{phantom_results}. 
	
	In general, the proposed CNR yields the lowest RE for all tubes with the exception of tube C that is best measured by FWHM. Also, an important aspect to notice is the small RE obtained for all tubes when measuring the lumen radius with the proposed technique.
		
	\begin{table}[t!]
		\caption{Mean relative error (RE) obtained for wall thickness (WT) measurement of the eight phantom tubes when using the proposed method (CNR) and traditional algorithms (FWHM and ZCSD). The smaller REs are reported in bold. For completeness, the last column reports the relative error obtained when measuring tubes' lumen with CNR (lumen not provided by traditional methods). All results are in \%.}
		\centering
		\resizebox{0.6\textwidth}{!}{
		\begin{tabular}{ccccc}
			\hline
			\textbf{Tube} & \textbf{CNR WT} & \textbf{FWHM WT} & \textbf{ZCSD WT} & \textbf{CNR Lumen}\\
			\hline
			A & \textbf{-2.5} & -115.1 & -126.1 & 4.6 \\
			B & \textbf{17.5} & -25.9 & -33.6 & -6.3 \\
			C & -28.9 & \textbf{-23.6} & -46.2 & 7.6 \\
			D & \textbf{-12.8} & -474.0 & -79.0 & 10.4 \\
			E & \textbf{-13.0} & -4884.6 & -20.5 & 9.8 \\
			F & \textbf{2.4} & -55.3 & -59.8 & -5.2 \\
			G & \textbf{5.4} & -18.06& -29.2 & -7.8 \\
			H & \textbf{18.5} & -30.7 & -40.3 & -5.2 \\ 
			\hline
		\end{tabular}
		}	
		\label{phantom_results}
	\end{table}
	
	\subsection{In-vivo Indirect Evaluation}
		\begin{figure}[t!]
		\centering
		\includegraphics[width=1.\columnwidth]{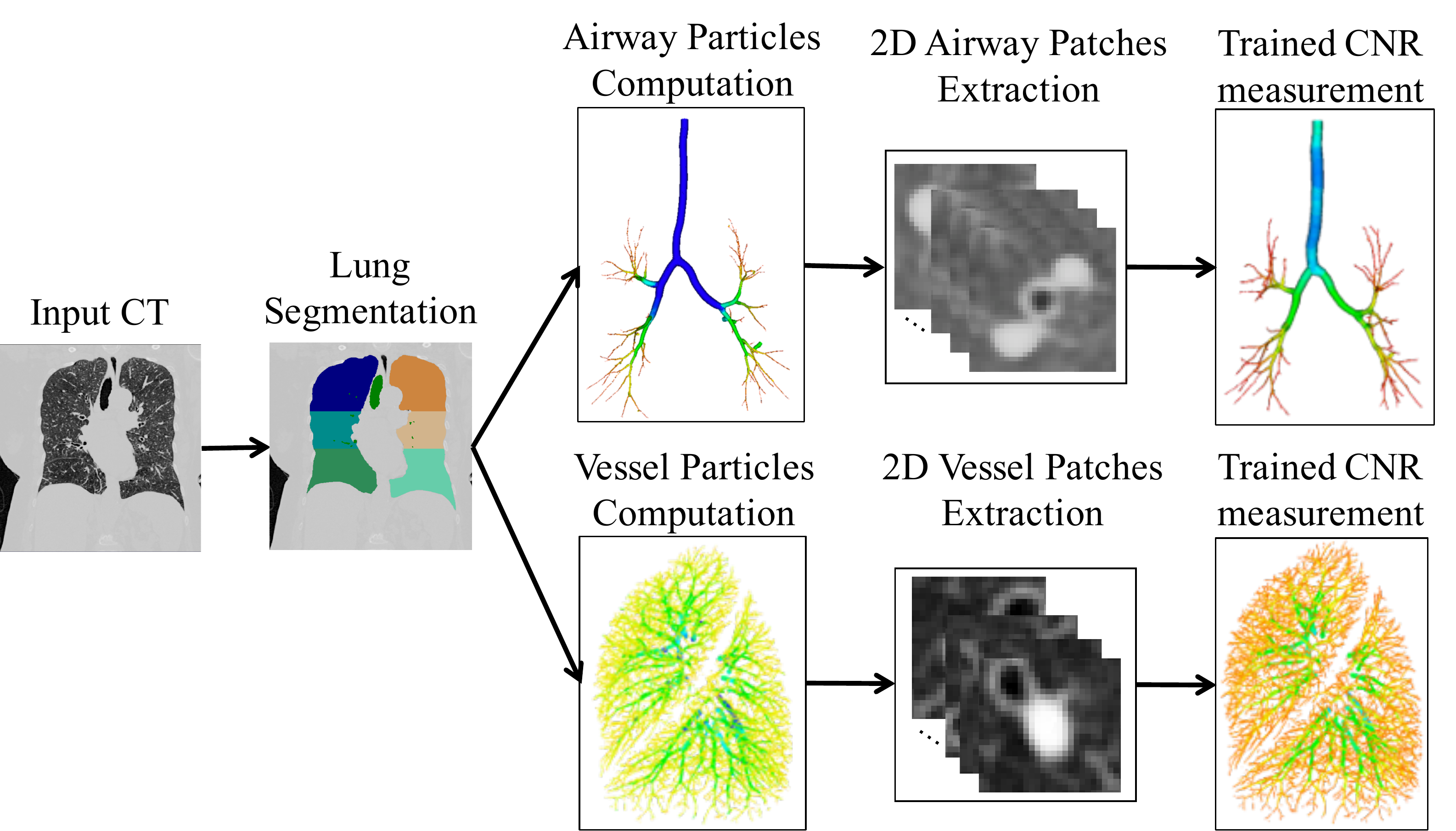} 
		\caption{Processing pipeline used for airway and vessel morpholgy assessment from chest CT images. The lung region is first segmented to then generate airway and vessel particles. From each particle point, axial-reformatted patches of 16$\times$16$\times$16 are then extracted and used as input of the trained CNR to measure athe structure physical dimensions.}
		\label{deployment_scheme} 
	\end{figure}
	
	For in-vivo evaluation, airway and vessel morphology has to be computed from CT images. In Fig. \ref{deployment_scheme}, a scheme of the processing pipeline to obtain bronchial and vascular measurement from the segmented tree is presented. For each particle, the corresponding structure size is computed.
	
	\subsubsection{Consistency to CT acquisition parameters}
	
	The CCC values obtained measuring airways and vessels when varying a single CT parameter (kernel, FOV, dose, reconstruction) in comparison to the corresponding reference image (STD) are presented in Tab. \ref{ICCtable}. In general, very high results are obtained, although, as expected, the measurement of wall thickness is the most complicated one. 
	
\begin{figure*}[t!]
		\centering
		\begin{tabular}{ccc}
			\centering
			\includegraphics[width=0.3\textwidth]{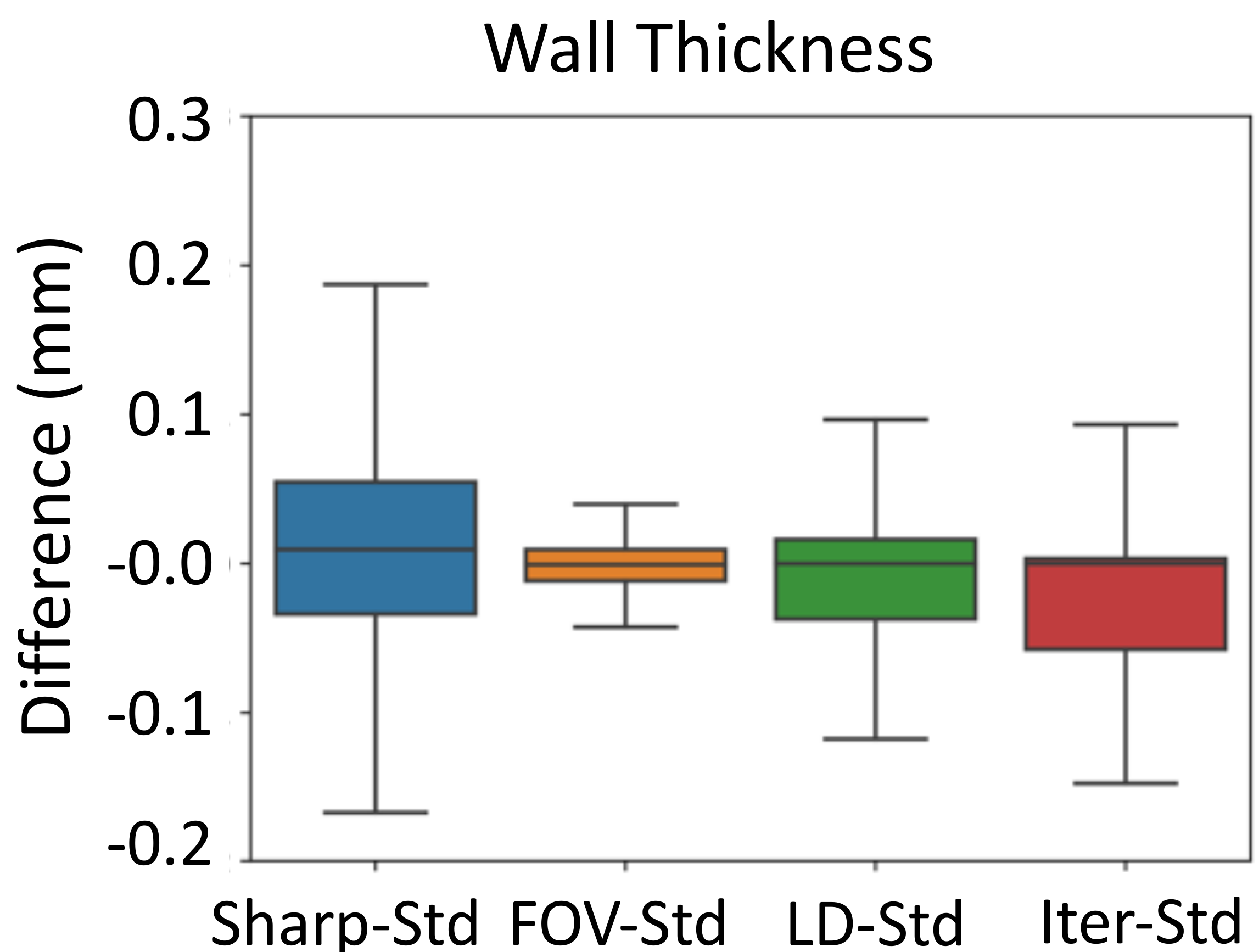}
			&
			\includegraphics[width=0.29\textwidth]{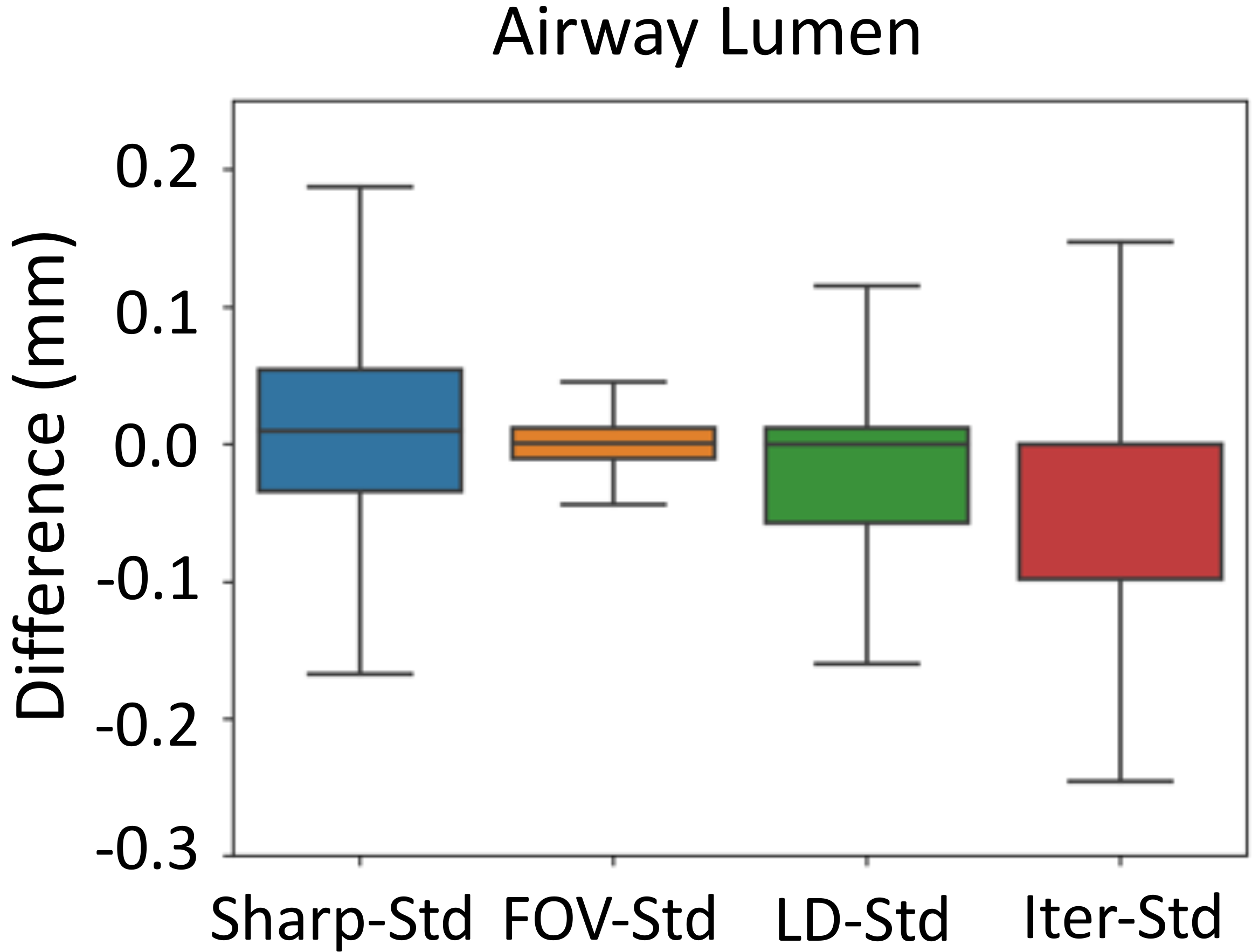} 
			&
			\includegraphics[width=0.3\textwidth]{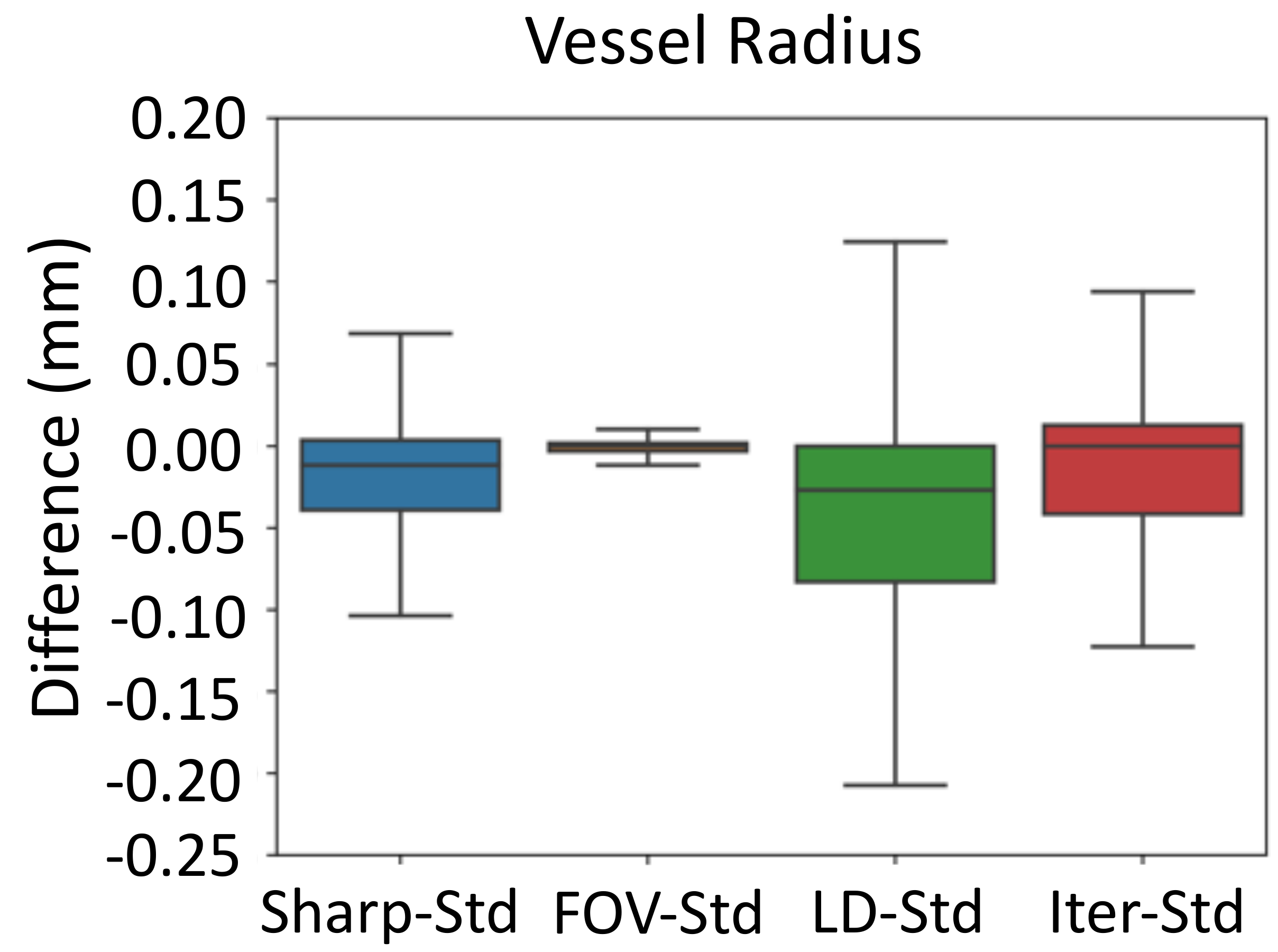} \\
			(a) & (b) & (c) \\
		\end{tabular}
		\caption{Box-plots of the difference to the reference image (STD, acquired with high dose, standard kernel, and bigger field of view) for (a) wall thickness, (b) airway lumen, and (c) vessel radius obtained when varying kernel (Sharp-STD, blue), field of view (FOV-STD, orange), dosage (LD-STD, green), and low dose reconstruction method (Iter-STD, red).}
		\label{boxplotComparison} 
	\end{figure*}

	In accordance to these results, the ICC obtained for all classes in comparison to the reference images was 97.5\%. The box-plots comparing the different parameter variations to the STDs are presented in Fig. \ref{boxplotComparison}. For completeness, bias and limits of agreement of the Bland-Altman (BA) analysis for all structures in comparison to measurements obtained on the reference image (STD) are shown in Tab. \ref{BA_diff}. 
	
	\begin{table}[t!]
		\caption{Concordance correlation coefficient (CCC) values (in \%) obtained measuring wall thickness (WT), airway lumen radius (ALR), and vessel radius (VR) when varying scan protocols. STD stands for reference scan (standard high dose, bigger FOV). SHARP are sharp high dose CTs. FOV are standard high dose CTs with smaller field of view, and LD and ITER are low dose images taken with standard and iterative reconstruction, respectively.}
		\centering
		\begin{tabular}{c c c c}
			& \multicolumn{3}{c}{\textbf{CCC (\%)}} \\
			\hline
			& WT & ALR & VR \\
			\hline
			STD-SHARP & 96.7 & 99.6 & 99.0\\
			STD-FOV & 99.3 & 99.7 & 99.9 \\
			STD-LD  & 90.8 & 94.2 & 95.3 \\
			STD-ITER & 87.3 & 91.3 & 96.4 \\
			\hline
		\end{tabular}
		\label{ICCtable}
	\end{table}

	\begin{table*}[th!]
		\caption{Bias and limits of agreement from a Bland-Altman analysis for (a) wall thickness, (b) airway lumen, and (c) vessel radius for the difference (in mm) to the reference image (STD, acquired with high dose, standard kernel, and bigger field of view) when varying kernel (Sharp-STD), field of view (FOV-STD), dosage (LD-STD), and low dose reconstruction method (Iter-STD). SD is used for standard deviation.}
		\centering
		\resizebox{1.\textwidth}{!}{
		\begin{tabular}{ccc}
			\textbf{Wall Thickness (mm)} & \textbf{Airway Lumen (mm)} & \textbf{Vessel Radius (mm)} \\
			\begin{tabular}{c c c}
				\hline
				&  Bias & [-1.96 SD, +1.96 SD] \\
				\hline
				Sharp-Std & -0.04 & [-0.23, 0.31]  \\ 
				FOV-Std & 0.00 & [-0.09, 0.10] \\
				LD-Std & 0.03 & [-0.45, 0.51] \\  
				Iter-Std & 0.03 & [-0.30, 0.37] \\
				\hline
			\end{tabular} 
			&
			\begin{tabular}{c c c}
					\hline
				&  Bias & [-1.96 SD, +1.96 SD] \\
				\hline
				Sharp-Std & 0.01 & [-0.21, 0.20]  \\ 
				FOV-Std & 0.00 & [-0.10, 0.10] \\
				LD-Std & 0.04 & [-0.42, 0.5] \\  
				Iter-Std & 0.10 & [-0.56, 0.76] \\
				\hline
			\end{tabular}
			&
			\begin{tabular}{c c c}
					\hline
				&  Bias & [-1.96 SD, +1.96 SD] \\
				\hline
				Sharp-Std & 0.02 & [-0.08, 0.13] \\ 
				FOV-Std & 0.00 & [-0.03, 0.03] \\
				LD-Std & 0.05 & [-0.18, 0.28] \\  
				Iter-Std & 0.01 & [-0.20, 0.23] \\
				\hline
			\end{tabular}
			 \\
			(a) & (b) & (c)			
		\end{tabular}
		}
		\label{BA_diff}
	\end{table*}
		
\begin{figure*}[t]
		\centering
		\begin{tabular}{ccc}
			\centering
			\includegraphics[width=0.3\textwidth]{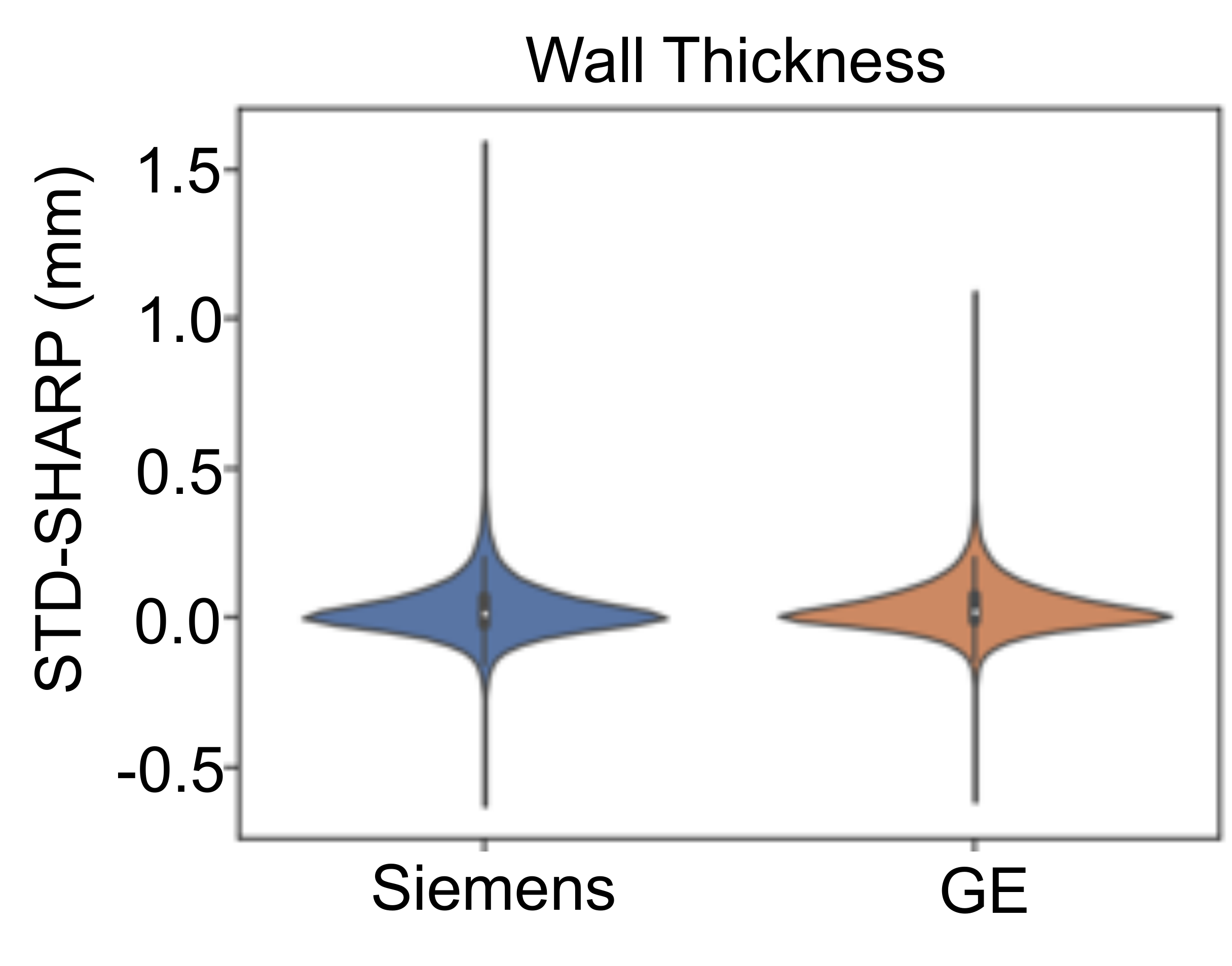}
			&
			\includegraphics[width=0.3\textwidth]{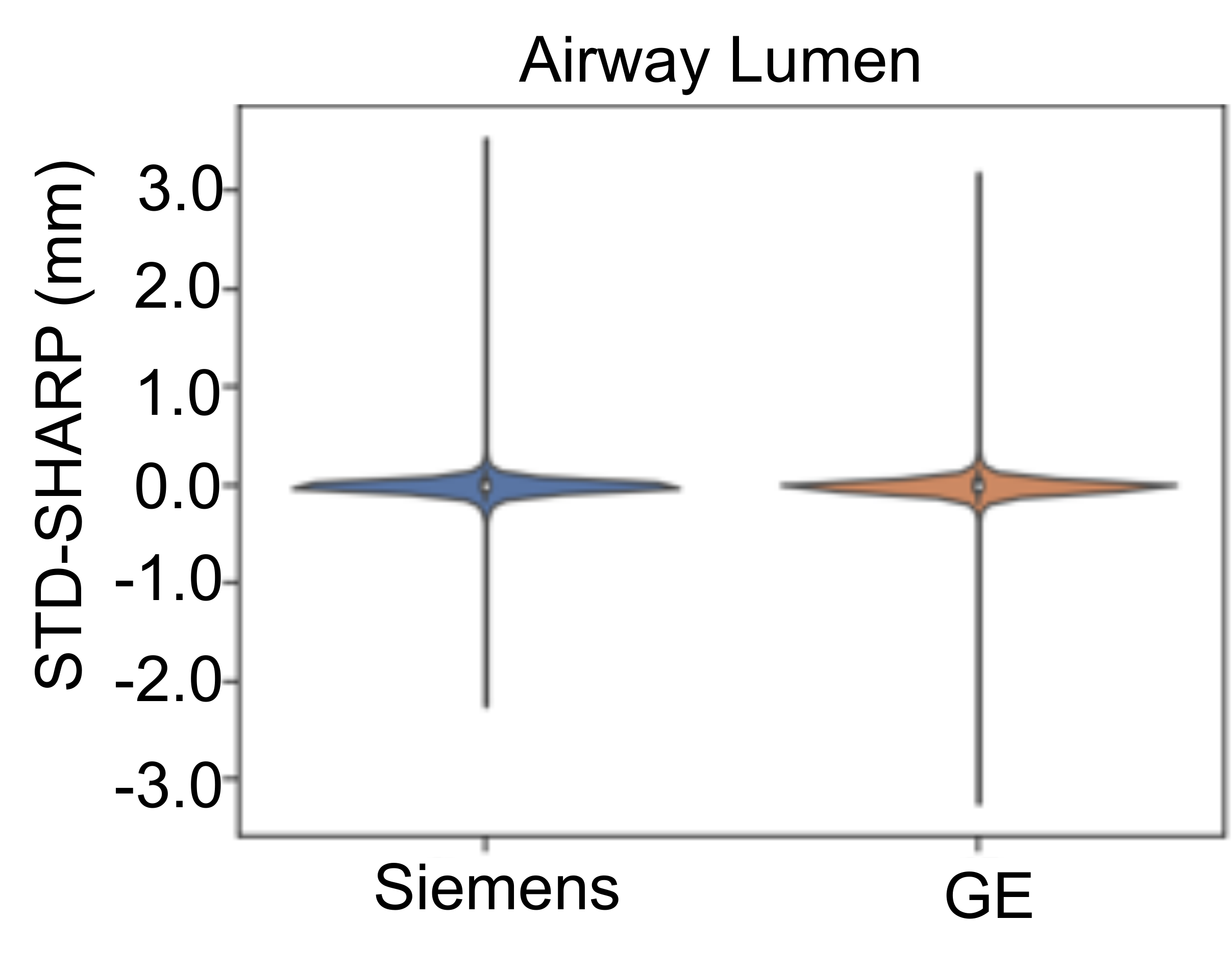} 
			&
			\includegraphics[width=0.32\textwidth]{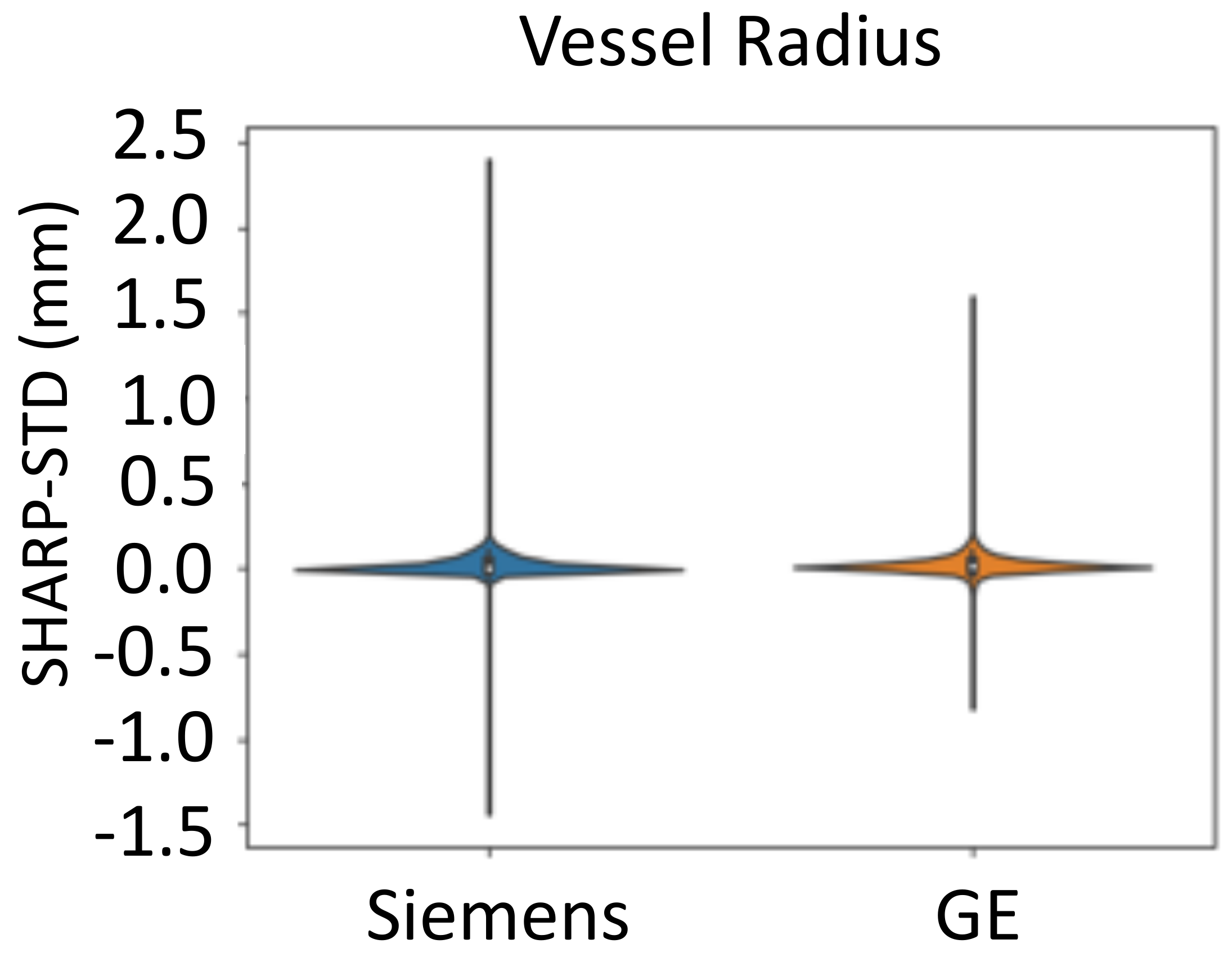} \\
			(a) & (b) & (c) \\
		\end{tabular}
		\caption{Violin-plot of difference in the threes structures obtained when varying the kernel (standard, STD, vs SHARP) for Siemens (blue) and GE (orange) images. Only intra-parenchymal airways were considered}.
		\label{GESiemens} 
	\end{figure*}

	Fig. \ref{GESiemens} shows the violin-plot of the measurements made on sharp images in comparison to the reference scan taken with GE or Siemens scanners. For this analysis, only intra-parenchymal airways (with sizes in the range used for training the network) were considered. As shown, the scanner type is not affecting the measure, as similar differences (mean close to zero) between the images taken with standard and sharp kernels are obtained with the two scanners for all structures.
	
\begin{figure*}[t]
		\centering
		\begin{tabular}{cccc}
			\centering
			\includegraphics[width=0.22\textwidth]{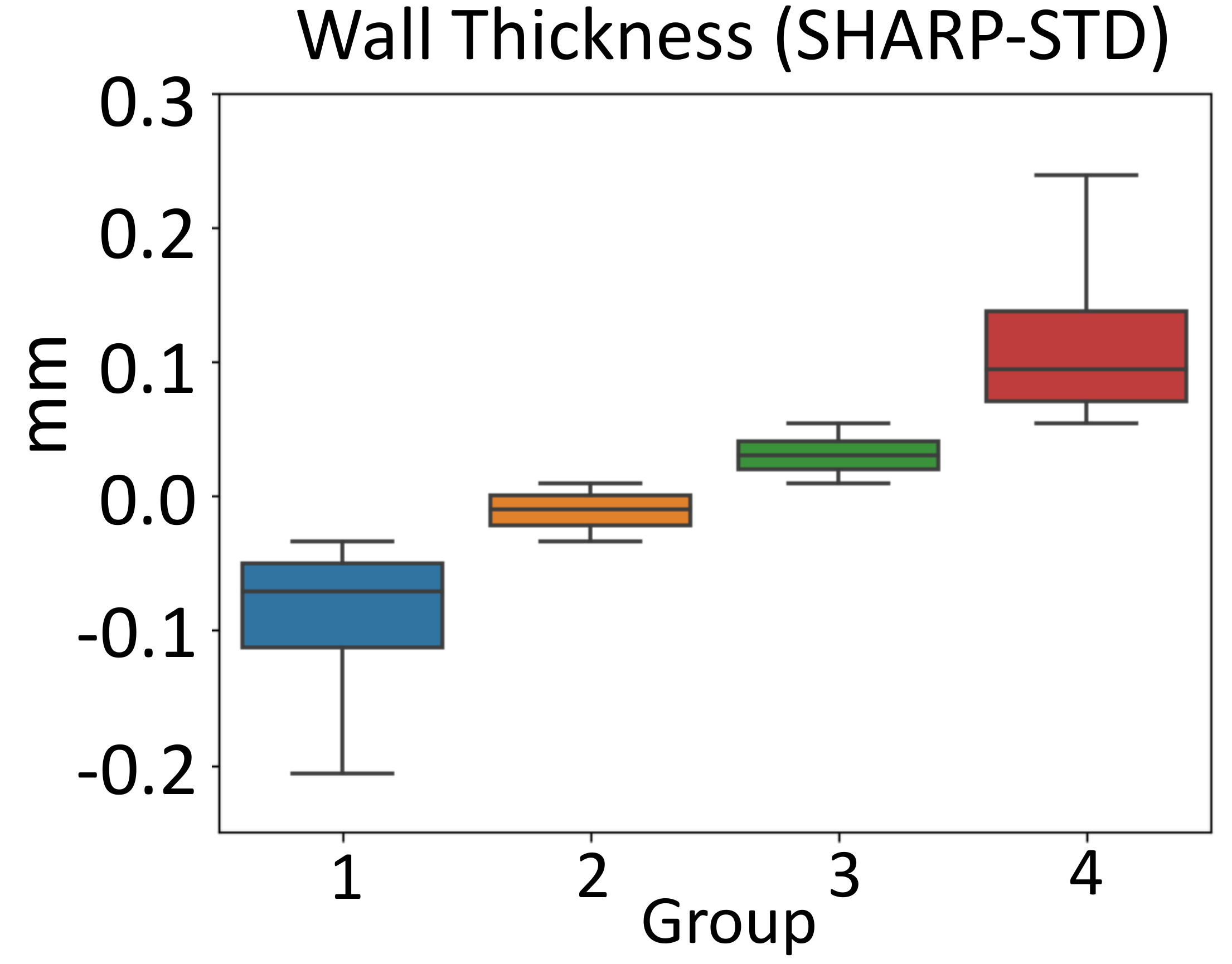}
			&
			\includegraphics[width=0.23\textwidth]{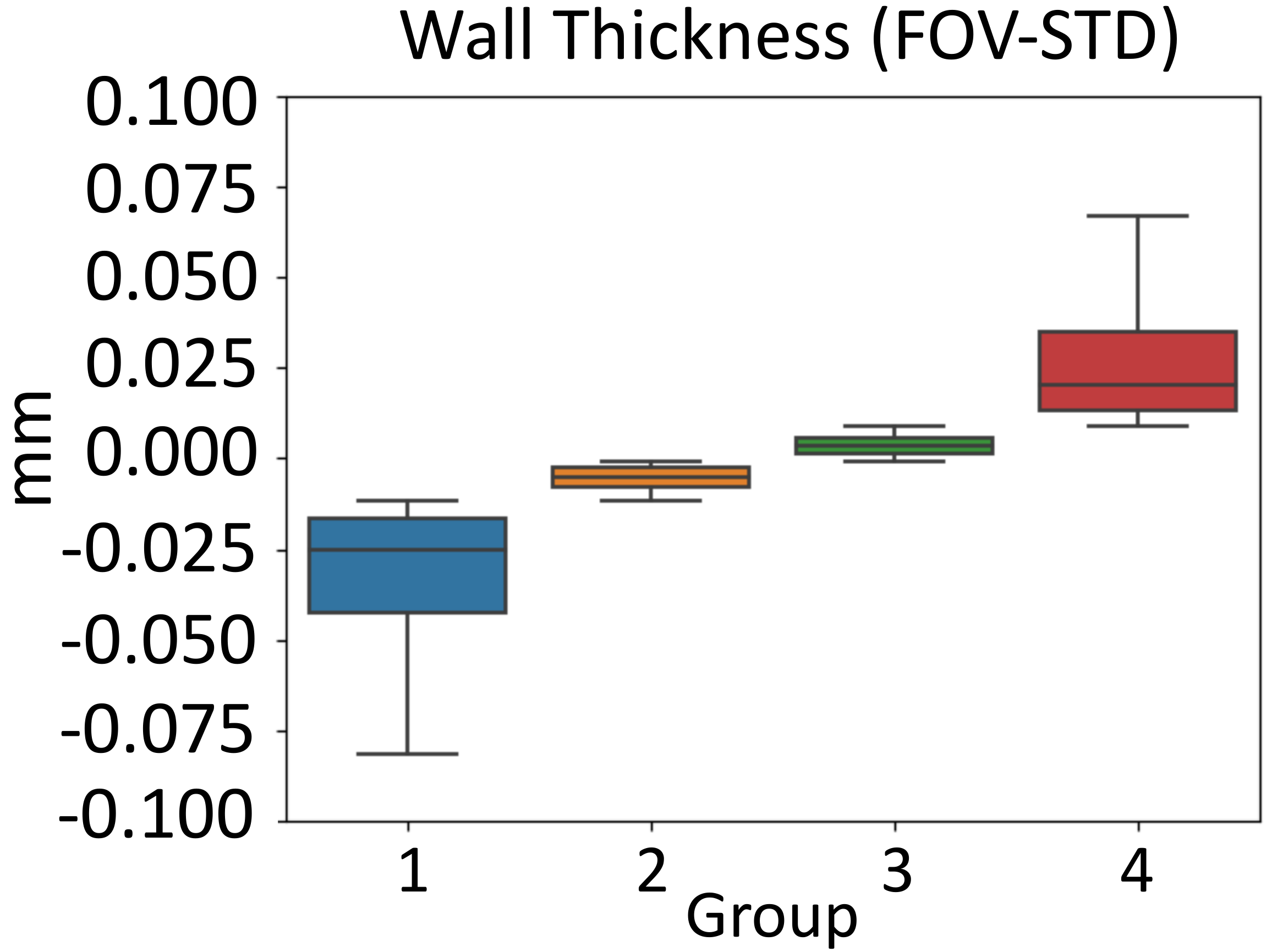} 
			&
			\includegraphics[width=0.22\textwidth]{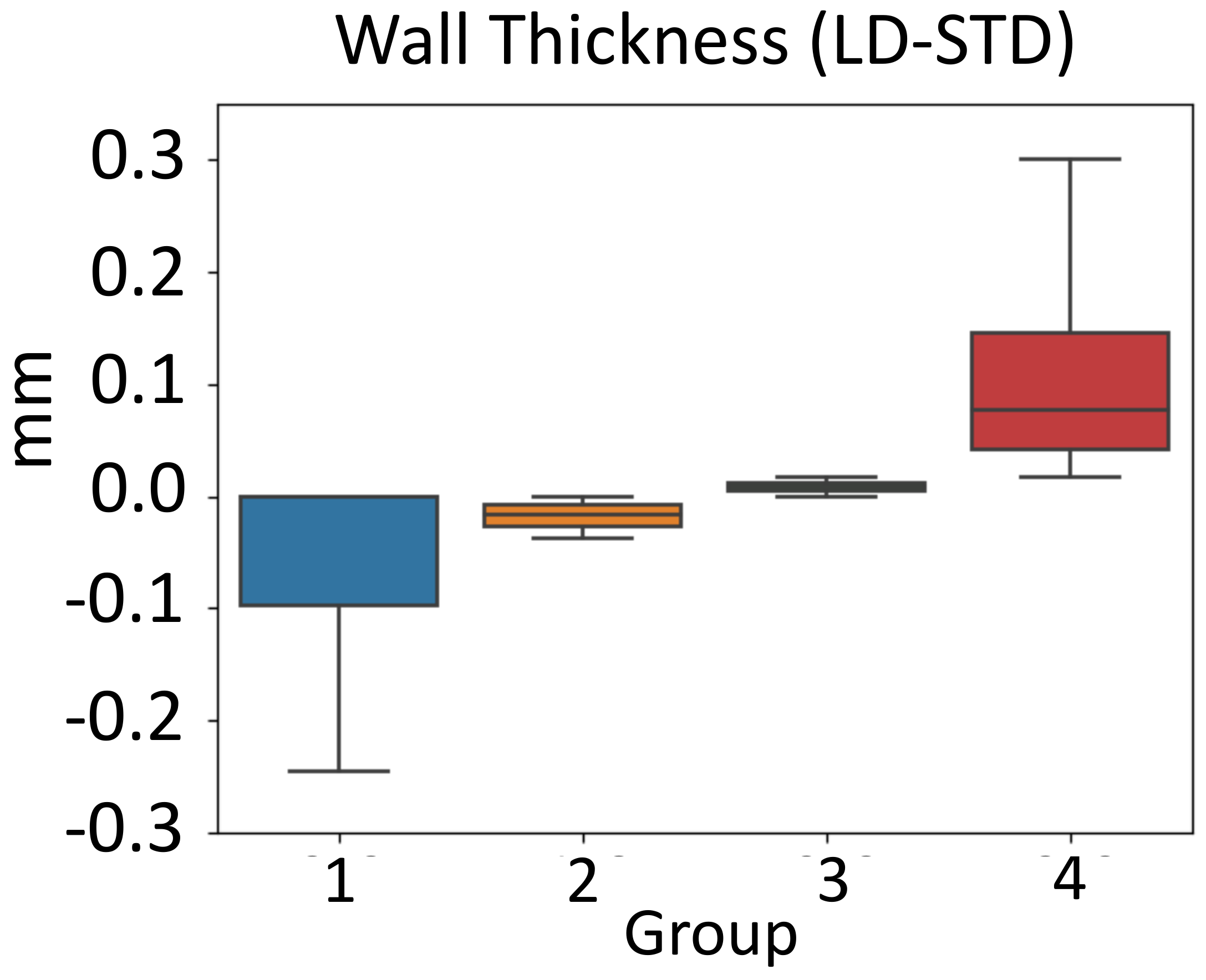} 
			&
			\includegraphics[width=0.22\textwidth]{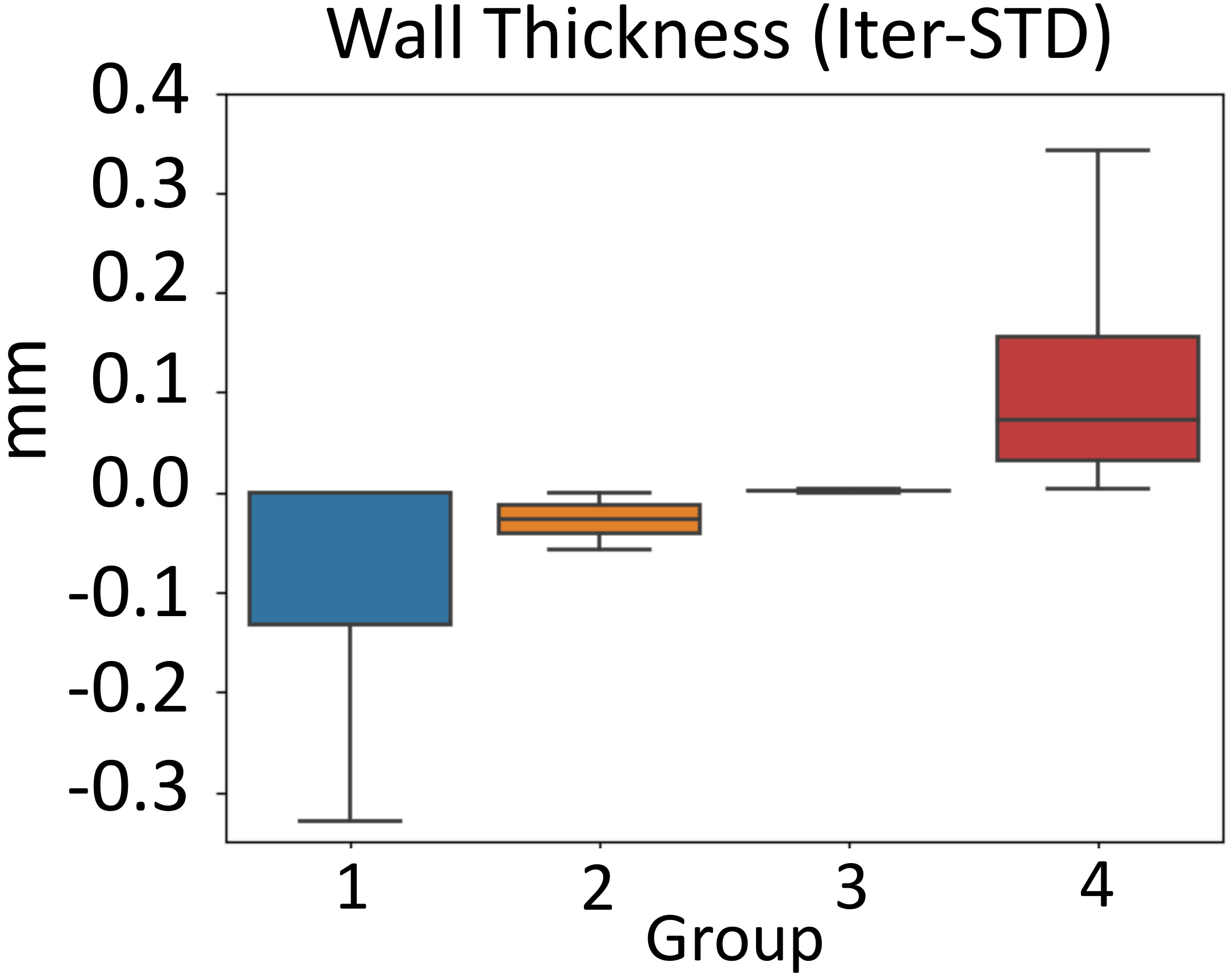}  \\
			
			\includegraphics[width=0.22\textwidth]{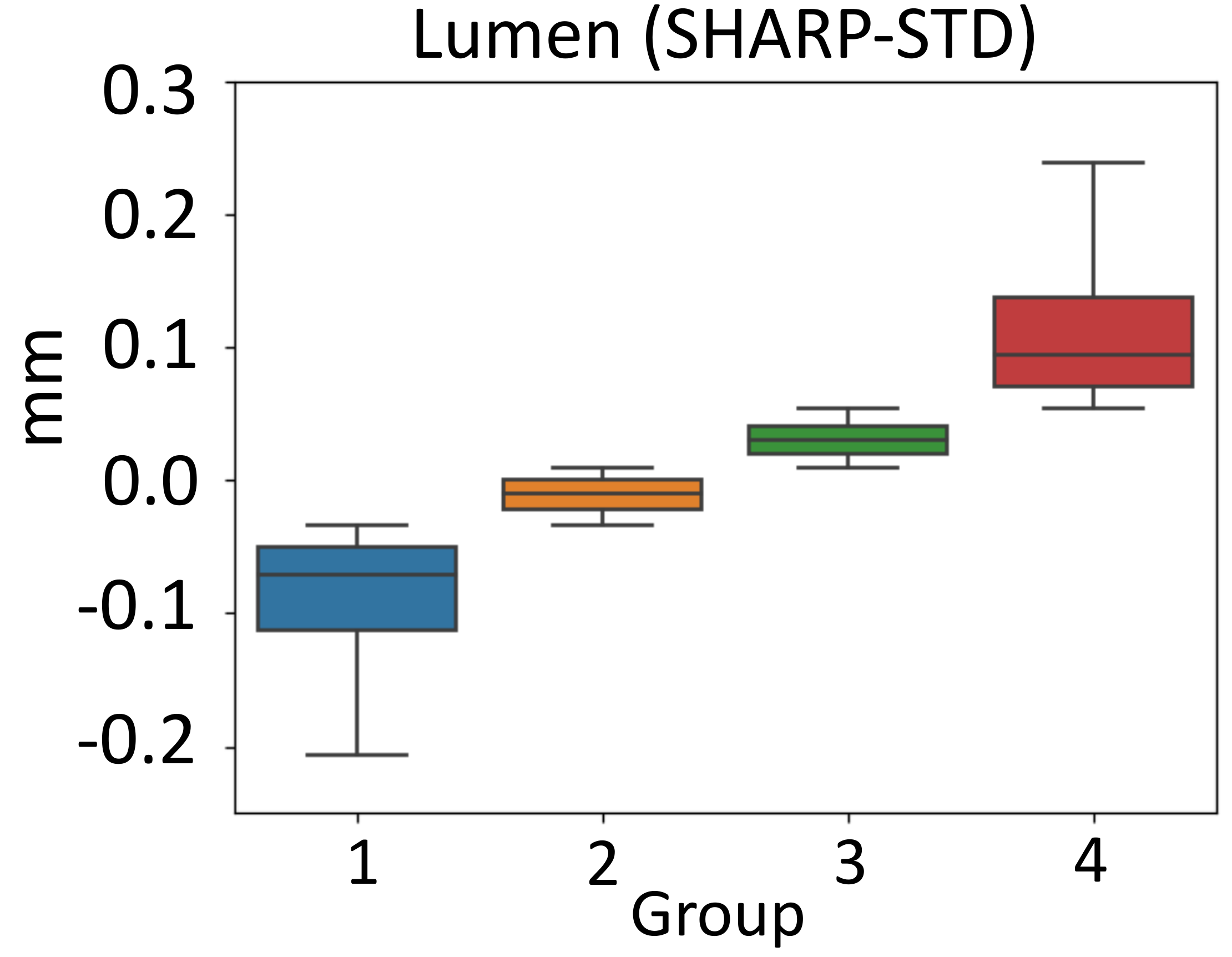}
			&
			\includegraphics[width=0.23\textwidth]{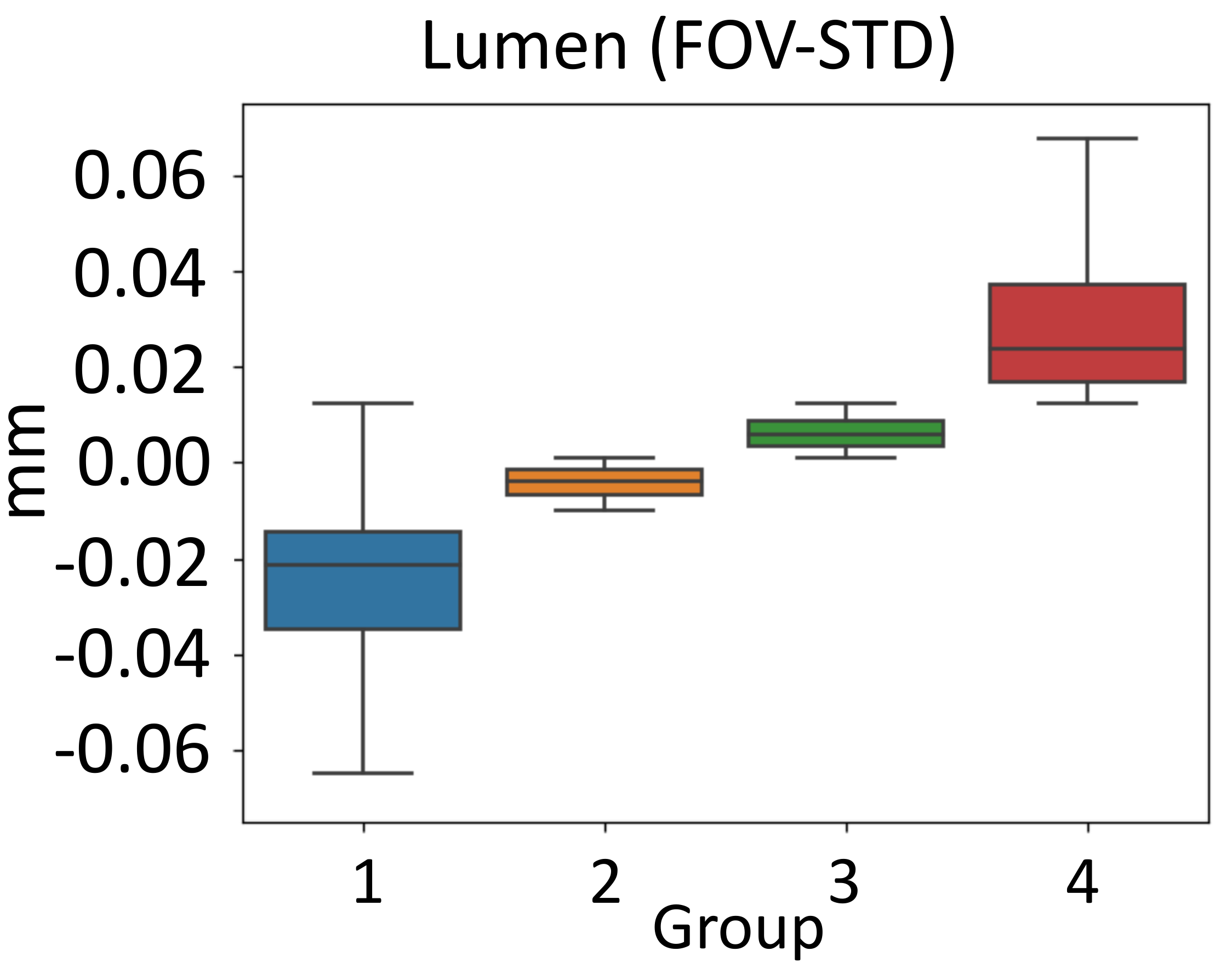} 
			&
			\includegraphics[width=0.22\textwidth]{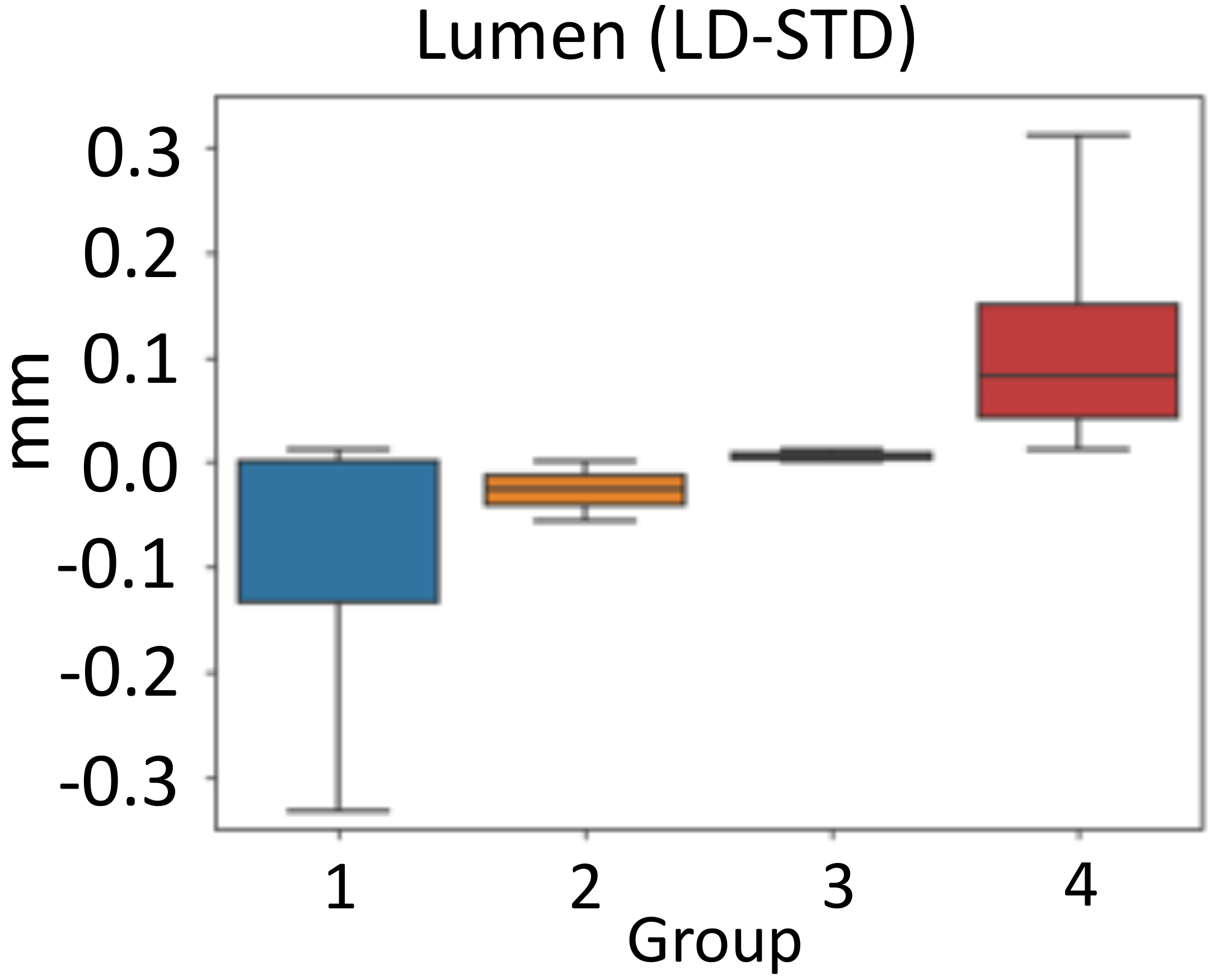} 
			&
			\includegraphics[width=0.22\textwidth]{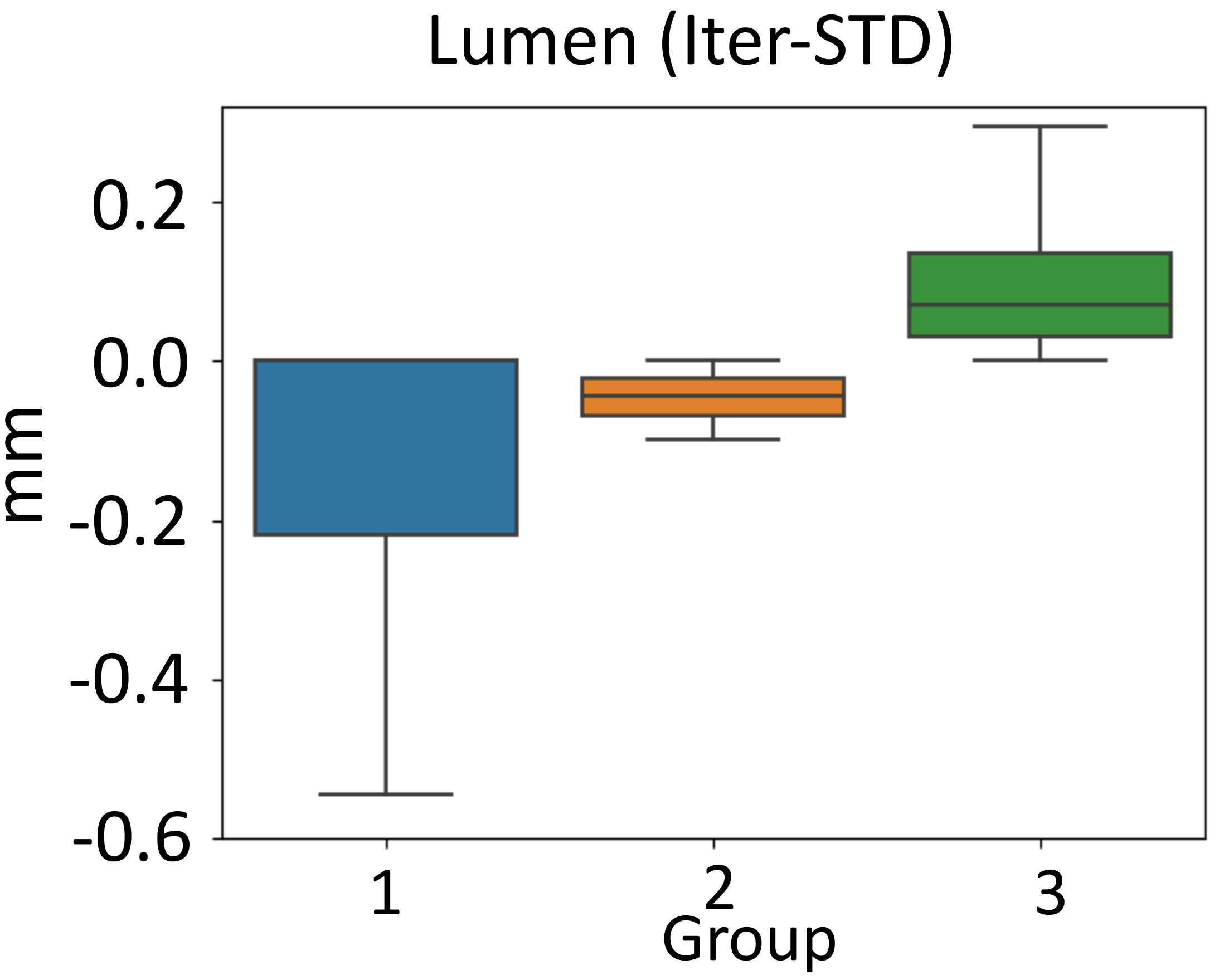}  \\
			
			\includegraphics[width=0.22\textwidth]{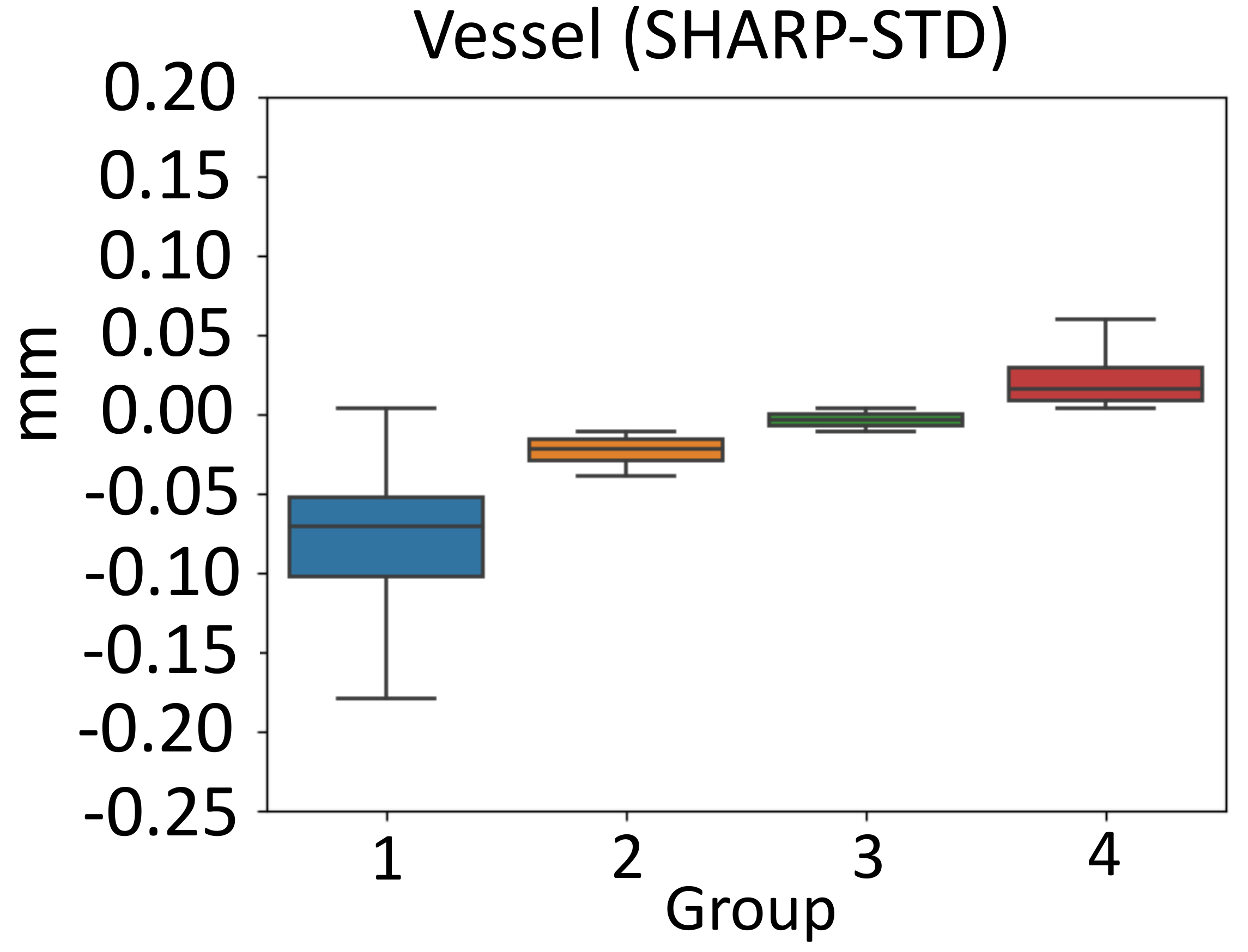}
			&
			\includegraphics[width=0.23\textwidth]{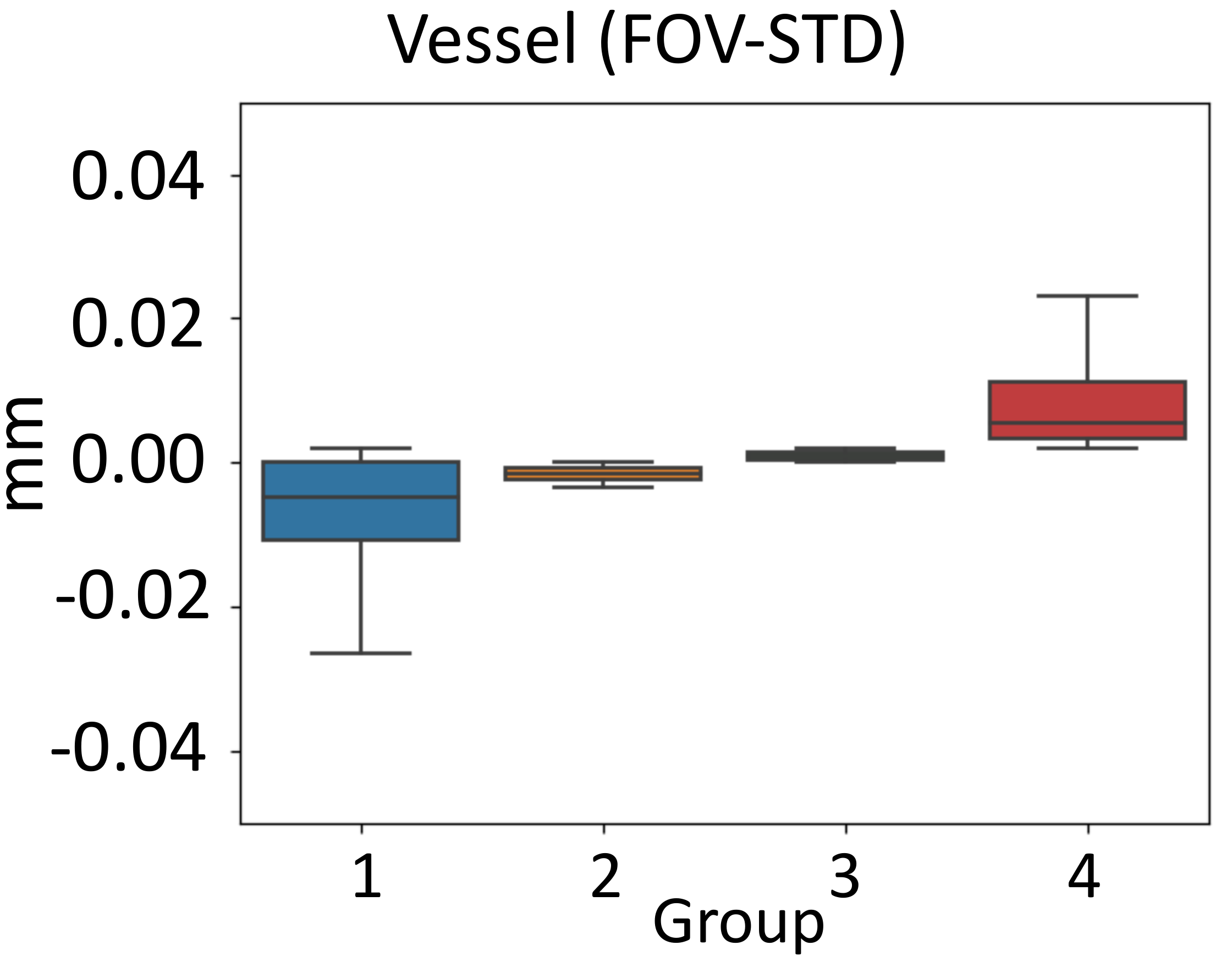} 
			&
			\includegraphics[width=0.23\textwidth]{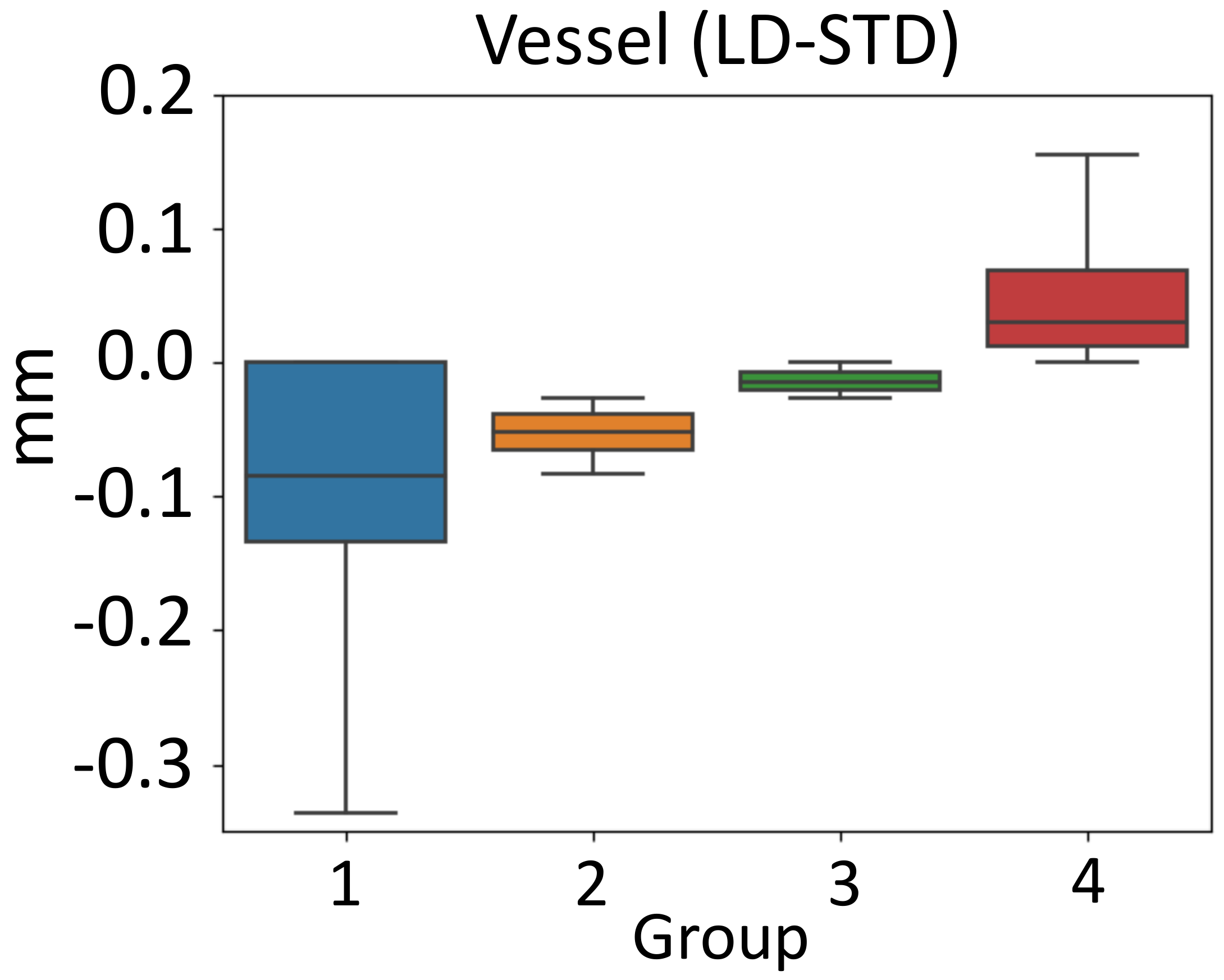} 
			&
			\includegraphics[width=0.23\textwidth]{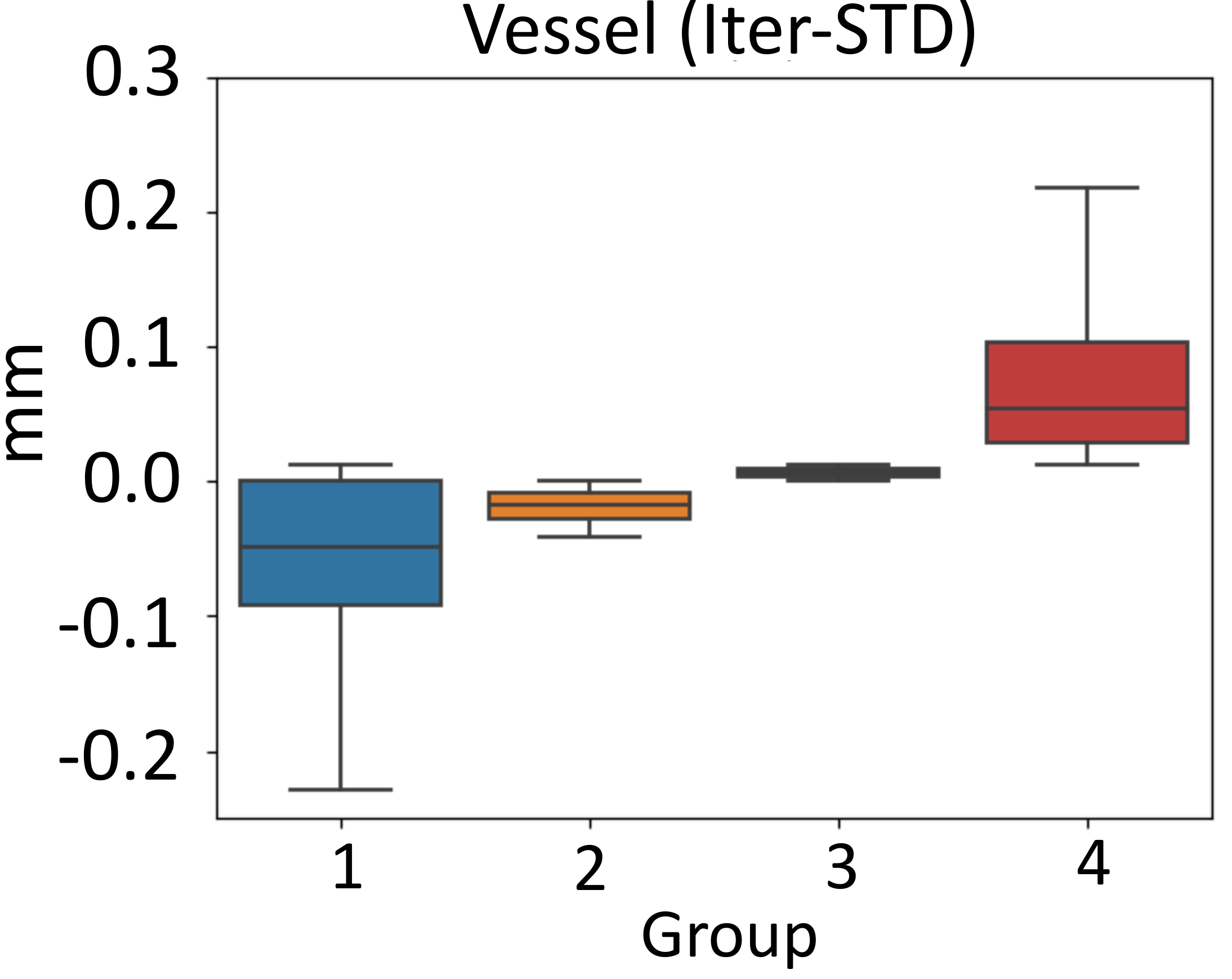}  \\
			
			(a) & (b) & (c) & (d)\\
		\end{tabular}
		\caption{Box-plots of the difference to the reference image (STD, acquired with high dose, standard kernel, and bigger field of view) when measuring wall thickness (top row), airway lumen (middle row), and vessel radius (bottom row), and varying (a) kernel, (b) field of view (FOV), (c) dose, and (d) reconstruction. 
		Four quartiles generated based on the size of the structure interest are used. Number of observations was 253,086 and 3,060,035 for airways and vessels, respectively.}
		\label{SizeComparison} 
	\end{figure*}

	In order to further demonstrate the reliability of the proposed technique, we also considered the difference in the measurement of airway lumen, airway wall thickness, and vessel radius when varying the scanner parameters in comparison to those obtained on STD, sub-divided in four quartiles based on the size of structure of interest. The results are shown in Fig. \ref{SizeComparison}, while a paired Tukey’s t-test analysis of the results showed no significant difference (p$<$0.001) between the groups for all variations.
	
	\subsubsection{FEV1\% in correlation to Pi10}
	
	As shown in Tab. \ref{TabInVivo}a, the Pearson’s correlation coefficients between Pi10 and FEV1\%pred for the presented method (CNR) and VIDA Diagnostics in airway patches extracted from clinical CTs were -0.51 (95\%CI -0.53, -0.48) and -0.33 (-0.36, -0.29), respectively. The correlation between Pi10 and fSAD was 0.401 (CI: 0.38, 0.439) for our approach and 0.0862 (CI: 0.051, 0.1213) for VIDA. In a multivariate analysis, the association of our method with fSAD was positive (beta=15.93 p<0.001), while the measurements obtained with VIDA and fSAD had a weaker and negative association (beta=-11.28, p$<$0.001).
	
	\subsubsection{DLCO in correlation to TBV, BV5, and BV10}
	\begin{table}[t!]
		\caption{Results from indirect in-vivo analysis for (a) airway and (b) vessel. In (a), the Pearson's coefficient for correlation between the Pi10 computed with VIDA and CNR and FEV1\% is reported. For this experiment, 3,038 clinical cases randomly extracted from the COPDGene Phase 1 study were used. Conversely, (b) shows the Pearson's coefficient for the correlation between particle scale and CNR radius and total blood volume (TBV), blood volume of vessels of less than $5 mm^2$ (BV5) and blood volume of vessels of less than $10 mm^2$ (BV10). Here, 1,958 clinical cases randomly extracted from the COPDGene Phase 2 study were considered.}
		\label{TabInVivo}
		\centering
		\begin{tabular}{c c}
			\resizebox{0.3\columnwidth}{!}{%
				\begin{tabular}{c c}
					\hline
					& Correlation [95\% CI]\\
					\hline
					VIDA & -0.33 [-0.36, -0.29]  \\
					CNR & \textbf{-0.51 [-0.53, -0.48]} \\
					\hline
				\end{tabular}
			}
		&
			\resizebox{0.6\columnwidth}{!}{%
				\begin{tabular}{c c c c}
					\hline
					& TBV Corr [95\% CI] & BV5 Corr [95\% CI] & BV10 Corr [95\% CI]\\
					\hline
					Scale & 0.34 [0.29, 0.37] & 0.43 [0.39, 0.47] & 0.40 [0.37, 0.44] \\
					CNR & \textbf{0.45 [0.41, 0.48]} & \textbf{0.45 [0.42, 0.49]} & \textbf{0.47 [0.44, 0.50]} \\
					\hline
				\end{tabular}
			}\\
		(a) & (b)
		\end{tabular}		
	\end{table}
	
	Tab. \ref{TabInVivo}b shows the Pearson’s correlation coefficients obtained between DLCO adjusted for site altitude and TBV, BV5 and BV10 when measuring vessels with the proposed method (CNR) and the particle's scale. Correlations to DLCO of 0.45 (95\%CI 0.41, 0.48), 0.45 (95\%CI 0.42, 0.49), and 0.47 (95\%CI 0.44, 0.50) were obtained with CNR for TBV, BV5, and BV10, respectively, compared to 0.34 (95\%CI 0.39, 0.37), 0.43 (95\%CI 0.39, 0.47), and 0.40 (95\%CI 0.37, 0.44) given by the scale measurements.
	
	\section{Discussion}
	
	In this paper, we presented a novel method for automatic morphology assessment of airways and vessels from CT scans. The use of a neural network in combination with a generative model refined by SimGAN and the customized loss function represent the innovative aspects of this work. 
	
	One of the fundamental limitations for the development of subvoxel sizing methods is the ability to generate realistic and high-resolution representations of 3D bronchial and vascular trees. Realistic anthropomorphic phantoms are complex to generate as several parameters need to be adjusted in order to properly define the topology of the structures of interest and their relationship in the 3D space \cite{jimenez2016automatic}. Moreover, these methods are not ideal for morphology assessment, as an exact size of the structure is difficult to obtain. An important contribution of this paper is given by the introduction of a 2D generative model, which does not require parameter adjustments, and allows for the generation of patches at will with the exact physical dimensions known a priori.
	
	Results obtained with synthetic patches showed a low absolute RE across all measurements. Although a direct comparison is not possible, considering the absolute RE for airways of 1.0 mm the presented CNR for wall thickness assessment outperforms (absolute RE $\sim$6\%) the method proposed in \cite{nakano2000,nakano2002development}, where the wall thickness measured on plastic tubes of 1.0 mm yield to an absolute RE of around 10\%. Also, a comparison to two traditional algorithms for airway measurement shows that our method improves the state-of-the-art, especially for small and complex airways. 
	
	One possible explanation to the bad performance of traditional methods is that they are based on sub-voxel detection of an edge, which has been already described in the literature as a tough problem \cite{reinhardt1997accurate,estepar2006accurate}. This is intimately related to the Nyquist limit theorem and the inability to accurately resolve the size of pulse function whose size is closed to the scanner resolution (i.e. the sampling period of the signal). The problem is further complicated by the variation of the spectral response imposed by different reconstruction kernels and the variation of noise due to different radiation doses.
	
	A validation with structures of different sizes, levels of noise, and smoothing showed that the proposed CNR is not affected by those components, and only sizes below image resolution may determine a small increase of the prediction error. This is also confirmed by the results obtained from clinical cases when varying scanner protocols. Although a direct validation is not possible, due to the complexity to obtain a precise ground truth, the CNR provides very similar results over all scanner variations, demonstrating the high precision of CNR for structure measurement. 
	
\begin{figure*}[t!]
		\centering
		\begin{tabular}{cccc}
			\centering
			\includegraphics[width=0.23\textwidth]{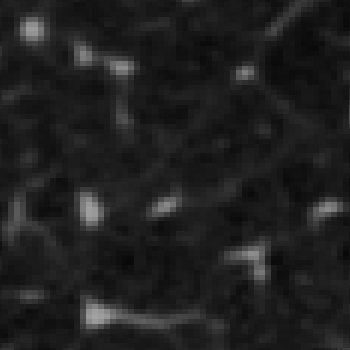}
			&
			\includegraphics[width=0.23\textwidth]{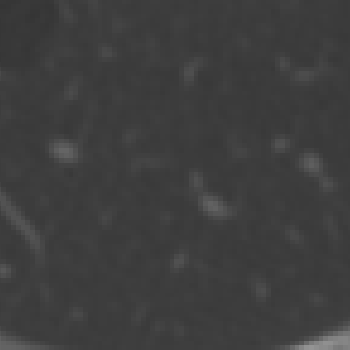}
			&
			\includegraphics[width=0.23\textwidth]{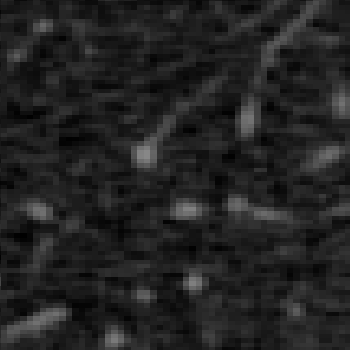}
			&
			\includegraphics[width=0.23\textwidth]{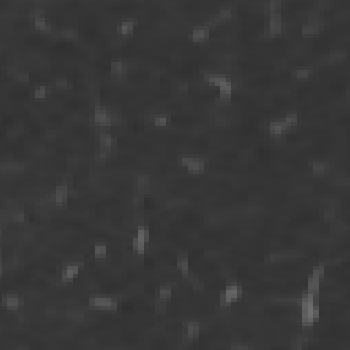} \\
			\includegraphics[width=0.23\textwidth]{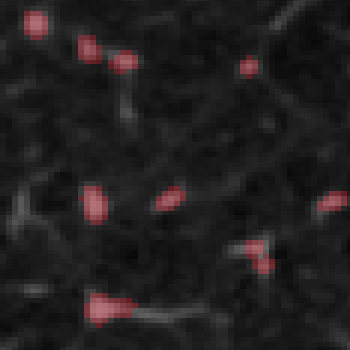}
			&
			\includegraphics[width=0.23\textwidth]{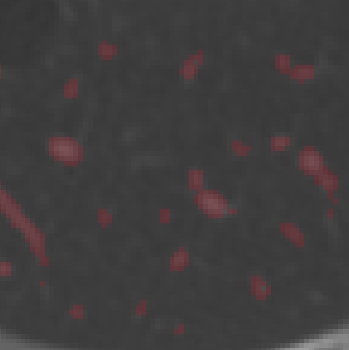}
			&
			\includegraphics[width=0.23\textwidth]{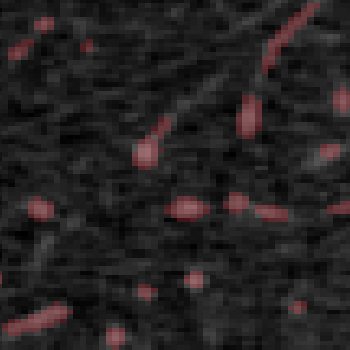}
			&
			\includegraphics[width=0.23\textwidth]{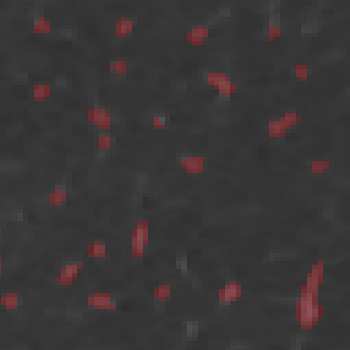} \\
			(a) & (b) & (c) & (d) \\
		\end{tabular}
		\caption{Four examples of vessel radius measurement using the proposed CNR. The first row shows the enlarged part of four CTs of different subjects with varying scanning parameters: (a) STD scan; (b) STD scan with small FOV; (c) STD with low dosage; (d) low dose with iterative reconstruction. The second row shows the same image overlaying the segmentation stenciled from vessel particles with size rescaled based on the radius provided by CNR.}
		\label{QualitativeExampleVessels} 
	\end{figure*}

	As a qualitative example, Fig. \ref{QualitativeExampleVessels} shows the enlarged part of the CT image of four subjects, each of which taken with different scan parameters (STD, lower FOV, LD, and ITER), overlaying the vessel segmentation scaled based on the radius provided by CNR. To this end, we first computed the vessel particles segmentation using \cite{kindlmann2009sampling} and for each particle, we generated a cylindrical stencil with the radius scaled according to an equation that relates the CNR vessel size and the CT's PSF to the actual vessel radius. From this example, the accuracy and precision of the presented algorithm regardless of the scan protocol can be appreciated.
	
	Considering the results on synthetic patches, an increased RE was expected for small wall thicknesses and high smoothing levels (Fig. \ref{RENoiseSmooth}b), due to a blurring effect that may confound the network. Furthermore, a higher RE was also expected for wall thicknesses bigger than 2.0 mm (Fig. \ref{RelativeErrorSize}b), as the network was trained with wall values lower than 1.7 mm, in accordance to the physiological values presented in \cite{weibel1965morphometry}. Indeed, we would have expected an even higher error, furtherly demonstrating the reliability of the proposed CNR even for structures it was not trained with. An important aspect to take into account is that airways with wall thickness bigger than 2.0 mm are very unlikely in humans and would be an indicator of strong pathological conditions, which were not aimed in this work.
	
	The results obtained from the phantom validation are promising, showing that the proposed technique properly measures even small and thin airways, as in case of tube D. Although a variance is present in the RE obtained for the wall thickness measurement of all tubes, this variance is smaller than the one obtained using traditional methods that for some tubes seem to get really confounded. In addition to showing that the proposed algorithm outperforms traditional methods, these results also confirm the reliability of the proposed CNR in accurately measuring airway wall thickness and lumen size regardless of the starting conditions.
	
	The indirect validation on clinical cases showed very encouraging results, indicating that the CNR can be used to accurately assess the bronchial and vascular system morphology regardless of different starting conditions and scanner brand. The obtained CCC indicates that a standard low dose (LD) and low dose with iterative reconstruction (ITER) are the most confounding factors in comparison to STD. This was an expected result as the lower resolution of the images causes blurring effects and noise artifacts that may affect the measurement, especially for small and sensible structures such as the wall thickness.
	
	Results from indirect physiological validation showed that CNR is robust and outperforms standard methods. While for bronchial assessment the correlation obtained with CNR is significantly different (p$<$ 0.001) from VIDA measurements, better explaining lung function decline in smokers, for the venous system the difference to using the particle scale is not statistically significant. However, there is a clear tendency that shows that the CNR approach improves the association between blood volume and DLCO.
	
\begin{figure*}[t!]
		\centering
		\begin{tabular}{ccc}
			\centering
			\includegraphics[width=0.38\textwidth]{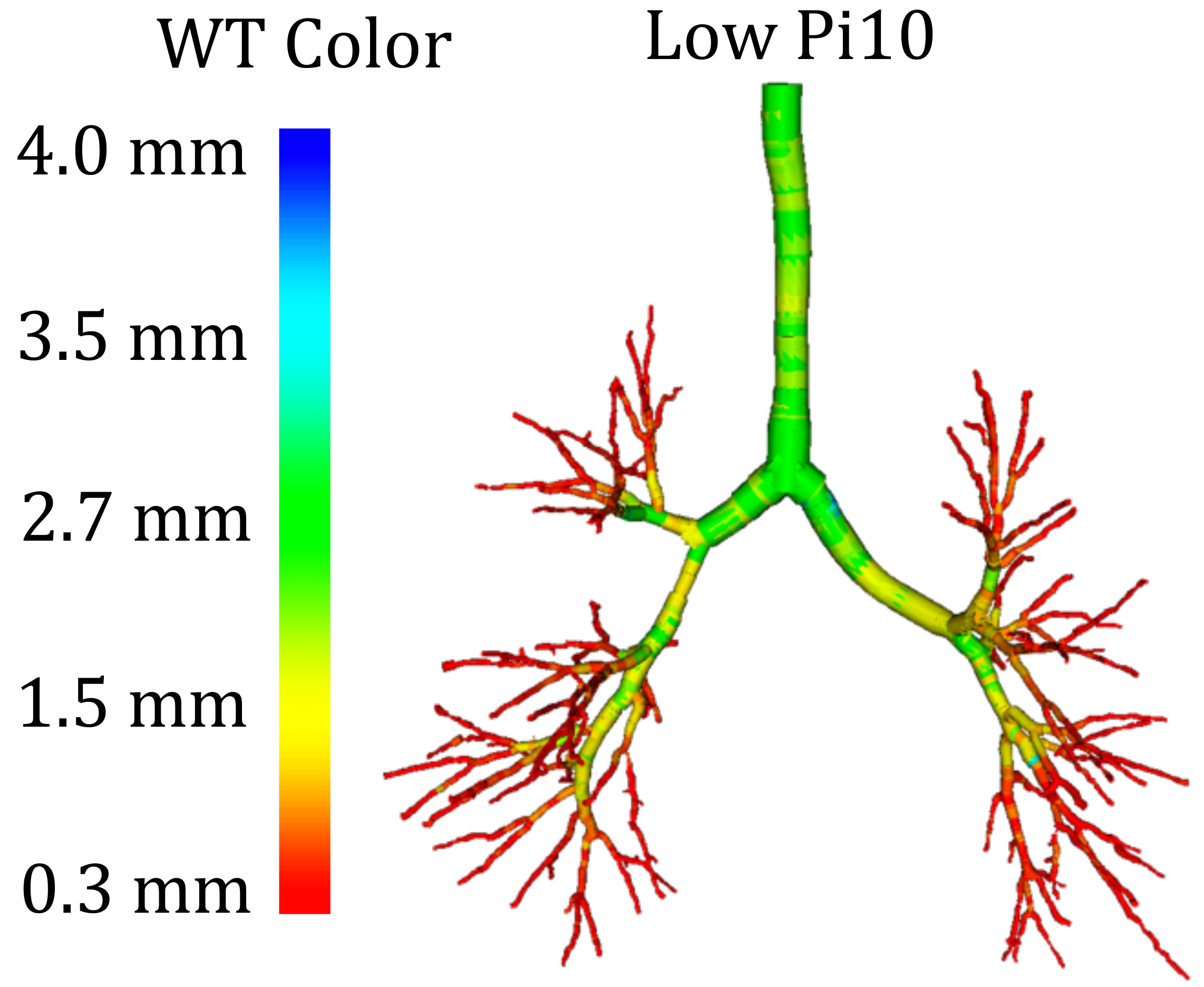}
			&
			\includegraphics[width=0.22\textwidth]{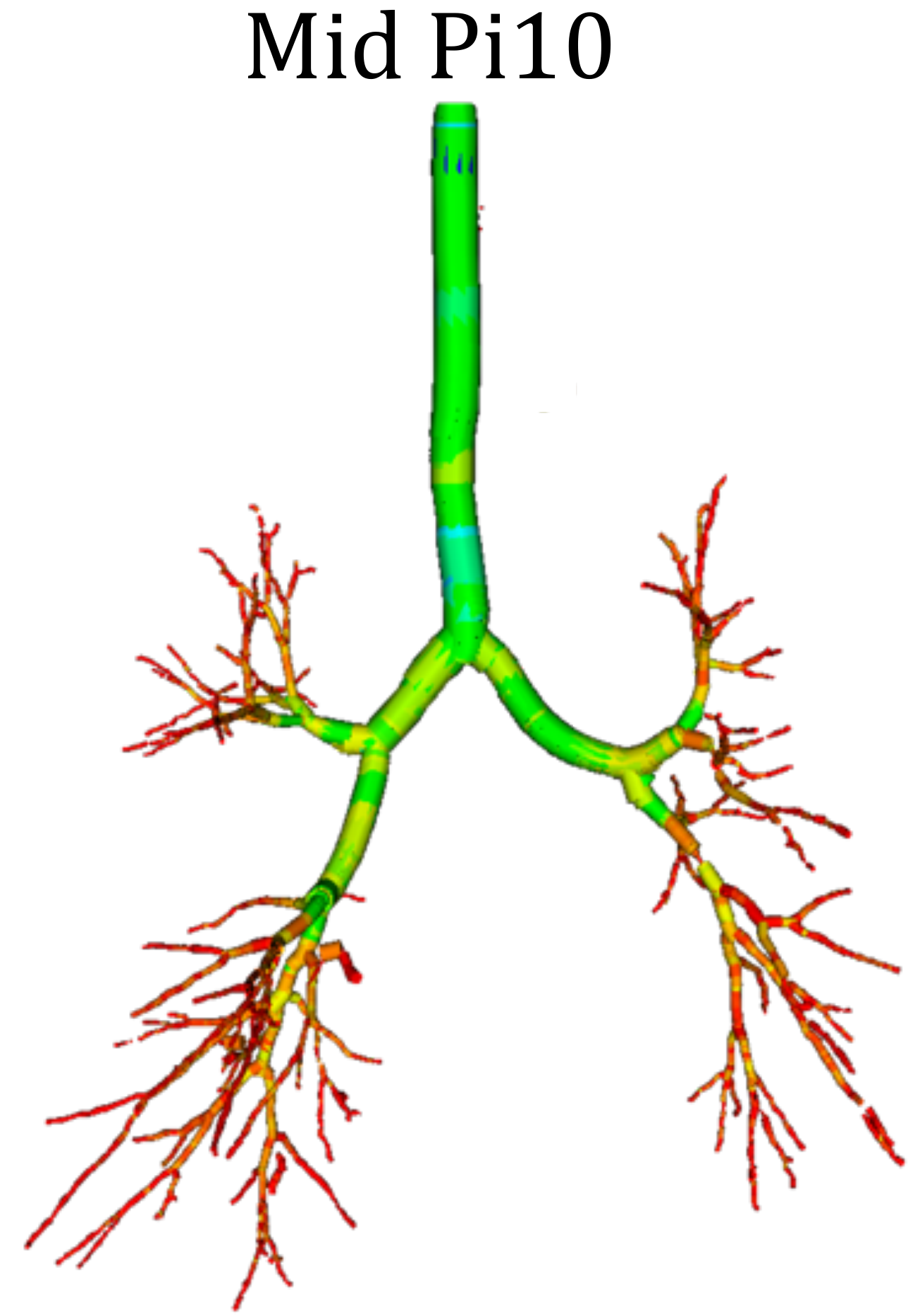} 
			&
			\includegraphics[width=0.225\textwidth]{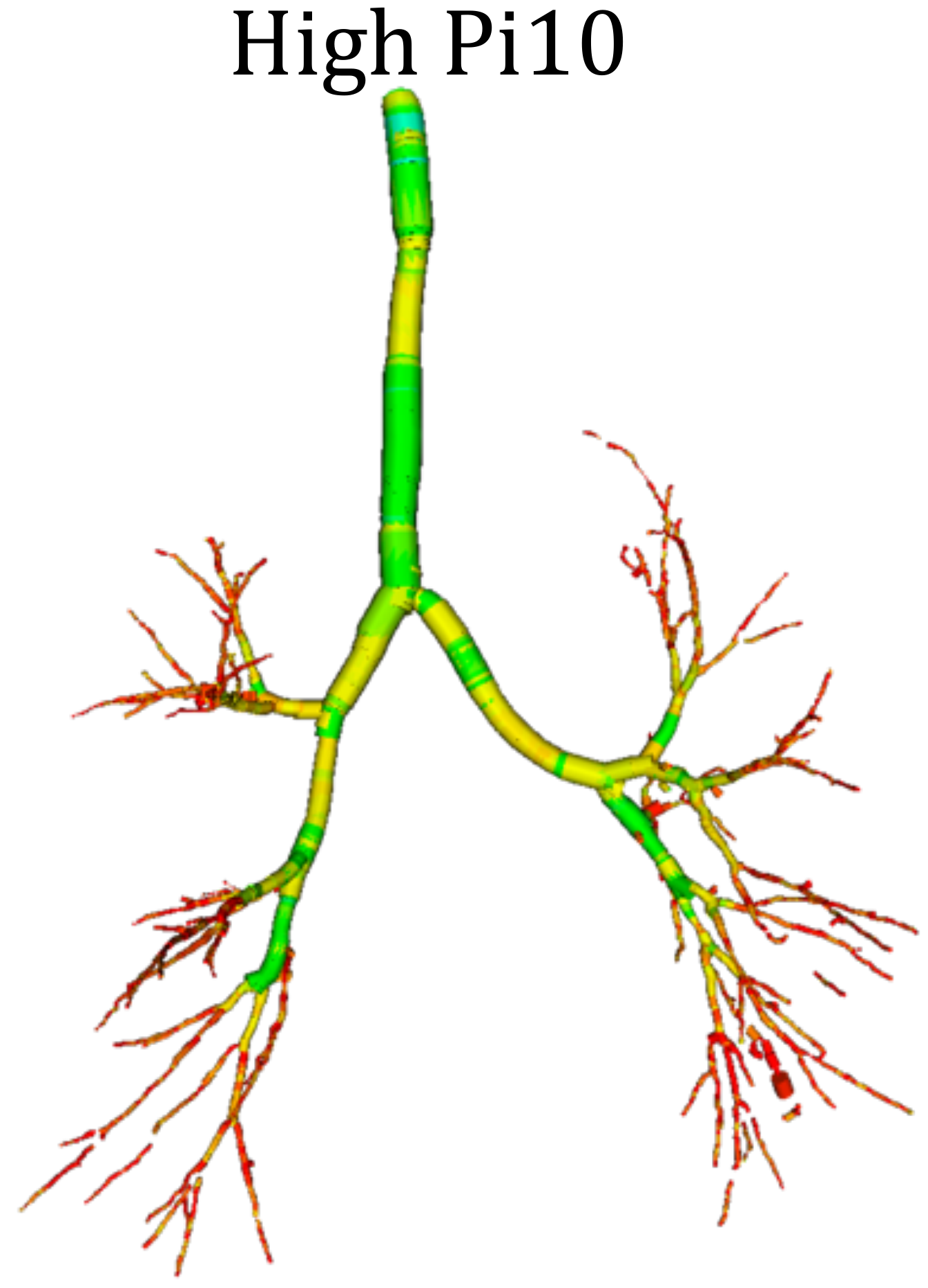} \\
		\end{tabular}
		\begin{tabular}{cc}
			\includegraphics[width=0.49\textwidth]{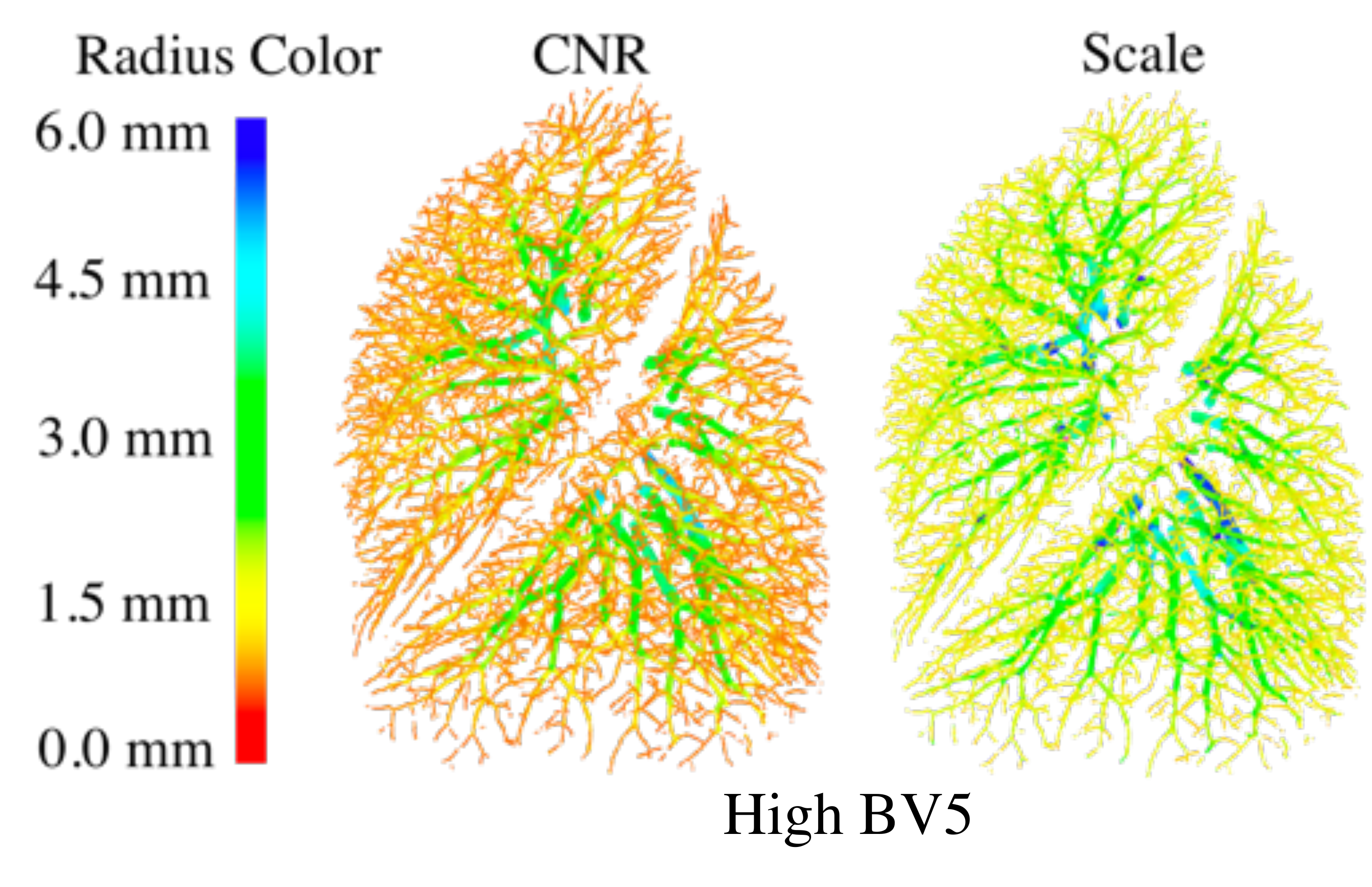}
			&
			\includegraphics[width=0.43\textwidth]{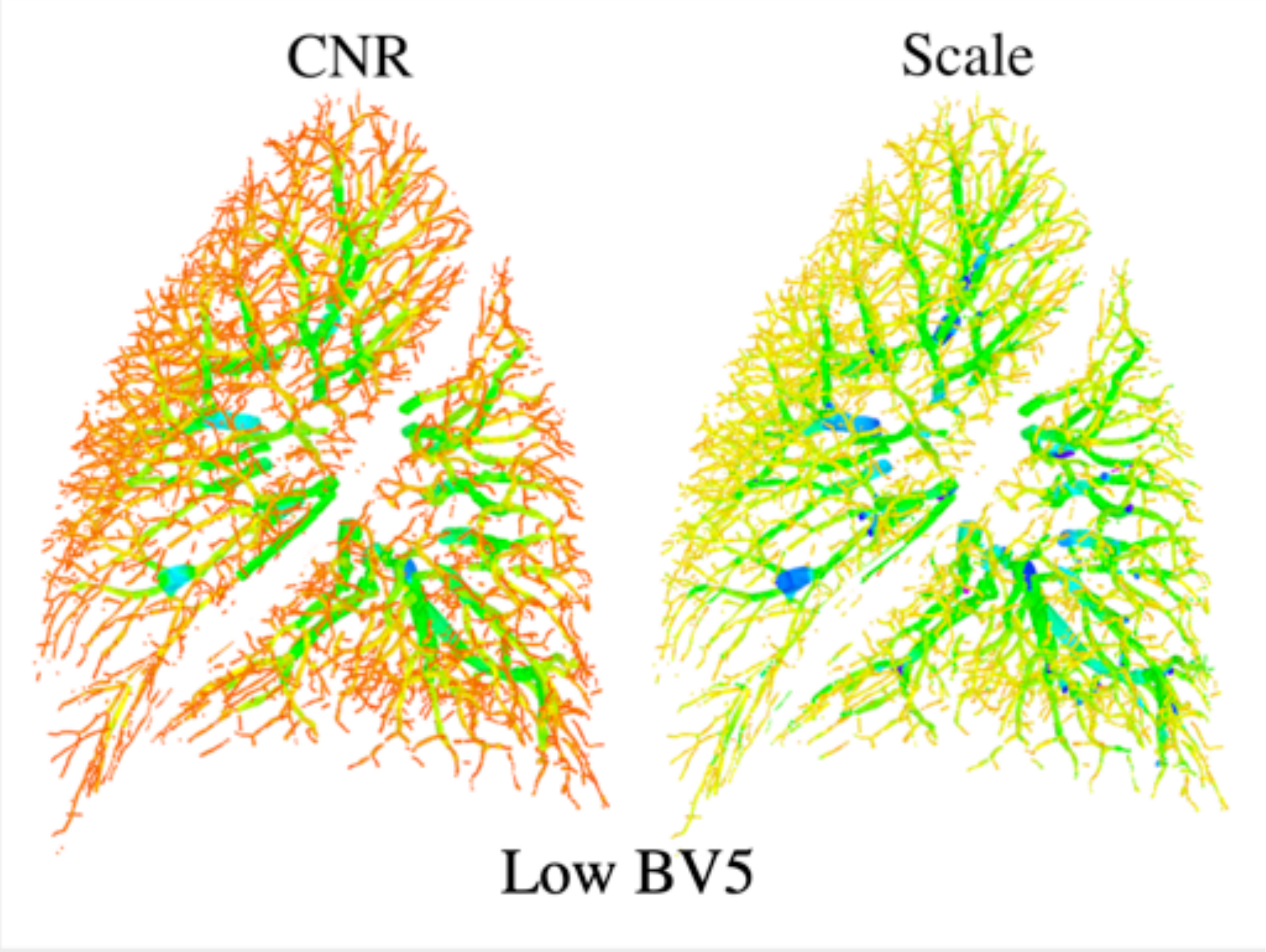}
			\\
		\end{tabular}
		\caption{Examples of airway lumen and wall thickness (top) and vessel radius (bottom) measurement. For airways, each particle size is provided by the lumen measured with CNR, while the color shows the wall thickness (WT) value. Three clinical cases have been selected from the COPDGene study to have three different level of Pi10. For vessels, two clinical cases from the COPDGene study have been chosen with high and low BV5. A comparison between the radius as provided by CNR and by the particle scale \cite{estepar2012computational}, is shown. }
		\label{Pi10airways} 
	\end{figure*}

	As a final demonstration of the reliability of the proposed technique, Fig. \ref{Pi10airways} shows the rendering of bronchial (top) and vascular (bottom) trees, of clinical cases from the COPDGene study with different levels of Pi10 (low, medium, high, as measured by the VIDA  workstation) and BV5 (low and high, measured by using the particle scale technique presented in \cite{estepar2012computational}). For airways, the physical size of each segment is given by the lumen size provided by CNR, while the color of each point represents the wall thickness. For vessels, the different colors represent the radius size, and a comparison between CNR and the particle scale method \cite{kindlmann2009sampling,estepar2012computational} is shown. From this example, we can claim that the measurements provided by CNR respect the natural trend of airway lumen and wall to get smaller when going distally into the lung and that CNR also provides a more realistic measurement than traditional approaches. 
	
\begin{figure*}[t!]
		\centering
		\begin{tabular}{c|c}
			\centering
			\includegraphics[width=0.23\textwidth]{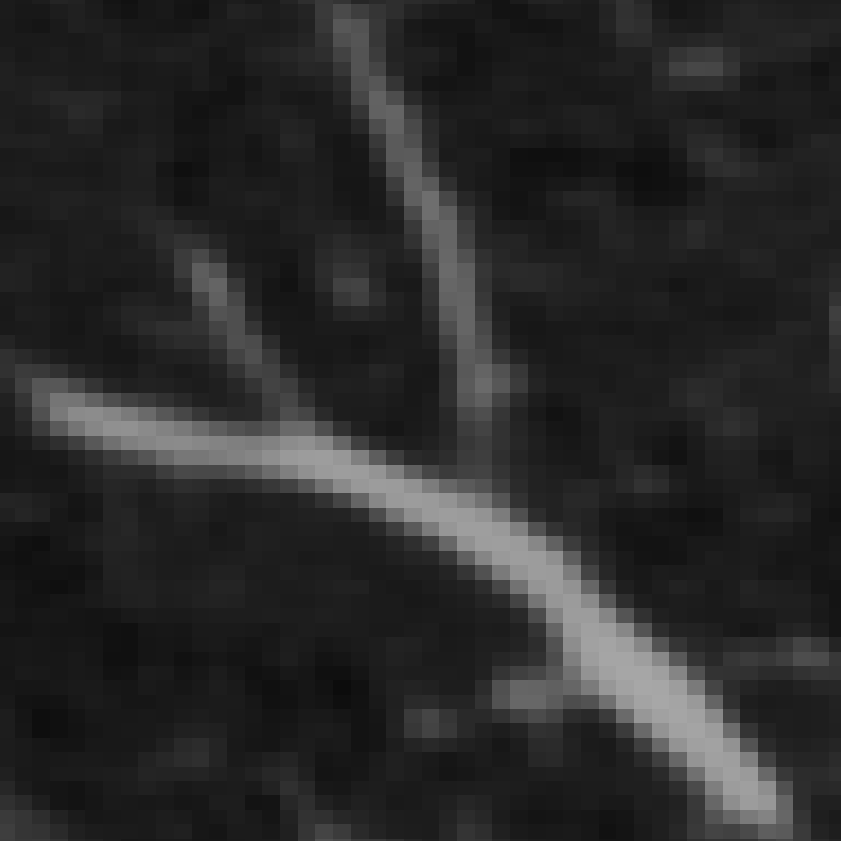}
			\includegraphics[width=0.23\textwidth]{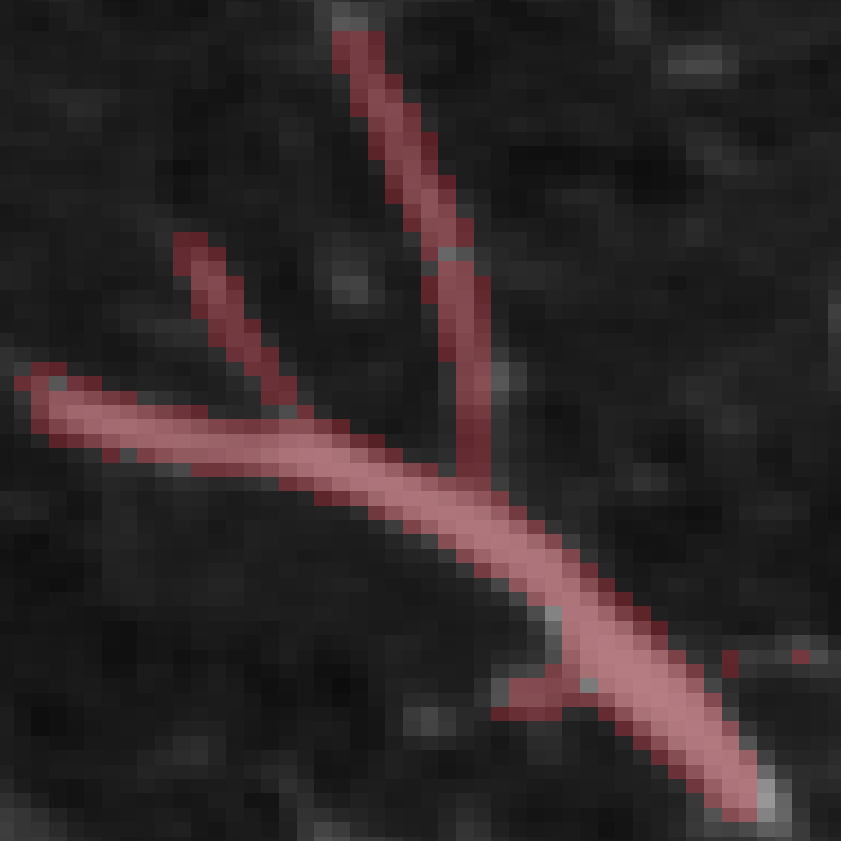} 
			&
			\includegraphics[width=0.23\textwidth]{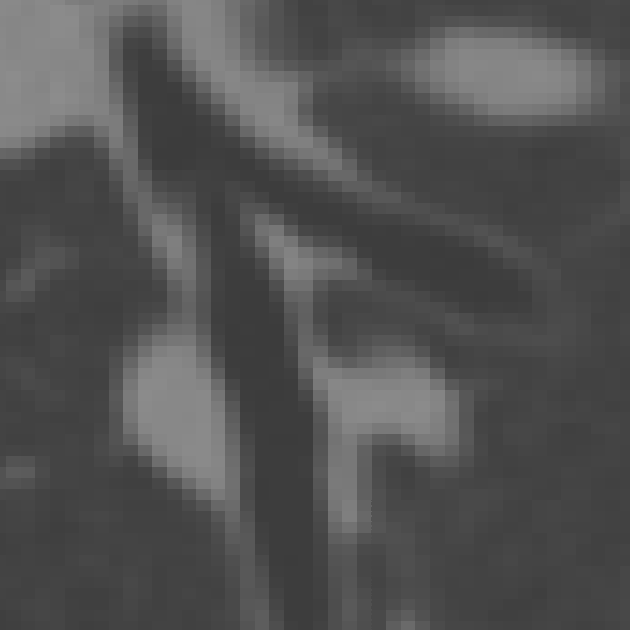}
			\includegraphics[width=0.23\textwidth]{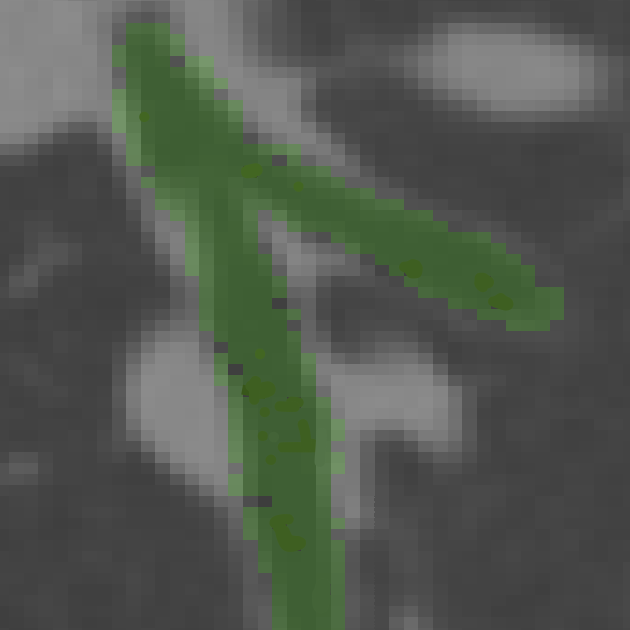}\\
			(a) & (b) \\
		\end{tabular}
		\caption{Two examples of measurement of (a) vessel and (b) airway branching points. Each column presents an enlarged part of the original CT at a branching point location and the same area overlaying the results of the segmentation with the stencil scaled based on the CNR measurement.}
		\label{BranchingPoints} 
	\end{figure*}

	Finally, an important aspect to take into account is that, although no simulated branching points have been provided during the training phase, the CNR accurately measure airways and vessels in those complicated points. An example is shown in Fig. \ref{BranchingPoints}, where two enlarged parts of a CT showing a branching point of a vessel (left) and an airway (right) overlaying the stenciled segmentation, scaled by the CNR measurements, are presented. 
	
	\section{Conclusions}
	
	In this paper, we present an innovative airway and vessel generative model to train a CNR for bronchial and vascular morphological assessment that represents a novel contribution to an unsolved problem that is limiting the application of image-based techniques for the phenotyping of airway and vessels. 
	
	Up-to-date not much work has been proposed in the literature for the assessment of airway and vessel morphology from CT images. The results obtained with the technique proposed in this work on synthetic images, on the airway phantom as well as by indirect in-vivo validation, showed that our method accurately and precisely measures the size of the different structures and may potentially be used for future early diagnosis of lung disorders and to study the level of resistance and obstruction in small airways and vessels.
	
	An important contribution of this paper is given by the introduction of a 2D generative model that, in combination with a generative adversarial network, creates axial-reformatted synthetic structures at will with known physical dimension to train a CNR. The same concept might be extended to other complex problems for which the biological and image-based knowledge is given, such as pulmonary fissures or cerebral vessels.
	
	For future work, we plan to improve the generation of the synthetic model by reducing the level of approximation of the PSF and additive noise. Also, a new adversarial method that may help neural networks to be more domain-invariant when using synthetic data has been recently proposed \cite{ganin2016domain}. We are planning on implement this idea and compare results to those obtained in the present paper. Finally, a method to properly validate image-to-image filtering with GANs will be also investigated. 
	
	\bibliographystyle{unsrt}
	\bibliography{mybibfile_media}

\begin{thebibliography}{10}

\bibitem{niimi2000airway}
Akio Niimi et~al.
\newblock Airway wall thickness in asthma assessed by computed tomography:
  relation to clinical indices.
\newblock {\em AJRCCM}, 162(4):1518--1523, 2000.

\bibitem{santos2003enhanced}
Salud Santos et~al.
\newblock Enhanced expression of vascular endothelial growth factor in
  pulmonary arteries of smokers and patients with moderate chronic obstructive
  pulmonary disease.
\newblock {\em AJRCCM}, 167(9):1250--1256, 2003.

\bibitem{barr2007impaired}
R~G Barr et~al.
\newblock Impaired flow-mediated dilation is associated with low pulmonary
  function and emphysema in ex-smokers: the emphysema and cancer action project
  (emcap) study.
\newblock {\em AJRCCM}, 176(12):1200--1207, 2007.

\bibitem{hogg2013small}
J~C Hogg et~al.
\newblock {Small airway obstruction in COPD: new insights based on micro-CT
  imaging and MRI imaging}.
\newblock {\em Chest}, 143(5):1436--1443, 2013.

\bibitem{charbonnier:2019fw}
Jean-Paul Charbonnier, Esther Pompe, Camille Moore, Stephen Humphries, Bram van
  Ginneken, Barry Make, Elizabeth Regan, James~D Crapo, Eva~M van Rikxoort, and
  David~A Lynch.
\newblock {Airway wall thickening on CT: Relation to smoking status and
  severity of COPD}.
\newblock {\em Respiratory medicine}, 146:36--41, 2019.

\bibitem{awadh1998airway}
Nasser Awadh et~al.
\newblock Airway wall thickness in patients with near fatal asthma and control
  groups: assessment with high resolution computed tomographic scanning.
\newblock {\em Thorax}, 53(4):248--253, 1998.

\bibitem{jacobson1967vascular}
G~Jacobson et~al.
\newblock Vascular changes in pulmonary emphysema: The radiologic evaluation by
  selective and peripheral pulmonary wedge angiography.
\newblock {\em American J of Roentgenology}, 100(2):374--396, 1967.

\bibitem{cordasco1968newer}
E.~M. Cordasco et~al.
\newblock Newer aspects of the pulmonary vasculature in chronic lung disease: a
  comparative study.
\newblock {\em Angiology}, 19(7):399--407, 1968.

\bibitem{rudyanto2014comparing}
Rina~D Rudyanto, Sjoerd Kerkstra, Eva~M Van~Rikxoort, Catalin Fetita,
  Pierre-Yves Brillet, Christophe Lefevre, Wenzhe Xue, Xiangjun Zhu, Jianming
  Liang, {\.I}lkay {\"O}ks{\"u}z, et~al.
\newblock Comparing algorithms for automated vessel segmentation in computed
  tomography scans of the lung: the vessel12 study.
\newblock {\em Medical image analysis}, 18(7):1217--1232, 2014.

\bibitem{bian2018small}
Zijian Bian, Jean-Paul Charbonnier, Jiren Liu, Dazhe Zhao, David~A Lynch, and
  Bram van Ginneken.
\newblock Small airway segmentation in thoracic computed tomography scans: A
  machine learning approach.
\newblock {\em Physics in Medicine \& Biology}, 63(15):155024, 2018.

\bibitem{schwab1993dynamic}
R~J Schwab et~al.
\newblock Dynamic imaging of the upper airway during respiration in normal
  subjects.
\newblock {\em J of Applied Physiology}, 74(4):1504--1514, 1993.

\bibitem{hackx2015chronic}
Maxime Hackx, Elodie Gyssels, Tiago Severo~Garcia, Isabelle De~Meulder,
  St{\'e}phane Alard, Marie Bruyneel, Alain Van~Muylem, Vincent Ninane, and
  Pierre~Alain Gevenois.
\newblock Chronic obstructive pulmonary disease: Ct quantification of airway
  dimensions, numbers of airways to measure, and effect of bronchodilation.
\newblock {\em Radiology}, 277(3):853--862, 2015.

\bibitem{nakano2000}
Y~Nakano et~al.
\newblock Computed tomographic measurements of airway dimensions and emphysema
  in smokers: correlation with lung function.
\newblock {\em AJRCCM}, 162(3):1102--1108, 2000.

\bibitem{nakano2002development}
Yasutaka Nakano, Kenneth~P Whittall, Steve~E Kalloger, Harvey~O Coxson, Julia
  Flint, Peter~D Pare, and John~C English.
\newblock Development and validation of human airway analysis algorithm using
  multidetector row ct.
\newblock In {\em Medical Imaging 2002: Physiology and Function from
  Multidimensional Images}, volume 4683, pages 460--469. International Society
  for Optics and Photonics, 2002.

\bibitem{reinhardt1997accurate}
Joseph~M Reinhardt et~al.
\newblock Accurate measurement of intrathoracic airways.
\newblock {\em IEEE TMI}, 16(6):820--827, 1997.

\bibitem{estepar2006accurate}
R~San Jos\'{e}~Est\'{e}par et~al.
\newblock Accurate airway wall estimation using phase congruency.
\newblock In {\em MICCAI}, pages 125--134. Springer, 2006.

\bibitem{conradi2010measuring}
Susan~H Conradi, Barbara~A Lutey, Jeffrey~J Atkinson, Wei Wang, Robert~M
  Senior, and David~S Gierada.
\newblock Measuring small airways in transverse ct images: Correction for
  partial volume averaging and airway tilt.
\newblock {\em Academic radiology}, 17(12):1525--1534, 2010.

\bibitem{lutey2013accurate}
Barbara~A Lutey, Susan~H Conradi, Jeffrey~J Atkinson, Jie Zheng, Kenneth~B
  Schechtman, Robert~M Senior, and David~S Gierada.
\newblock Accurate measurement of small airways on low-dose thoracic ct scans
  in smokers.
\newblock {\em Chest}, 143(5):1321--1329, 2013.

\bibitem{tschirren2005intrathoracic}
Juerg Tschirren, Eric~A Hoffman, Geoffrey McLennan, and Milan Sonka.
\newblock Intrathoracic airway trees: segmentation and airway morphology
  analysis from low-dose ct scans.
\newblock {\em IEEE transactions on medical imaging}, 24(12):1529, 2005.

\bibitem{petersen2011optimal}
J~Petersen et~al.
\newblock Optimal graph based segmentation using flow lines with application to
  airway wall segmentation.
\newblock In {\em Bienn. International Conference on Inf. Processing in Medical
  Imaging}, pages 49--60. Springer, 2011.

\bibitem{li2005optimal}
Kang Li, Xiaodong Wu, Danny~Z Chen, and Milan Sonka.
\newblock Optimal surface segmentation in volumetric images-a graph-theoretic
  approach.
\newblock {\em IEEE transactions on pattern analysis and machine intelligence},
  28(1):119--134, 2005.

\bibitem{liu2012optimal}
Xiaomin Liu, Danny~Z Chen, Merryn~H Tawhai, Xiaodong Wu, Eric~A Hoffman, and
  Milan Sonka.
\newblock Optimal graph search based segmentation of airway tree double
  surfaces across bifurcations.
\newblock {\em IEEE transactions on medical imaging}, 32(3):493--510, 2012.

\bibitem{estepar2012computational}
R~San Jos\'{e}~Est\'{e}par et~al.
\newblock Computational vascular morphometry for the assessment of pulmonary
  vascular disease based on scale-space particles.
\newblock In {\em ISBI}, pages 1479--1482. IEEE, 2012.

\bibitem{zhai2019automatic}
Zhiwei Zhai, Marius Staring, Irene Hern{\'a}ndez~Gir{\'o}n, Wouter~JH Veldkamp,
  Lucia~J Kroft, Maarten~K Ninaber, and Berend~C Stoel.
\newblock Automatic quantitative analysis of pulmonary vascular morphology in
  ct images.
\newblock {\em Medical physics}, 2019.

\bibitem{lecun2015deep}
Y~LeCun et~al.
\newblock Deep learning.
\newblock {\em nature}, 521(7553):436, 2015.

\bibitem{shrivastava2017learning}
A~Shrivastava et~al.
\newblock Learning from simulated and unsupervised images through adversarial
  training.
\newblock In {\em IEEE CVPR}, pages 2107--2116, 2017.

\bibitem{nardelli2018accurate}
P~Nardelli et~al.
\newblock Accurate measurement of airway morphology on chest {CT} images.
\newblock In {\em Image Analysis for Moving Organ, Breast, and Thoracic
  Images}, pages 335--347. Springer, 2018.

\bibitem{galban2012computed}
Craig~J Galb{\'a}n, Meilan~K Han, Jennifer~L Boes, Komal~A Chughtai, Charles~R
  Meyer, Timothy~D Johnson, Stefanie Galb{\'a}n, Alnawaz Rehemtulla, Ella~A
  Kazerooni, Fernando~J Martinez, et~al.
\newblock Computed tomography--based biomarker provides unique signature for
  diagnosis of copd phenotypes and disease progression.
\newblock {\em Nature medicine}, 18(11):1711, 2012.

\bibitem{weibel1965morphometry}
E~R Weibel.
\newblock {\em Morphometry of the human lung}.
\newblock Springer, 1965.

\bibitem{san2008three}
Raul San Jos\'{e}~Est\'{e}par et~al.
\newblock Three-dimensional airway measurements and algorithms.
\newblock {\em Proceedings of the Am Thoracic Soc}, 5(9):905--909, 2008.

\bibitem{schwarzband2005point}
G~Schwarzband and N~Kiryati.
\newblock The point spread function of spiral ct.
\newblock {\em Physics in Medicine \& Biology}, 50(22):5307, 2005.

\bibitem{VegasSanchezFerrero:2017vt}
Gonzalo Vegas-S{\'a}nchez-Ferrero et~al.
\newblock {Statistical characterization of noise for spatial standardization of
  CT scans: Enabling comparison with multiple kernels and doses.}
\newblock {\em MedIA}, 40:44--59, August 2017.

\bibitem{regan2011genetic}
E~A Regan et~al.
\newblock Genetic epidemiology of copd (copdgene) study design.
\newblock {\em COPD}, 7(1):32--43, 2011.

\bibitem{ross2009lung}
J~C Ross et~al.
\newblock Lung extraction, lobe segmentation and hierarchical region assessment
  for quantitative analysis on high resolution computed tomography images.
\newblock In {\em MICCAI}, pages 690--698. Springer, 2009.

\bibitem{nardelli2017ct}
P~Nardelli et~al.
\newblock {CT Image Enhancement for Feature Detection and Localization}.
\newblock In {\em MICCAI}, pages 224--232. Springer, 2017.

\bibitem{kindlmann2009sampling}
G~L Kindlmann et~al.
\newblock Sampling and visualizing creases with scale-space particles.
\newblock {\em IEEE TVCG}, 15(6), 2009.

\bibitem{chollet2015keras}
F~Chollet et~al.
\newblock Keras.
\newblock \url{https://keras.io}, 2015.

\bibitem{abadi2016tensorflow}
M~Abadi et~al.
\newblock Tensorflow: Large-scale machine learning on heterogeneous distributed
  systems.
\newblock {\em arXiv preprint arXiv:1603.04467}, 2016.

\bibitem{avants2009advanced}
B.~B. Avants et~al.
\newblock Advanced normalization tools (ants).
\newblock {\em Insight J}, 2:1--35, 2009.

\bibitem{estepar2013computed}
San~Jos\'{e} Est\'{e}par et~al.
\newblock Computed tomographic measures of pulmonary vascular morphology in
  smokers and their clinical implications.
\newblock {\em AJRCCM}, 188(2):231--239, 2013.

\bibitem{jimenez2016automatic}
Daniel Jimenez-Carretero, Raul San Jose~Estepar, Mario~Diaz Cacio, and Maria~J
  Ledesma-Carbayo.
\newblock Automatic synthesis of anthropomorphic pulmonary ct phantoms.
\newblock {\em PloS one}, 11(1):e0146060, 2016.

\bibitem{ganin2016domain}
Y~Ganin et~al.
\newblock Domain-adversarial training of neural networks.
\newblock {\em The J of Machine Learning Research}, 17(1):2096--2030, 2016.

\end{thebibliography}
	
\end{document}